\documentclass[12pt]{article}
\pdfoutput=1

\usepackage{epsfig}
\usepackage{psfrag}
\usepackage{latexsym}
\usepackage{indentfirst}
\usepackage{fancyhdr}
\usepackage{amssymb}
\usepackage{amsmath}
\usepackage{amsfonts}
\usepackage{pifont}
\usepackage{cite}
\usepackage{bbold}
\usepackage{color}
\usepackage{colordvi}
\usepackage{fancybox}
\usepackage[footnotesize]{caption2}
\usepackage{graphicx}
\usepackage[center,footnotesize,hang]{subfigure}
\usepackage{bbm}
\usepackage{url}
\usepackage{multirow}
\usepackage{array}
\usepackage{arydshln}
\usepackage{ulem}
\usepackage{enumitem}
\newcommand{\PreserveBackslash}[1]{\let\temp=\\#1\let\\=\temp}
\newcolumntype{C}[1]{>{\PreserveBackslash\centering}p{#1}}
\newcolumntype{R}[1]{>{\PreserveBackslash\raggedleft}p{#1}}
\newcolumntype{L}[1]{>{\PreserveBackslash\raggedright}p{#1}}
\allowdisplaybreaks  

\newcommand{\bq}{\begin{eqnarray}}
\newcommand{\nq}{\end{eqnarray}}

\newcommand{\cmark}{\ding{51}}%
\newcommand{\xmark}{\ding{55}}%

\makeatletter
\@addtoreset{equation}{section}
\makeatother

\begin{document}

\title{\hfill ~\\[0mm]
        \textbf{Lepton Mixing in $A_5$ Family Symmetry and Generalized CP}}

\date{}

\author{\\[1mm]Cai-Chang Li\footnote{E-mail: {\tt lcc0915@mail.ustc.edu.cn}}~,~Gui-Jun Ding\footnote{E-mail: {\tt dinggj@ustc.edu.cn}}\\ \\
\it{\small Department of Modern Physics, University of Science and
    Technology of China,}\\
  \it{\small Hefei, Anhui 230026, China}\\[4mm] }
\maketitle

\begin{abstract}

We study lepton mixing patterns which can be derived from the $A_5$ family symmetry and generalized CP. We find five phenomenologically interesting mixing patterns for which one column of the PMNS matrix is $(\sqrt{\frac{5+\sqrt{5}}{10}},\frac{1}{\sqrt{5+\sqrt{5}}},\frac{1}{\sqrt{5+\sqrt{5}}})^{T}$ (the first column of the golden ratio mixing), $(\sqrt{\frac{5-\sqrt{5}}{10}},\frac{1}{\sqrt{5-\sqrt{5}}},\frac{1}{\sqrt{5-\sqrt{5}}})^{T}$ (the second column of the golden ratio mixing), $(1,1,1)^{T}/\sqrt{3}$ or $(\sqrt{5}+1,-2,\sqrt{5}-1)^{T}/4$. The three lepton mixing angles are determined in terms of a single real parameter $\theta$, and agreement with experimental data can be achieved for certain values of $\theta$. The Dirac CP violating phase is predicted to be trivial or maximal while Majorana phases are trivial. We construct a supersymmetric model based on $A_5$ family symmetry and generalized CP. The lepton mixing is exactly the golden ratio pattern at leading order, and the mixing patterns of case III and case IV are reproduced after higher order corrections are considered.

\end{abstract}
\thispagestyle{empty}
\vfill

\newpage
\setcounter{page}{1}

\section{\label{sec:1}Introduction}

In the standard three flavor neutrino oscillation paradigm, lepton flavor mixing is described by the so-called Pontecorvo-Maki-Nakagawa-Sakata (PMNS) matrix $U_{PMNS}$ which is a $3\times3$ unitary matrix~\cite{Agashe:2014kda}. $U_{PMNS}$ contains three mixing angles $\theta_{12}$, $\theta_{13}$, $\theta_{23}$ and one Dirac CP violating phase $\delta_{CP}$. There are two more Majorana CP phases if neutrinos are Majorana particles. With the measurement of the last mixing angle $\theta_{13}$ by Daya Bay~\cite{An:2012eh}, RENO~\cite{Ahn:2012nd} and Double Chooz~\cite{Abe:2011fz}, all three lepton mixing angles have been measured with good accuracy in neutrino oscillation experiments~\cite{Capozzi:2013csa,Forero:2014bxa,Gonzalez-Garcia:2014bfa}. Recently T2K has reported a slight preference for $\delta_{CP}$ close to $3\pi/2$~\cite{Abe:2013hdq}, when the data are combined with the measurements of the reactor experiments. The present global fit to neutrino data also indicates nontrivial values of $\delta_{CP}$~\cite{Capozzi:2013csa,Forero:2014bxa,Gonzalez-Garcia:2014bfa}. However, the values of the both Majorana phases are unknown so far. Search for leptonic CP violation via the determination of $\delta_{CP}$ is one of the major goals of future long-baseline experiments such as the proposed LBNE~\cite{Adams:2013qkq}, LBNO~\cite{::2013kaa} and HyperKamiokande~\cite{Kearns:2013lea}.

Now it is established that both neutrino and charged lepton mass matrices have residual flavor symmetries determined by lepton flavor mixing, and vice versa residual flavor symmetries in the mass matrices can determine the lepton mixing matrix up to Majorana phases and permutations of rows and columns~\cite{Lam:2007qc}. Inspired by the fact, it is assumed that the residual flavor symmetries arise from a underlying flavor symmetry group $G_{f}$ which is usually chosen to be a finite and non-abelian subgroup of $U(3)$. In the past years, much effort has been devoted to the discussion of lepton flavor mixing from a discrete flavor symmetry $G_{f}$ and its breaking~\cite{Altarelli:2010gt}. It is surprising that the mixing patterns achievable in this way are quite limited, the PMNS matrix can only be of the trimaximal form to accommodate the experimental data and the Dirac phase is trivial~\cite{Fonseca:2014koa}.

Beside residual flavor symmetries, neutrino and charged lepton mass matrices have residual CP symmetries~\cite{Chen:2014wxa,Everett:2015oka}. Analogous to residual flavor symmetries, residual CP symmetries also impose strong constraints on the mass matrices and therefore allow us to reconstruct the lepton mixing matrix~\cite{Chen:2014wxa}. A simple example is the well-known $\mu-\tau$ reflection symmetry~\cite{Harrison:2002kp} which predicts maximal atmospheric mixing angle $\theta_{23}$ and maximal Dirac CP phase. It is natural to conjecture that there is a CP symmetry $H_{CP}$ (also called generalized CP symmetry) at high energy scale, which is broken down to the residual CP symmetries at low energy. Note that the effects of CP symmetry on the fermion mass matrix have been discussed several decades ago~\cite{Ecker:1981wv,Grimus:1995zi}.

Recently it is proposed to predict the lepton mixing angles and CP phases by combining a discrete flavor symmetry $G_{f}$ with a CP symmetry $H_{CP}$~\cite{Feruglio:2012cw,Holthausen:2012dk}. $H_{CP}$ has to be compatible with $G_{f}$ such that the possible forms of the CP transformations are strongly constrained. It has been proved that the mathematical structure of the group comprising $G_f$ and $H_{CP}$ is in general a semi-direct product $G_{f}\rtimes H_{CP}$~\cite{Feruglio:2012cw}.
In this framework, the flavor symmetry $G_{f}$ is broken down to different abelian subgroups $G_{\nu}$ and $G_{l}$ in the neutrino and charged lepton sectors respectively, and $H_{CP}$ is broken into residual CP symmetry $H^{\nu}_{CP}$ and  $H^{l}_{CP}$ respectively. The mismatch between the remnant symmetries $G_{\nu}\rtimes H^{\nu}_{CP}$ and $G_{l}\rtimes H^{l}_{CP}$ generates the PMNS matrix. Neutrinos are generically assumed to be Majorana particles. As a consequence, $G_{\nu}$ can only be a $K_4$ or $Z_2$ subgroup of $G_{f}$. In the case that $G_{\nu}=K_4$ and $G_{l}$ is capable of distinguishing the three generations (i.e.,$G_{l}$ can not be smaller than $Z_3$), all lepton mixing parameters including the Majorana phases would be completely fixed by residual symmetries once the CP symmetry is considered. In this way, both Dirac and Majorana CP violating phases are found to be conserved in the context of $\Delta(6n^2)$ family symmetry combined with generalized CP~\cite{King:2014rwa}. Recently a bottom up analysis of the remnant $K_4$ flavor symmetry and CP symmetry in the neutrino sector has been performed~\cite{Everett:2015oka}. On the other hand, if $G_{\nu}=Z_2$ and a CP symmetry is preserved in the neutrino sector, only one column of the PMNS matrix can be fixed and all lepton mixing parameters depend on one single real parameter $\theta$. Along this line, the family symmetries $A_{4}$~\cite{Ding:2013bpa}, $S_{4}$~\cite{Feruglio:2012cw, Ding:2013hpa,Li:2014eia,Feruglio:2013hia, Luhn:2013lkn, Li:2013jya}, $T^\prime$~\cite{Girardi:2013sza}, $\Delta(48)$~\cite{Ding:2013nsa}, $\Delta(96)$~\cite{Ding:2014ssa}, $\Delta(3n^{2})$~\cite{Hagedorn:2014wha} and $\Delta(6n^{2})$~\cite{Hagedorn:2014wha,Ding:2014ora} which are combined with the corresponding generalized CP symmetries have been investigated already. It is found that CP phases can only be trivial or maximal in simple family symmetries $A_{4}$~\cite{Ding:2013bpa} and $S_{4}$~\cite{Feruglio:2012cw, Ding:2013hpa,Feruglio:2013hia, Luhn:2013lkn, Li:2013jya,Li:2014eia} while $\Delta(48)$~\cite{Ding:2013nsa} and $\Delta(96)$~\cite{Ding:2014ssa} (also $\Delta(3n^{2})$ and $\Delta(6n^{2})$~\cite{Hagedorn:2014wha,Ding:2014ora}) family symmetries admit mixing patterns in which all CP phases nontrivially depend on the parameter $\theta$. In addition, some models with both flavor and CP symmetries have been constructed~\cite{Ding:2013hpa,Li:2014eia,Feruglio:2013hia, Luhn:2013lkn, Li:2013jya,Ding:2013nsa}. Last but not least, if remnant symmetries in the neutrino and charged lepton sectors are $K_{4}\rtimes H^{\nu}_{CP}$ and $Z_{2}\times H^{l}_{CP}$ respectively, then the PMNS matrix is also predicted in terms of the parameter $\theta$ and one row instead of one column would be fixed~\cite{Li:2014eia,Ding:2014ora}.

It is known that the flavor symmetry group should be of the von Dyck type~\cite{Hernandez:2012ra}. The finite von Dyck groups include $S_3$, $A_4$, $S_4$, $A_5$ and dihedral groups~\cite{johnson}. Since $S_3$ and dihedral groups don't have irreducible three dimensional representations, they are not suitable as flavor symmetry otherwise two mixing angles would vanish. The phenomenological consequences of $A_4$ and $S_4$ flavor symmetries combined with generalized CP have been studied~\cite{Ding:2013bpa,Feruglio:2012cw, Ding:2013hpa,Feruglio:2013hia, Luhn:2013lkn, Li:2013jya,Li:2014eia}. In the present work, we shall investigate the $A_5$ flavor symmetry and CP symmetry. We shall perform a model independent analysis of possible lepton flavor mixing obtained from breaking of the original symmetry $A_5\rtimes H_{CP}$. We find five phenomenologically interesting mixing patterns summarized in Table~\ref{tab:best_fitting}. The three mixing angles turn out to depend on only one free parameter $\theta$ and good agreement with their measured values can be achieved for certain values of $\theta$, the Dirac CP phase is conserved or maximal and the Majorana CP phases are trivial. Furthermore, we construct a model based on $A_{5}\rtimes H_{CP}$. The lepton mixing is exactly the golden ratio (GR) texture at leading order (LO). A non-zero $\theta_{13}$ is generated by the next-to-leading-order (NLO) corrections, and the mixing patterns of cases III and IV discussed in the model independent analysis are generated.

The layout of the rest of this paper is as follows. In section~\ref{sec:2}, the physical CP transformations compatible with the $A_5$ family symmetry are found. In section~\ref{sec:3}, we perform a model independent analysis of possible lepton mixing patterns achievable from the underlying symmetry group $A_{5}\rtimes H_{CP}$. In section~\ref{sec:4}, we present our $A_{5}\rtimes H_{CP}$ model, the LO structure, vacuum alignment and the NLO corrections are discussed. Section~\ref{sec:Conclusion} concludes the paper. In Appendix~\ref{sec:A}, we review the group theory of $A_5$ and the Clebsch-Gordan coefficients in our working basis are reported. In Appendix~\ref{sec:B}, we present the possible mixing patterns arising from the $A_5$ flavor symmetry without CP symmetry, where the residual flavor
symmetry in the neutrino sector is either Klein or $Z_2$ subgroup of $A_5$. Compared with section~\ref{sec:3}, we see that generalized CP is really a powerful method of predicting CP phases as well as lepton mixing angles.

\section{\label{sec:2} Approach}

Both family symmetry and CP symmetry acts on the flavor space in a non-trivial way, and the interplay between them should be carefully treated. In order to consistently combine a family symmetry $G_{f}$ with a CP symmetry which is represented by unitary CP transformation matrix $X$, $X$ must be related to an automorphism $u: G_{f}\rightarrow G_{f}$. To be precise, the CP transformation $X$ should be a solution to the consistency equation~\cite{Feruglio:2012cw,Holthausen:2012dk}
\begin{equation}
\label{eq:cons_equ}X\rho^{\ast}(g)X^{-1}=\rho\left(u(g)\right),\qquad \forall g\in G_{f}\,,
\end{equation}
where $\rho$ is a representation of $G_{f}$ with $\rho:G\rightarrow GL(N, \mathbb{C})$, and it is generally reducible. We can easily check that the automorphism associated with $\rho(h)X$ for any $h\in G_{f}$ is an composition of $u$ and an inner automorphism $\mu_{h}:$ $g\rightarrow hgh^{-1}$ with $h, g\in G_{f}$~\cite{Ding:2014ssa,Li:2014eia}. Therefore the effects of inner automorphism can be easily included, and it is equivalent to a family symmetry transformation. As a consequence, we could firstly focus on the  the outer automorphism of $G_f$. Furthermore, it has been shown that $u$ has to a class-inverting automorphism for $X$ to be a physical CP transformation~\cite{Chen:2014tpa}. In other words, $u$ should map each irreducible representation $\mathbf{r}$ of $G_{f}$ into its own complex conjugate. Hence the consistency condition in Eq.~\eqref{eq:cons_equ} takes a more restricted form:
\begin{equation}
\label{eq:cons_equ_restr}X_{\mathbf{r}}\rho^{\ast}_{\mathbf{r}}(g)X^{-1}_{\mathbf{r}}=\rho_{\mathbf{r}}\left(u(g)\right),\qquad \forall g\in G_{f}\,,
\end{equation}
where the subscript ``$\mathbf{r}$'' refers to the representation space acted on. The CP transformation $X$ in Eq.~\eqref{eq:cons_equ} is given by the direct sum of the $X_{\mathbf{r}}$ corresponding to the particle content of the model. Notice that the consistency conditions of Eq.~\eqref{eq:cons_equ_restr} can also be derived from the requirement that the Lagrangian is invariant under both CP symmetry and flavor symmetry~\cite{Branco:2015hea}.

In the present work, we are interested in the family symmetry $G_{f}=A_5$. The group theory of $A_5$, its representation and all the Clebsch-Gordan coefficients are reported in Appendix~\ref{sec:A}. The structure of the automorphism group of $A_5$ is quite simple and is very clear in mathematica.
\begin{eqnarray}
\nonumber&&\textrm{Z}(A_5)\cong Z_1,\qquad \textrm{Aut}(A_5)\cong S_5,\\
&&\textrm{Inn}(A_5)\cong A_5,\qquad \textrm{Out}(A_5)\cong Z_2\,,
\end{eqnarray}
where $\textrm{Z}(A_5)$, $ \textrm{Aut}(A_5)$, $\textrm{Inn}(A_5)$ and $\textrm{Out}(A_5)$ denote the center, automorphism group, inner automorphism group and outer automorphism group of $A_5$ respectively. We see that the outer automorphism group of $A_5$ is isomorphic to $Z_2$. Consequently there is only one non-trivial outer automorphism $\mathfrak{u}$ with
\begin{equation}
S\stackrel{\mathfrak{u}}{\longmapsto}S,\qquad T\stackrel{\mathfrak{u}}{\longmapsto}\left(ST^3\right)^2\,.
\end{equation}
The order of $\mathfrak{u}$ is really 2, i.e., $\mathfrak{u}^2=id$, where $id$ represents the trivial automorphism $id(g)=g$, $\forall g\in A_5$. One can straightforwardly check that $\mathfrak{u}$ acts on the $A_5$ conjugacy classes as follows
\begin{equation}
1C_1\stackrel{\mathfrak{u}}{\longmapsto} 1C_1,\qquad 15C_2\stackrel{\mathfrak{u}}{\longmapsto} 15C_2,\qquad 20C_3\stackrel{\mathfrak{u}}{\longmapsto} 20C_3,\qquad 12C_5\stackrel{\mathfrak{u}}{\longleftrightarrow} 12C^{\prime}_5\,.
\end{equation}
It interchanges the classes $12C_5$ and $12C^{\prime}_5$. Since the inverse of each $A_5$ conjugacy class is equal to itself, $\mathfrak{u}$ is not a class-inverting automorphism, and the corresponding CP transformation is unphysical. In terms of representations, the two different three-dimensional irreducible representations $\mathbf{3}$ and $\mathbf{3}^{\prime}$ are exchanged not mapped into their conjugate under the action of $\mathfrak{u}$. The generalized CP symmetry related with $\mathfrak{u}$ can only be consistently defined if fields transforming as $\mathbf{3}$ and $\mathbf{3}^{\prime}$ are absent in a model. As a result, we conclude that only the CP transformation associated with the trivial outer automorphism (i.e., the inner automorphism) can be compatibly imposed on the theory with $A_5$ family symmetry.

Now we consider the representative inner automorphism $\mu_{T^3ST^2ST^3S}$: $(S,T)\rightarrow (S,T^4)$. The corresponding generalized CP transformation $X^{0}_{\bf{r}}$ is fixed by the consistency equations:
\begin{eqnarray}
\nonumber &&X^{0}_{\mathbf{r}}\rho^{*}_{\mathbf{r}}(S)(X^{0}_{\mathbf{r}})^{-1}=\rho_{\mathbf{r}}(S),\\ &&X^{0}_{\mathbf{r}}\rho^{*}_{\mathbf{r}}(T)(X^{0}_{\mathbf{r}})^{-1}=\rho_{\mathbf{r}}(T^4)\,.
\end{eqnarray}
From the representation matrices given in Appendix~\ref{sec:A}, we see that for any representation
\begin{equation}
\rho^{*}_{\bf{r}}(S)=\rho_{\bf{r}}(S),\qquad \rho^{*}_{\bf{r}}(T)=\rho_{\bf{r}}(T^4)\,.
\end{equation}
Therefore $X^{0}_{\bf{r}}$ is an identity matrix up to an overall phase, i.e.,
\begin{equation}
X^{0}_{\bf{r}}=1\,.
\end{equation}
Including the contribution of the remaining inner automorphisms in the manner stated below Eq.~\eqref{eq:cons_equ}, the most general CP transformation consistent with $A_5$ family symmetry is of the form
\begin{equation}
\label{eq:CP_transformations}
X_{\bf{r}}=\rho_{\bf{r}}(g)X^{0}_{\bf{r}}=\rho_{\bf{r}}(g),  \qquad g\in A_{5}\,.
\end{equation}
This means that the generalized CP transformation consistent with $A_5$ is of the same form as the family group transformation in our working basis while they act on the a field multiplet in different ways:
$\varphi(x)\stackrel{g}{\longmapsto}\rho_{\bf{r}}(g)\varphi(x), ~ g\in A_{5}$ versus $\varphi(x)\stackrel{CP}{\longmapsto}X_{\bf r}\varphi^{*}(x_P)=\rho_{\bf r}(g)\varphi^{*}(x_P)$, where $x_{P}=(t,-\vec{x})$.

In this work, the phenomenological implications of $A_5$ family symmetry combined with the generalized CP symmetry would be investigated in a systematical and comprehensive way. The parent symmetry is $A_5\rtimes H_{CP}$ at high energy scale, where the element of $H_{CP}$ is the CP transformation compatible with $A_5$ and its explicit form is given by Eq.~\eqref{eq:CP_transformations}. In this setup, lepton mixing can be predicted from $A_5\rtimes H_{CP}$ breaking into different remnant symmetries $G_{l}\rtimes H^{l}_{CP}$ and $G_{\nu}\rtimes H^{\nu}_{CP}$ in the charged lepton and neutrino masses respectively, where $G_{l}$, $G_{\nu}$ and $H^{l}_{CP}$, $H^{\nu}_{CP}$ denote residual family symmetries and residual CP symmetries respectively. It is notable that the predictions for the lepton flavor mixing only depend on the assumed symmetry breaking patterns and are independent of the details of a specific implementation scheme, such as the possible additional symmetries of the model and the involved flavon fields and their assignments etc. In practice, the three generations of left-handed leptons doublets are embedded into the faithful three-dimensional representation $\mathbf{3}$ of $A_5$. Since $\mathbf{3}^{\prime}$ is related to $\mathbf{3}$ by the outer automorphism $\mathfrak{u}$, the results would be the same and no additional results would be found, if we assign the three left-handed leptons to the representation $\mathbf{3}^{\prime}$ instead. The requirement that $G_{l}\rtimes H^{l}_{CP}$ is preserved by the charged lepton mass term implies that the hermitian combination $m^{\dagger}_{l}m_{l}$ must be invariant under the remnant symmetry $G_{l}\rtimes H^{l}_{CP}$, i.e.,
\begin{subequations}
\begin{eqnarray}
\label{eq:Ch_flavour}&&\rho^{\dagger}_{\mathbf{3}}(g_{l})m^{\dagger}_{l}m_{l}\rho_{\mathbf{3}}(g_{l})=m^{\dagger}_{l}m_{l},  \qquad g_{l}\in G_{l}\,, \\
\label{eq:Ch_CP}&&X^{\dagger}_{l\mathbf{3}}m^{\dagger}_{l}m_{l}X_{l\mathbf{3}}=(m^{\dagger}_{l}m_{l})^{*},  \qquad X_{l\mathbf{3}} \in H^{l}_{CP}\,,
\end{eqnarray}
\end{subequations}
where the mass matrix $m_{l}$ is defined in the convention $\overline{l}_{R}m_{l}l_{L}$. Once $G_{l}$ and $H^{l}_{CP}$ are specified, the most general form of $m^{\dagger}_{l}m_{l}$ can be straightforwardly constructed from Eqs.~(\ref{eq:Ch_flavour}, \ref{eq:Ch_CP}). In the present work, we shall assume neutrinos are Majorana particles. In the same fashion, requiring that $G_{\nu}\rtimes H^{\nu}_{CP}$ is a symmetry of the neutrino mass matrix $m_{\nu}$ implies that $m_{\nu}$ should be invariant under the action of $G_{\nu}\rtimes H^{\nu}_{CP}$,
\begin{subequations}
\begin{eqnarray}
\label{eq:nu_flavour} && \rho_{\mathbf{3}}^{T}(g_{\nu})m_{\nu}\rho_{\mathbf{3}}(g_{\nu})=m_{\nu}, \qquad g_{\nu}\in G_{\nu}\,,   \\
\label{eq:nu_CP} && X_{\nu\mathbf{3}}^Tm_{\nu}X_{\nu\mathbf{3}}=m^{*}_{\nu}, \qquad X_{\nu\mathbf{3}}\in H^{\nu}_{CP}\,,
\end{eqnarray}
\end{subequations}
which allow us to derive the explicit form of $m_{\nu}$. Since both remnant family symmetry and remnant CP symmetries are still preserved after symmetry breaking, they should be compatible with each other. That is to say  consistency equation similar to Eq.~\eqref{eq:cons_equ_restr} has to be fulfilled,
\begin{subequations}
\begin{eqnarray}
\label{eq:consistency_remnant_nu}&&X_{\nu}\rho^{*}(g_{\nu_i})X^{-1}_{\nu}=\rho(g_{\nu_j}),\qquad
g_{\nu_i},g_{\nu_j}\in G_{\nu},\\
\label{eq:consistency_remnant_ch}&&X_{l}\rho^{*}(g_{l_i})X^{-1}_{l}=\rho_(g_{l_j}),\qquad~~\,
g_{l_i},g_{l_j}\in G_{l}\,.
\end{eqnarray}
\end{subequations}
The prediction for the PMNS matrix can be obtained by further diagonalizing the reconstructed mass matrices $m^{\dagger}_{l}m_{l}$ and $m_{\nu}$. Please see Ref.~\cite{Chen:2014wxa} for an alternative way of directly extracting the PMNS matrix from the representation matrices of the remnant symmetries without resorting to the mass matrices. As the order of neutrino and charged lepton masses is indeterminate in our framework, it is only possible to determine the PMNS matrix up to independent row and column permutations.

From the remnant symmetry invariant conditions of Eqs.~(\ref{eq:Ch_flavour}, \ref{eq:Ch_CP}), we can see that $X_{l\mathbf{r}}$ and $\rho_{\mathbf{r}}(g_{l})X_{l\mathbf{r}}$ with $g_{l}\in G_{l}$ lead to the same constraint on $m^{\dagger}_{l}m_{l}$. Furthermore, the residual CP transformation $X_{l{\bf r}}$ should be a symmetric matrix otherwise the charged lepton masses would be restricted to be partially degenerate~\cite{Li:2014eia,Chen:2014wxa}. The same comments apply to $X_{\nu{\bf{r}}}$ and $\rho_{{\bf{r}}}(g_{\nu})X_{\nu\mathbf{r}}$ with $g_{\nu}\in G_{\nu}$. Notice that the same result for PMNS matrix would be obtained~\cite{Li:2014eia,Ding:2013bpa,Ding:2014ssa}, if a pair of subgroups $\{G^{\prime}_{l}, G^{\prime}_{\nu}\}$ is conjugated to the pair of subgroups $\{G_{l}, G_{\nu}\}$ under an element of $A_5$, i.e.,
\begin{equation}
\label{eq:conjugate}
G^{\prime}_{l}=gG_{l}g^{-1},\qquad G^{\prime}_{\nu}=gG_{\nu}g^{-1}, \qquad g\in A_{5}\,.
\end{equation}
The reason is that remnant CP symmetries determined by restricted consistency condition of Eqs.~(\ref{eq:consistency_remnant_nu}, \ref{eq:consistency_remnant_ch}) are strongly correlated in the two cases such that lepton mass matrices $\{m^{\prime\dagger}_{l}m^{\prime}_{l}, m^{\prime}_{\nu}\}$ for the new primed residual symmetry are related to  $\{m^{\dagger}_{l}m_{l}, m_{\nu}\}$ by a similarity transformation $\rho_{\bf 3}(g)$~\cite{Li:2014eia,Ding:2013bpa,Ding:2014ssa}. In this way, it is sufficient to only discuss the independent pairs of $\{G_{l}, G_{\nu}\}$ which are not related by group conjugation and subsequently all possible residual CP compatible with the residual family symmetry should be included .

\section{\label{sec:3}Lepton mixing from remnant symmetries of $A_5\rtimes H_{CP}$ }

Neutrino are assumed to be Majorana particles here, therefore the remnant flavor symmetry $G_{\nu}$ must be a Klein four $K_{4}\cong Z_2\times Z_2$ subgroup or a single $Z_2$ subgroup of $A_5$. $G_{l}$ can be any abelian subgroups of $A_5$ with order equal or greater than 3. A complete or partial degeneracy of the charged lepton mass spectrum would be produced if $G_{l}$ had a non-abelian character. In the case of $G_{\nu}=K_4$, the lepton mixing matrix $U_{PMNS}$ is fully determined by the mismatch between the remnant family symmetry $G_{l}$ and $G_{\nu}$. As shown in Appendix~\ref{sec:B},  $U_{PMNS}$ can take four possible forms such as the golden ratio mixing, democratic mixing and so on. However, none of them is compatible with experimental data. Then we turn to the scenario of $G_{\nu}=Z_2$. With this setting, $U_{PMNS}$ is partially constrained, and only one column of the lepton mixing matrix is fixed up to reordering and rephasing of the elements. The explicit forms of the fixed column vectors for all the independent residual flavor symmetries  are summarized in Table~\ref{tab:fix_column_row}. We find that four cases are viable: $\left(G_{l}, G_{\nu}\right)=\left(Z^{T}_5, Z^{S}_2\right)$, $(Z^{T}_5, Z^{T^3ST^2ST^3}_{2})$, $(Z^{T^3ST^2S}_{3}, Z^{ST^2ST^3S}_{2})$ and $(K^{(ST^2ST^3S,~TST^4)}_{4}, Z^{S}_{2})$ lead to the mixing column vectors
$(-\sqrt{\frac{\kappa}{\sqrt{5}}}, \frac{1}{\sqrt{2\sqrt{5}\,\kappa}}, \frac{1}{\sqrt{2\sqrt{5}\,\kappa}})^T$, $(\sqrt{\frac{1}{\sqrt{5}\,\kappa}},\sqrt{\frac{\kappa}{2\sqrt{5}}},\sqrt{\frac{\kappa}{2\sqrt{5}}})^T$, $(\frac{1}{\sqrt{3}},\frac{1}{\sqrt{3}},\frac{1}{\sqrt{3}})^T$ and $(\frac{\kappa}{2},-\frac{1}{2},\frac{\kappa-1}{2})^T$ respectively, where $\kappa=(1+\sqrt{5})/2$ is the golden ratio. The phenomenological implications of each case are explored in Appendix~\ref{sec:B}, and the lepton mixing matrix $U_{PMNS}$ turns out to depend on two free parameters up to indeterminant Majorana phases. We see that the measured values of the three mixing angles can be accommodated very well, but the allowed values of Dirac CP phase $\delta_{CP}$ scatter in a quite large range. Furthermore, the breaking patterns with $\left(G_{l}, G_{\nu}\right)=\left(Z_2, K_4\right)$ are studied as well. Accordingly a row of the lepton mixing matrix $U_{PMNS}$ is determined to be $\frac{1}{2}\left(\kappa, 1, \kappa-1\right)$ or $\left(1, 0, 0\right)$ which are not in the experimentally preferred regions.

In order to be able to predict the values of CP phases, we extend the $A_5$ family symmetry to include the generalized CP. In the following, we shall perform a thorough analysis of lepton mixing patterns for the possible residual symmetries $G_{l}\rtimes H^{l}_{CP}$ and $G_{\nu}\rtimes H^{\nu}_{CP}$ in the charged lepton and neutrino sectors, where the remnant family symmetries $G_{l}$ and $G_{\nu}$ would be restricted to the four viable cases listed in Table~\ref{tab:fix_column_row}, and the remnant CP symmetries $H^{l}_{CP}$ and $H^{\nu}_{CP}$ are determined by consistency condition of Eqs.~(\ref{eq:consistency_remnant_nu},\ref{eq:consistency_remnant_ch}). In this setup, $U_{PMNS}$ as well as all mixing angles and all CP phases generically depend on a free parameters $\theta$ whose value can be fixed by the measured value of $\theta_{13}$. As a consequence, all observables are strongly correlated. For the concerned $A_5$ family symmetry, the Dirac phase would be predicted to be trivial or maximal while both Majorana phases are trivial after generalized CP symmetry is imposed. In order to evaluate how well the predicted mixing patterns agree with the experimental data on mixing angles, we shall perform a usual $\chi^2$ analysis which uses the global fit results of Ref.~\cite{Capozzi:2013csa}. We begin to discuss all possible cases one by one.

\subsection{\label{sec:3.2.1}$G_{l}=Z^{T}_{5}$, $G_{\nu}=Z^{S}_{2}$ }

In this case, the parent symmetry $A_{5}\rtimes H_{CP}$ is broken down to $Z^{T}_{5}\rtimes H^{l}_{CP}$ and $Z^{S}_{2}\times H^{\nu}_{CP}$ subgroups in the charged lepton and neutrino sectors, respectively. The residual CP symmetry $H^{l}_{CP}$ must be consistent with the residual flavor symmetry $Z^{T}_{5}$ in the charged lepton sector. That is to say the element $X_{l{\bf r}}$ of $H^{l}_{CP}$ should fulfill the consistency equation of Eq.~\eqref{eq:consistency_remnant_ch},
\begin{equation}
\label{eq:Ch_consistent_one}
X_{l\bf{r}}\rho^{*}_{\bf{r}}(T) X^{-1}_{l\bf{r}}=\rho_{\bf{r}}(g^{\prime}), \qquad g^{\prime}\in Z^{T}_{5}\,.
\end{equation}
Then we find only 10 choices out of the 60 CP transformations of $H_{CP}$ listed in Eq.~\eqref{eq:CP_transformations} are acceptable
\begin{eqnarray}
\label{eq:CP_trans_Z5T}
\hskip-0.17in \nonumber H^{l}_{CP}=&&\big\{\rho_{\bf r}(1),\rho_{\bf r}(T),\rho_{\bf r}(T^2),\rho_{\bf r}(T^3),\rho_{\bf r}(T^4),\rho_{\bf r}(ST^2ST^3S),\rho_{\bf r}((T^2S)^{2}T^3S),\\
&& ~~\rho_{\bf r}(T^{3}ST^2ST^3S),\rho_{\bf r}(T^{4}ST^2ST^3S),\rho_{\bf r}(ST^{3}ST^2S)\big\}\,.
\end{eqnarray}
As shown in Eq.~\eqref{eq:Ch_flavour}, the residual family symmetry $Z^{T}_{5}$ impose the following constraint on the charged lepton mass matrix:
\begin{equation}\label{eq:Ch_flavour_one}
\rho^{\dagger}_{\bf{3}}(T)m^{\dagger}_{l}m_{l}\rho_{\bf{3}}(T)=m^{\dagger}_{l}m_{l}\,.
\end{equation}
In our working basis, the representation matrix of the generator $T$ is diagonal with $\rho_{\bf{3}}(T)=\textrm{diag}(1,\omega_{5},\omega^{4}_{5})$. Consequently the hermitian combination $m^{\dagger}_{l}m_{l}$ of charged lepton mass matrix is also diagonal, i.e.,
\begin{equation}
m^{\dagger}_{l}m_{l}=\text{diag}\left(m^2_{e},m^2_{\mu},m^2_{\tau}\right)\,,
\end{equation}
where $m_e$, $m_{\mu}$ and $m_{\tau}$ represent the electron, muon and tau masses respectively. Furthermore, we can check that the remnant CP invariant condition of Eq.~\eqref{eq:Ch_CP} is automatically satisfied for $X_{l\mathbf{r}}=\rho_{\bf{r}}(1), \rho_{\bf{r}}(T), \rho_{\bf{r}}(T^{2}), \rho_{\bf{r}}(T^{3}), \rho_{\bf{r}}(T^{4})$. However, the mass degeneracy $m_{\mu}=m_{\tau}$ arises for the remaining values $X_{l\bf{r}}=\rho_{\bf r}(ST^2ST^3S)$, $\rho_{\bf r}((T^2S)^{2}T^3S)$, $\rho_{\bf r}(T^{3}ST^2ST^3S)$, $\rho_{\bf r}(T^{4}ST^2ST^3S)$, $\rho_{\bf r}(ST^{3}ST^2S)$. The reason is that all remnant CP
transformations except $\rho_{\bf r}(T^{3}ST^2ST^3S)$ are not symmetric. Generally speaking, any remnant CP transformation must be a symmetric matrix to avoid degenerate masses~\cite{Chen:2014wxa,Li:2014eia}. This case is obviously not viable, and will be disregarded hereafter.

Now we turn to the neutrino sector. The residual CP transformations $X_{\nu\bf{r}}$ of $H^{\nu}_{CP}$ is specified by the consistency condition:
\begin{equation}
\label{eq:nu_consistent_one}
X_{\nu\bf{r}}\rho^{*}_{\bf{r}}(S)X^{-1}_{\nu\bf{r}}=\rho_{\bf{r}}(S)\,,
\end{equation}
which can be easily obtained by applying the general consistency condition of Eq.~\eqref{eq:consistency_remnant_nu}. We see that the CP transformation $X_{\nu\bf{r}}$ commutes with flavor symmetry transformation $\rho_{\bf{r}}(S)$, and therefore remnant symmetry is $Z^{S}_2\times H^{\nu}_{CP}$ in the neutrino sector in this case. Notice that the semi-direct product structure between residual flavor and CP symmetries generally reduces to a direct product if the residual flavor symmetry is a $Z_2$ subgroup~\cite{Ding:2013hpa,Ding:2013bpa}. It is easy to check that $X_{\nu\bf{r}}$ can only take 4 possible values,
\begin{equation}
\label{eq:CP_transformations_Z2S}
H^\nu_{CP}=\{\rho_{\bf r}(1), \rho_{\bf r}(S), \rho_{\bf r}(T^3ST^2ST^3), \rho_{\bf r}(T^3ST^2ST^3S)\}\,.
\end{equation}
The neutrino mass matrix $m_{\nu}$ respects the residual symmetry $Z^{S}_2\times H^{\nu}_{CP}$, satisfying
\begin{eqnarray}
\nonumber&& \rho_{\mathbf{3}}^{T}(S)m_{\nu}\rho_{\mathbf{3}}(S)=m_{\nu}\,,\\
\label{eq:nu_flavour_one}&&X^{T}_{\nu{\bf 3}}m_{\nu}X_{\nu{\bf 3}}=m^{\ast}_{\nu},\qquad X_{\nu{\bf 3}}\in H^{\nu}_{CP}\,.
\end{eqnarray}
We find that the most general neutrino mass matrix invariant under the residual family symmetry $Z^{S}_2$, takes the following form
\begin{equation}
\label{eq:nu_general_mass_one}
m_{\nu}=\alpha
\left(\begin{array}{ccc}
 1 & 0 & 0 \\
 0 & 0 & 1 \\
 0 & 1 & 0
\end{array}\right)
+\frac{\beta}{\sqrt{2}}
\left(\begin{array}{ccc}
 -2 \sqrt{2} & 3 & 3 \\
 3 & 0 & \sqrt{2} \\
 3 & \sqrt{2} & 0
\end{array}\right)
+\gamma
\left(\begin{array}{ccc}
 2 & ~0~ & 0 \\
 0 & ~3~ & -1 \\
 0 & ~-1~ & 3
\end{array}\right)
+\delta
\left(\begin{array}{ccc}
 0 &~ -\sqrt{2} ~& \sqrt{2} \\
 -\sqrt{2} &~ -2\kappa ~& 0 \\
 \sqrt{2} &~ 0 ~& 2\kappa
\end{array}\right)\,,
\end{equation}
where $\alpha$, $\beta$, $\gamma$ and $\delta$ are generally complex parameters, and they are further constrained to be real or pure imaginary by residual CP. This neutrino mass matrix $m_{\nu}$ can be simplified into a quite simple form by performing a golden ratio transformation,
\begin{equation}
m^{\prime}_{\nu}=U^{T}_{GR}m_{\nu}U_{GR}=
\left(\begin{array}{ccc}
 \alpha -(3\kappa-1) \beta +2 \gamma  & 0 & 0 \\
 0 & \alpha +(3 \kappa-2) \beta +2 \gamma  & 2\sqrt{2+\kappa}~ \delta  \\
 0 & 2\sqrt{2+\kappa}~\delta  & -\alpha - \beta +4 \gamma
\end{array}\right)\,,
\end{equation}
where
\begin{equation}
\label{eq:golden_ratio}
U_{GR}=\left(\begin{array}{ccc}
-\sqrt{\frac{\kappa}{\sqrt{5}}}  & ~\sqrt{\frac{1}{\sqrt{5}\,\kappa}} ~ & 0  \\
\sqrt{\frac{1}{2\sqrt{5}\,\kappa}}   & ~ \sqrt{\frac{\kappa}{2\sqrt{5}}} ~  & -\frac{1}{\sqrt{2}}  \\
\sqrt{\frac{1}{2\sqrt{5}\,\kappa}}   & ~ \sqrt{\frac{\kappa}{2\sqrt{5}}} ~  & \frac{1}{\sqrt{2}}
  \end{array}\right)\;,
\end{equation}
is the golden ratio mixing pattern~\cite{Datta:2003qg} which can be naturally derived in $A_5$ models~\cite{Everett:2008et}. The neutrino mass matrix $m^{\prime}_{\nu}$ is further diagonalized by a unitary rotation $U^{\prime}_{\nu}$ in the (2,3)-plane,
\begin{equation}
U^{\prime T}_{\nu}m^{\prime}_{\nu}U^{\prime}_{\nu}=\textrm{diag}(m_{1},m_{2},m_{3})\,.
\end{equation}
The next step is to explore the constraint of remnant CP on $m_{\nu}$. Two different phenomenological predictions arise for the four possibe $X_{\nu{\bf r}}$ shown in Eq.~\eqref{eq:CP_transformations_Z2S}, as $\rho_{\bf r}(S)X_{\nu{\bf r}}$ and $X_{\nu{\bf r}}$ lead to the same predictions.
\begin{description}[labelindent=-0.7em, leftmargin=0.1em]
\item[~~(\uppercase\expandafter{\romannumeral1})] $X_{\nu{\bf r}}=\rho_{\bf r}(1),\rho_{\bf r}(S)$

Obviously we have $m_{\nu}=m^{\ast}_{\nu}$ such that all the four parameters $\alpha$, $\beta$, $\gamma$ and $\delta$ are real. As a consequence, the neutrino mass matrix $m^{\prime}_{\nu}$ is a real symmetric matrix. The unitary transformation $U^{\prime}_{\nu}$ is of the form:
\begin{equation}
\label{eq:un_case1}
U^{\prime}_{\nu}=\left(\begin{array}{ccc}
 1  &  0  &  0  \\
 0  &  ~\cos\theta &  \sin\theta \\
 0  &  -\sin\theta &  \cos\theta
\end{array}\right)K_{\nu}\,.
\end{equation}
where $K_{\nu}$ is a diagonal phase matrix with elements equal to $\pm1$ or $\pm i$ which makes the neutrino masses $m_{1,2,3}$ positive. The effect of $K_{\nu}$ is a possible change of the Majorana phases by $\pi$, and it would be omitted hereinafter for the other cases. The parameter $\theta$ is given by
\begin{equation}
\tan{2\theta}=-\frac{4 \sqrt{2+\kappa}~ \delta }{2(\alpha-\gamma) +(3 \kappa-1)\beta}\,.
\end{equation}
The light neutrino mass eigenvalues are
\begin{eqnarray}
\nonumber &&m_{1}=|\alpha -(3 \kappa-1) \beta  +2 \gamma|\,, \\
\nonumber &&m_{2}=\frac{1}{2}\left|3(\kappa-1)\beta+6\gamma+\frac{2(\alpha-\gamma) +(3 \kappa-1) \beta   }{\cos{2\theta}}\right|\,, \\
&&m_{3}=\frac{1}{2}\left|3(\kappa-1)\beta+6\gamma-\frac{2(\alpha-\gamma) +(3 \kappa-1) \beta   }{\cos{2\theta}}\right|\,.
\end{eqnarray}
Given the diagonal charged lepton mass matrix, the lepton mixing matrix takes the form
\begin{equation}
\label{eq:PMNS_caseI}
U_{PMNS}=U_{GR}U^{\prime}_{\nu}=
\left(\begin{array}{ccc}
 -\sqrt{\frac{\kappa}{\sqrt{5}} } &~ \sqrt{\frac{1}{\sqrt{5}\kappa} } \cos\theta ~& \sqrt{\frac{1}{\sqrt{5}\kappa}} \sin\theta \\
 \sqrt{\frac{1}{2\sqrt{5}\kappa}} &~ \sqrt{\frac{\kappa}{2\sqrt{5}}}\cos\theta+\frac{\sin\theta}{\sqrt{2}} ~& \sqrt{\frac{\kappa}{2\sqrt{5}}}\sin\theta-\frac{\cos\theta}{\sqrt{2}} \\
 \sqrt{\frac{1}{2\sqrt{5}\kappa}} &~ \sqrt{\frac{\kappa}{2\sqrt{5}}}\cos\theta-\frac{\sin\theta}{\sqrt{2}} ~& \sqrt{\frac{\kappa}{2\sqrt{5}}}\sin\theta+\frac{\cos\theta}{\sqrt{2}}
\end{array}\right)K_{\nu}\,.
\end{equation}
One can straightforwardly extract the lepton mixing angles and CP phases as follows,
\begin{eqnarray}
\nonumber && \qquad\qquad \sin^{2}\theta_{13}=\frac{3-\kappa}{5}\sin^{2}\theta\;, \quad \sin^{2}\theta_{12}=\frac{1+\cos2\theta}{3+2\kappa+\cos2\theta}\,,\\
&&\sin^{2}\theta_{23}=\frac{1}{2}-\frac{\sqrt{2+\kappa}\sin2\theta}{3\kappa-1+(\kappa-1)\cos2\theta}\,, \quad \sin\delta_{CP}=\sin\alpha_{21}=\sin\alpha_{31}=0\,,
\end{eqnarray}
where $\delta_{CP}$ is the Dirac CP phase, $\alpha_{21}$ and $\alpha_{31}$ are the Majorana CP phases in the standard parameterization~\cite{Agashe:2014kda}. There is no CP violation in this case as the neutrino mass matrix is real. Expressing $\theta$ in terms of $\theta_{13}$, correlations among the three mixing angles follow immediately,
\begin{eqnarray}
\nonumber&&\sin^2\theta_{12}=\frac{3-\kappa}{5}-\frac{2+\kappa}{5}\tan^{2}\theta_{13}\,,\\
\label{eq:correlation_case_I}&&\sin^2\theta_{23}=\frac{1}{2}\pm\kappa\tan\theta_{13}\sqrt{1-(1+\kappa)\tan^2\theta_{13}}\,.
\end{eqnarray}
For the measured reactor mixing angles $\sin^2\theta_{13}\simeq0.0234$~\cite{Capozzi:2013csa}, we have $\sin^2\theta_{23}\simeq0.258$ or 0.742 which is outside of the experimentally favored $3\sigma$ region~\cite{Capozzi:2013csa} although $\sin^2\theta_{12}\simeq0.259$ is acceptable. As a consequence, this mixing pattern isn't viable. This point remains even after permutation of rows and columns is considered.

\item[~~(\uppercase\expandafter{\romannumeral2})] $X_{\nu{\bf r}}=\rho_{\bf r}(T^3ST^2ST^3), \rho_{\bf r}(T^3ST^2ST^3S)$

Solving the residual CP invariant condition in Eq.~\eqref{eq:nu_flavour_one}, we find $\alpha$, $\beta$ and $\gamma$ are real while $\delta$ is pure imaginary. The unitary diagonalization matrix $U^{\prime}_{\nu}$ is
\begin{equation}
\label{eq:un_case2}
U^{\prime}_{\nu}=\left(\begin{array}{ccc}
 1 & ~0 & ~0 \\
 0 &  ~\cos\theta     & ~\sin\theta \\
 0 & ~-i\sin\theta    & ~i\cos\theta
\end{array}\right)\,,
\end{equation}
where the diagonal matrix $K_{\nu}$ multiplied from the right-hand side has been omitted, and the rotation angle $\theta$ fulfills
\begin{equation}\label{eq:theta_caseII}
\tan{2\theta}=-\frac{4 i\sqrt{2+\kappa}~\delta }{3 (\kappa-1) \beta +6 \gamma }\,.
\end{equation}
The three neutrino masses are given by
\begin{eqnarray}\label{eq:nu_mass_caseII}
\nonumber && m_{1}=|\alpha -\left(3\kappa-1\right) \beta +2 \gamma|\,,\\
\nonumber && m_{2}=\frac{1}{2}\left|2 \alpha +(3 \kappa-1) \beta -2 \gamma+\frac{3\left((\kappa-1) \beta +2 \gamma\right)}{\cos{2\theta}}\right|\,,\\
&& m_{3}=\frac{1}{2}\left|2 \alpha +(3 \kappa-1) \beta -2 \gamma-\frac{3\left((\kappa-1) \beta +2 \gamma\right)}{\cos{2\theta}}\right|\,.
\end{eqnarray}
All the four parameters $\alpha$, $\beta$, $\gamma$ and $\delta$ are involved in the three neutrino masses. As a result, the measured mass squared differences $\delta m^2\equiv m^2_2-m^2_1$ and $\Delta m^2\equiv m^2_3-(m^2_1+m^2_2)/2$ can be easily accommodated~\cite{Capozzi:2013csa}, the absolute neutrino mass scale can not be fixed, and the neutrino mass spectrum can be either normal ordering (NO) or inverted ordering (IO). The PMNS matrix takes the following form:
\begin{equation}\label{eq:PMNS_caseII}
  U_{PMNS}=
\left(\begin{array}{ccc}
 -\sqrt{\frac{\kappa}{\sqrt{5}}} &~ \sqrt{\frac{1}{\sqrt{5}\kappa} } \cos\theta ~& \sqrt{\frac{1}{\sqrt{5}\kappa} } \sin\theta \\
 \sqrt{\frac{1}{2\sqrt{5}\,\kappa}} &~ \sqrt{\frac{\kappa}{2\sqrt{5}}}\cos\theta+\frac{i\sin\theta}{\sqrt{2}} ~& \sqrt{\frac{\kappa}{2\sqrt{5}}}\sin\theta-\frac{i\cos\theta}{\sqrt{2}} \\
 \sqrt{\frac{1}{2\sqrt{5}\,\kappa}} &~ \sqrt{\frac{\kappa}{2\sqrt{5}}}\cos\theta-\frac{i\sin\theta}{\sqrt{2}} ~& \sqrt{\frac{\kappa}{2\sqrt{5}}}\sin\theta+\frac{i\cos\theta}{\sqrt{2}}
\end{array}\right)\,.
\end{equation}
Note that the first column vector of this mixing pattern coincides with the first column of the GR mixing. The lepton mixing angles and CP phases can be read out as\footnote{In the case of $\sin2\theta=0$, either $\theta_{12}$ or $\theta_{13}$ vanishes, consequently the value of $\delta_{CP}$ can not be determined uniquely.}
\begin{eqnarray}\label{eq:mixing_parameters_caseII}
\nonumber && \sin^{2}\theta_{13}=\frac{3-\kappa}{5}\sin^{2}\theta\,, \qquad \sin^{2}\theta_{12}=\frac{1+\cos2\theta}{3+2\kappa+\cos2\theta}\,,\\
\label{eq:mixing_parameters_caseI} && \sin^{2}\theta_{23}=\frac{1}{2}\,, \quad \left|\sin\delta_{CP}\right|=1\,, \quad \sin\alpha_{21}=\sin\alpha_{31}=0\,.
\end{eqnarray}
Here we present the absolute value of $\sin\delta_{CP}$, since the sign of $\sin\delta_{CP}$ depends on the ordering of rows and columns. We see that both atmospheric angle $\theta_{23}$ and Dirac CP phase $\delta_{CP}$ are maximal while Majorana phases are conserved. Given the weak evidence of $\delta_{CP}\sim3\pi/2$ from T2K~\cite{Abe:2013hdq}, this pattern is slightly preferred. The prediction of maximal Dirac CP can be tested by next generation long-baseline neutrino oscillation experiments such as the proposed LBNE~\cite{Adams:2013qkq}, LBNO~\cite{::2013kaa} and HyperKamiokande~\cite{Kearns:2013lea}, which aim to search for leptonic CP violation. Moreover, the correlation between $\theta_{13}$ and $\theta_{12}$ is of the same form as that of case I, and it is plotted in Fig.~\ref{fig:general_mixing_caseII}. The results of the $\chi^2$ analysis are reported in Table~\ref{tab:best_fitting}. We see that the experimental data~\cite{Capozzi:2013csa} on lepton mixing angles can be accommodated very well. Notice that the solar mixing angle $\theta_{12}$ is predicted to be around the present $3\sigma$ lower bound. As far as we known, the JUNO experiment can measure $\theta_{12}$ with high accuracy~\cite{JUNO}. If significant deviations $\sin^2\theta_{12}$ from 0.259 was detected, this mixing pattern would be excluded.
\begin{figure}[t!]
\centering
\includegraphics[width=0.51\textwidth]{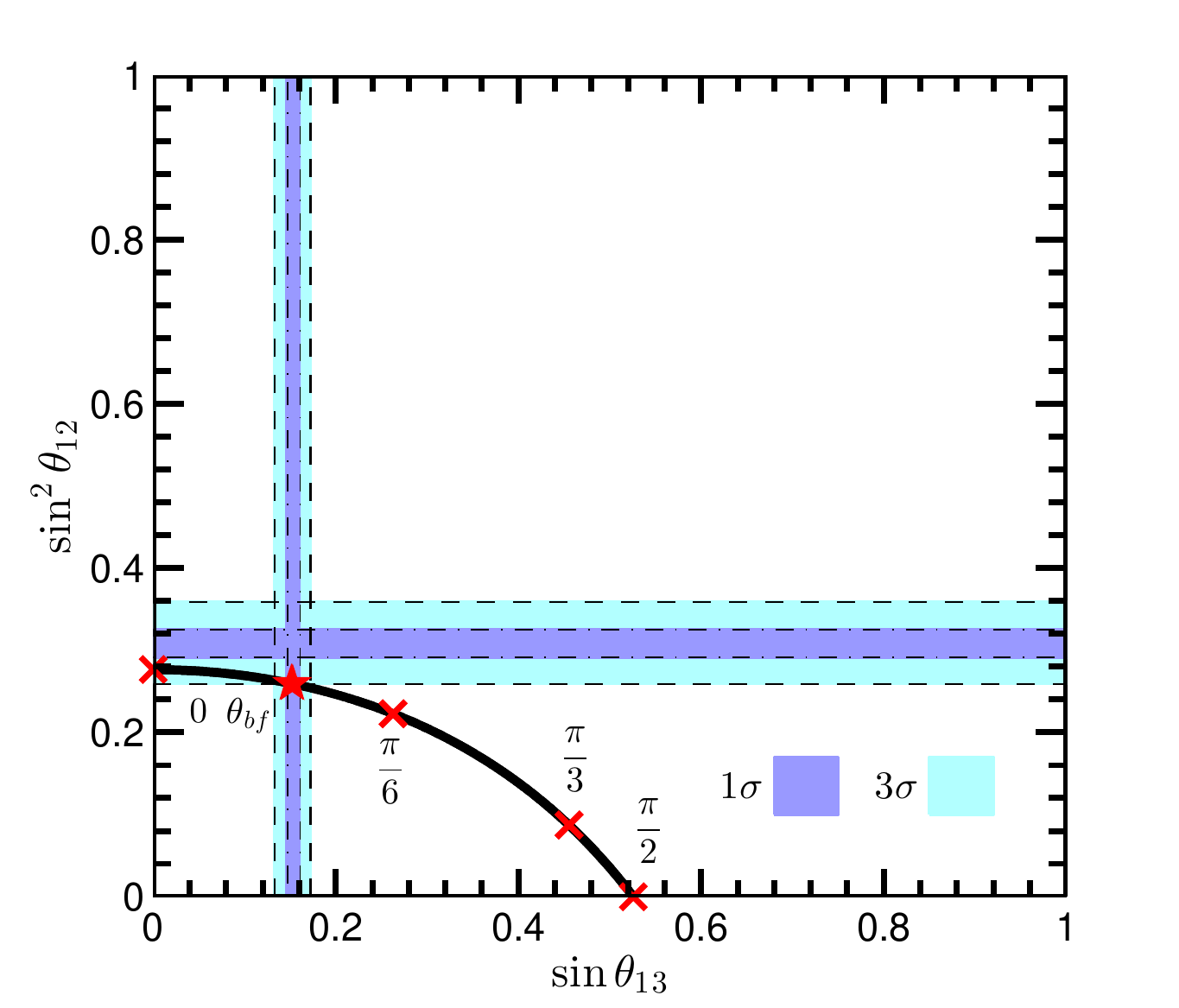}
\hskip-0.30in \includegraphics[width=0.51\textwidth]{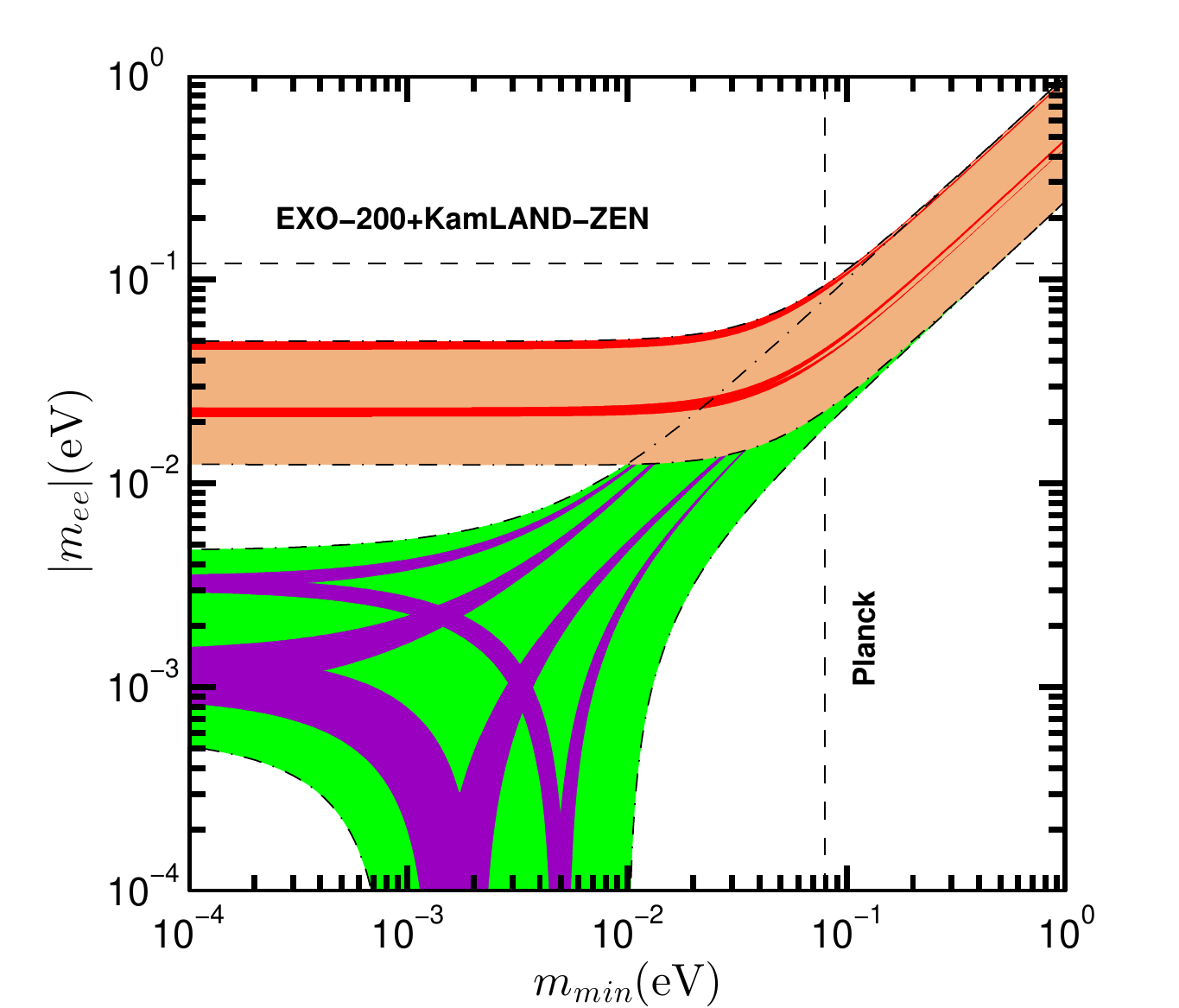}
\caption{\label{fig:general_mixing_caseII} The correlation between $\sin^{2}\theta_{12}$ and $\sin\theta_{13}$ (left panel) and the allowed values of the effective mass $|m_{ee}|$ (right panel) in case II. On the left panel, the best fitting value is labelled with a red pentagram, and the points for $\theta=0$, $\pi/6$, $\pi/3$ and $\pi/2$ are marked with a cross to guide the eye. The $1\sigma$ and $3\sigma$ ranges of the mixing angles are taken from Ref.~\cite{Capozzi:2013csa}. On the right panel, the orange and green bands denote the $3\sigma$ regions for normal ordering and inverted ordering mass spectrum respectively. The red and purple areas are the predictions for the lepton mixing matrix in Eq.~\eqref{eq:PMNS_caseII}. The present most strict bound $|m_{ee}|<(0.120-0.250)$ eV from
EXO-200~\cite{Auger:2012ar,Albert:2014awa} combined with KamLAND-ZEN~\cite{Gando:2012zm} is represented
by the horizontal dashed line, and the upper limit on $m_{min}$ from the latest Planck result $m_1+m_2+m_3<0.230$ eV at $95\%$ confidence level~\cite{Ade:2013zuv} is shown by vertical dashed line.}
\end{figure}
It is well-known that leptonic CP phases can play a crucial role in the rare process neutrinoless double beta ($(\beta\beta)_{0\nu}-$) decay. The dependence of the $(\beta\beta)_{0\nu}-$decay amplitude on the neutrino mixing parameters is characterized by the effective Majorana mass $\left|m_{ee}\right|$~\cite{Agashe:2014kda} with the definition:
\begin{equation}
\label{eq:mee}\left|m_{ee}\right|=\left|m_1\cos^2\theta_{12}\cos^2\theta_{13}+m_2\sin^2\theta_{12}\cos^2\theta_{13}e^{i\alpha_{21}}+m_3\sin^2\theta_{13}e^{i(\alpha_{31}-2\delta_{CP})}\right|\,.
\end{equation}
For the predicted mixing parameters in Eq.~\eqref{eq:mixing_parameters_caseI}, we have
\begin{equation}
\left|m_{ee}\right|=\frac{1}{\sqrt{5}}\left|\kappa m_1+\kappa^{-1}k_2m_2\cos^2\theta+\kappa^{-1}k_3m_3\sin^2\theta\right|\,,
\end{equation}
where $k_2, k_3=\pm1$ originates from the ambiguity of the matrix $K_{\nu}$. The prediction for the effective mass $|m_{ee}|$ with respect to the lightest neutrino mass is shown Fig.~\ref{fig:general_mixing_caseII}. We find that $|m_{ee}|$ is close to 0.022eV or the upper bound 0.045eV in case of IO neutrino mass spectrum, which are within the future sensitivity of planned $(\beta\beta)_{0\nu}-$decay experiments. However, in case of NO spectrum, $|m_{ee}|$ strongly depends on lightest neutrino mass $m_{min}$, and it can even be approximately vanishing for particular value of $m_{min}$.

\end{description}

\begin{table}[t!]
\centering
\footnotesize
\renewcommand{\tabcolsep}{0.8mm}
\begin{tabular}{|c|c|c|c|c|c|c|c|c|c|}
\hline \hline
 \multicolumn{2}{|c|}{} & \multicolumn{3}{|c|}{\texttt{Analytic expression}} & \multicolumn{5}{|c|}{\texttt{Best fitting}}  \\ \hline
 & & & & & & & & &  \\[-0.14in]
\multicolumn{2}{|c|}{} & $\sin^2\theta_{13}$ & $\sin^2\theta_{12}$ & $\sin^2\theta_{23}$ & $\theta_{bf}$ & $\chi^2_{min}$   & $\sin^2\theta_{13}$ & $\sin^2\theta_{12}$ & $\sin^2\theta_{23}$    \\ \hline
  & & & & & & & & &  \\[-0.14in]
 \multirow{3}{*}{II} & IO & \multirow{2}{*}{$\frac{3-\kappa}{5}\sin^{2}\theta$} & \multirow{2}{*}{$\frac{2\cos^{2}\theta}{3+2\kappa+\cos2\theta}$}  & $\frac{1}{2}$ & $0.295$ & $8.468$ & $0.0234$ & $0.259$ & $0.5$  \\
 & & & & & & & & &  \\[-0.14in] \cline{2-2} \cline{5-10}
& & & & & & & & &  \\[-0.14in]
& NO & & &  $\frac{1}{2}$ & $0.292$ & $11.88$ & $0.0229$ & $0.259$ & $0.5$  \\
& & & & & & & & &  \\[-0.14in] \hline
& & & & & & & & &  \\[-0.14in]
\multirow{7}{*}[12pt]{III} & \multirow{3}{*}{IO} & \multirow{9}{*}{$\frac{\kappa}{\sqrt{5}}\sin^{2}\theta$} & \multirow{9}{*}{$\frac{4-2\kappa}{5-2\kappa+\cos{2\theta}}$} & \multirow{7}{*}[12pt]{$\frac{1}{2}-\frac{\sqrt{3-\kappa}\sin2\theta}{3\kappa-2+\kappa\cos2\theta}$} & $0.182$ & $4.851$ & $0.0236$ & $0.283$ & $0.404$ ($\theta_{23}<45^{\circ}$)   \\
& & & & & & & & &  \\[-0.14in] \cline{6-10}
& & & & & & & & &  \\[-0.14in]
 & & & & & $2.958$ & $3.165$ &  $0.0240$ & $0.283$ & $0.597$ ($\theta_{23}>45^{\circ}$)  \\
 & & & & & & & & &  \\[-0.14in] \cline{2-2}\cline{6-10}
& & & & & & & & &  \\[-0.14in]
& \multirow{3}{*}{NO} & & & & $0.179$ & $4.087$ & $0.0230$ & $0.283$ & $0.406$ ($\theta_{23}<45^{\circ}$)  \\
 & & & & & & & & &  \\[-0.14in] \cline{6-10}
& & & & & & & & &  \\[-0.14in]
& & & & & $2.965$ & $24.88$ &  $0.0224$ & $0.283$ & $0.593$ ($\theta_{23}>45^{\circ}$)  \\
 & & & & & & & & &  \\[-0.14in] \cline{1-2} \cline{5-10}
& & & & & & & & &  \\[-0.14in]
\multirow{3}{*}{IV} & IO & &  &  $\frac{1}{2}$ & $0.183$ & $2.232$ & $0.0241$ & $0.283$ & $0.5$  \\
 & & & & & & & & &  \\[-0.14in] \cline{2-2} \cline{5-10}
& & & & & & & & &  \\[-0.14in]
& NO & & &  $\frac{1}{2}$  & $0.181$ & $5.802$ & $0.0235$ & $0.283$ & $0.5$  \\
& & & & & & & & &  \\[-0.14in] \hline
& & & & & & & & &  \\[-0.14in]
\multirow{3}{*}{V} & IO & \multirow{2}{*}{$\frac{1-\sin{2\theta}}{3}$} & \multirow{2}{*}{$\frac{1}{2+\sin{2\theta}}$}  &  $\frac{1}{2}$ & $0.976$ & $3.987$ & $0.0238$ & $0.341$ & $0.5$  \\
 & & & & & & & & &  \\[-0.14in] \cline{2-2} \cline{5-10}
& & & & & & & & &  \\[-0.14in]
& NO & & &  $\frac{1}{2}$ & $0.973$ & $7.480$ & $0.0233$ & $0.341$ & $0.5$  \\
& & & & & & & & &  \\[-0.14in] \hline
& & & & & & & & &  \\[-0.14in]
\multirow{6}{*}{VII} & \multirow{3}{*}{IO} & \multirow{6}{*}{$\frac{(\cos\theta-\kappa\sin\theta)^2}{4\kappa^{2}}$} & \multirow{6}{*}{$\frac{(\kappa\cos\theta+ \sin\theta)^2}{4\kappa^2-(\cos\theta- \kappa\sin\theta)^2}$} & $\frac{(\kappa^2\cos\theta-\sin\theta)^2}{4\kappa^2-(\cos\theta-\kappa\sin\theta)^2}$ & $0.286$ & $1.626$ &  $0.0242$ & $0.329$ & $0.486$ ($\theta_{23}<45^{\circ}$)   \\
& & & & & & & & &  \\[-0.14in] \cline{5-10}
& & & & & & & & &  \\[-0.14in]
 & & & & $\frac{\kappa^{2}(\cos\theta+\kappa\sin\theta)^2}{4\kappa^{2}-(\cos\theta- \kappa\sin\theta)^2}$ & $0.286$ & $1.751$ &  $0.0242$ & $0.329$ & $0.513$ ($\theta_{23}>45^{\circ}$)  \\
 & & & & & & & & &  \\[-0.14in] \cline{2-2}\cline{5-10}
& & & & & & & & &  \\[-0.14in]
 & \multirow{3}{*}{NO} &  &  & $\frac{(\kappa^2\cos\theta-\sin\theta)^2}{4\kappa^2-(\cos\theta-\kappa\sin\theta)^2}$ & $0.293$ & $3.503$ &  $0.0229$ & $0.330$ & $0.480$ ($\theta_{23}<45^{\circ}$)   \\
 & & & & & & & & &  \\[-0.14in] \cline{5-10}
& & & & & & & & &  \\[-0.14in]
 & & & & $\frac{\kappa^{2}(\cos\theta+\kappa\sin\theta)^2}{4\kappa^{2}-(\cos\theta- \kappa\sin\theta)^2}$ & $0.282$ & $6.958$ &  $0.0248$ & $0.329$ & $0.510$ ($\theta_{23}>45^{\circ}$)  \\
 & & & & & & & & &  \\[-0.14in]\hline \hline
\end{tabular}
\caption{\label{tab:best_fitting}Summary of the predictions for the lepton mixing angles and their best fitting values for all viable cases in the framework of $A_5\rtimes H_{CP}$. In case VII, the mixing patterns for $\theta_{23}$ in the first and second octant are related through the exchange of the second and third rows of the PMNS matrix. Notice that all the three CP phases are independent of $\theta$ in all cases: Dirac phase is trivial or maximal, and both Majorana phases are trivial.}
\end{table}

\subsection{\label{sec:3.2.2}$G_{l}=Z^{T}_{5}$, $G_{\nu}=Z^{T^3ST^2ST^3}_{2}$}

The charged lepton sector preserves the same remnant symmetry $Z^{T}_{5}\rtimes H^{l}_{CP}$ as that discussed in section \ref{sec:3.2.1}. Therefore the charged lepton mass is subject to the same constraint, and $m^{\dagger}_{l}m_{l}$ should be diagonal as well. In neutrino sector, the residual CP symmetry $H^{\nu}_{CP}$ has to be compatible with the residual family symmetry $G_{\nu}=Z^{T^3ST^2ST^3}_{2}$, i.e.,
\begin{equation}\label{eq:nu_consistent_Z28}
X_{\nu\bf{r}}\rho^{*}_{\bf{r}}(T^3ST^2ST^3)X^{-1}_{\nu\bf{r}}=\rho_{\bf{r}}(T^3ST^2ST^3)\,.
\end{equation}
It is easy to check that only 4 generalized CP transformations are acceptable,
\begin{equation}\label{eq:CP_transformations_Z28}
H^\nu_{CP}=\left\{\rho_{\bf r}(1), \rho_{\bf r}(S), \rho_{\bf r}(T^3ST^2ST^3), \rho_{\bf r}(T^3ST^2ST^3S)\right\}\,.
\end{equation}
Straightforward calculations demonstrate that the most general neutrino mass matrix invariant under $Z^{T^3ST^2ST^3}_{2}$ is of the form
\begin{equation}
\label{eq:nu_general_mass_two}
m_{\nu}=\alpha
\left(\begin{array}{ccc}
 1 & 0 & 0 \\
 0 & 0 & 1 \\
 0 & 1 & 0
\end{array}\right)
+\frac{\beta}{\sqrt{2}}
\left(\begin{array}{ccc}
 -2 \sqrt{2} & 3 & 3 \\
 3 & 0 & \sqrt{2} \\
 3 & \sqrt{2} & 0
\end{array}\right)
+\gamma
\left(\begin{array}{ccc}
 2 & ~0~ & 0 \\
 0 & ~3~ & -1 \\
 0 & ~-1~ & 3
\end{array}\right)
+\delta
\left(\begin{array}{ccc}
 0 & \sqrt{2}\kappa ~& -\sqrt{2}\kappa \\
 \sqrt{2}\kappa & -2 ~& 0 \\
 -\sqrt{2}\kappa & 0 ~& 2
\end{array}\right)\,,
\end{equation}
where the parameters $\alpha$, $\beta$, $\gamma$ and $\delta$ are generically complexes, and they are further constrained by the remnant CP. After performing a GR transformation, $m_{\nu}$ becomes
\begin{equation}
m^{\prime}_{\nu}=U^{T}_{GR}m_{\nu}U_{GR}=
\left(\begin{array}{ccc}
 \alpha -(3 \kappa-1 )\beta +2 \gamma  &~ 0 ~& 2\sqrt{2+\kappa}~ \delta  \\
 0 &~ \alpha +(3 \kappa-2) \beta  +2 \gamma  ~& 0 \\
 2\sqrt{2+\kappa}~ \delta  &~ 0 ~& -\alpha-\beta +4 \gamma
\end{array}\right)\,.
\end{equation}
In the following, we proceed to investigate the constraints imposed by the
remnant CP transformations shown in Eq.~\eqref{eq:CP_transformations_Z28}.  The four possible $X_{\nu\bf{r}}$ can be divided into two classes.

\begin{description}[labelindent=-0.7em, leftmargin=0.1em]

\item[~~(\uppercase\expandafter{\romannumeral3})] $X_{\nu{\bf r}}=\rho_{\bf r}(1),\rho_{\bf r}(T^3ST^2ST^3)$\\
In this case, the residual flavor and residual CP transformations are of the same form. As a result, the four parameters $\alpha$, $\beta$, $\gamma$ and $\delta$ are all real. The neutrino mass matrix $m^{\prime}_{\nu}$ can be diagonalized by a unitary transformation
\begin{equation}
U^{\prime}_{\nu}=\left(\begin{array}{ccc}
 ~\cos\theta  &~  0  ~&  \sin\theta  \\
  0  &~ 1 ~&  0 \\
  -\sin\theta  &~  0 ~&  \cos\theta
 \end{array}\right)\,,
\end{equation}
with
\begin{equation}
\tan{2\theta}=-\frac{4 \sqrt{2+\kappa} ~\delta }{2 (\alpha-\gamma) -(3\kappa-2) \beta }\,,
\end{equation}
The three neutrino masses are
\begin{eqnarray}\label{eq:nu_mass_caseIII}
\nonumber && m_{1}=\frac{1}{2}\left|-3 \kappa \beta  +6\gamma +\frac{2( \alpha-\gamma) -(3 \kappa-2) \beta }{\cos{2\theta}}\right|\,,\\
\nonumber && m_{2}=\left|\alpha+(3 \kappa-2) \beta  +2 \gamma \right|\,, \\
&& m_{3}=\frac{1}{2}\left|-3 \kappa \beta  +6\gamma -\frac{2( \alpha-\gamma) -(3 \kappa-2) \beta }{\cos{2\theta}}\right|\,.
\end{eqnarray}
The absolute neutrino mass scale can not be predicted. Then the PMNS matrix reads
\begin{equation}\label{eq:PMNS_caseIII}
U_{PMNS}=U_{GR}U^{\prime}_{\nu}=
\left(\begin{array}{ccc}
 -\sqrt{\frac{\kappa}{\sqrt{5}}} \cos\theta &~ \sqrt{\frac{1}{\sqrt{5}\,\kappa}} ~& -\sqrt{\frac{\kappa}{\sqrt{5}}} \sin\theta \\
 \frac{\cos\theta}{\sqrt{2\sqrt{5}\,\kappa}}+\frac{\sin\theta}{\sqrt{2}} &~ \sqrt{\frac{\kappa}{2\sqrt{5}}} ~& \frac{\sin\theta}{\sqrt{2\sqrt{5}\,\kappa}}-\frac{\cos\theta}{\sqrt{2}} \\
 \frac{\cos\theta}{\sqrt{2\sqrt{5}\,\kappa}}-\frac{\sin\theta}{\sqrt{2}} &~ \sqrt{\frac{\kappa}{2\sqrt{5}}} ~& \frac{\sin\theta}{\sqrt{2\sqrt{5}\,\kappa}}+\frac{\cos\theta}{\sqrt{2}}
\end{array}\right)\,.
\end{equation}
\begin{figure}[t!]
\centering
\includegraphics[width=0.51\textwidth]{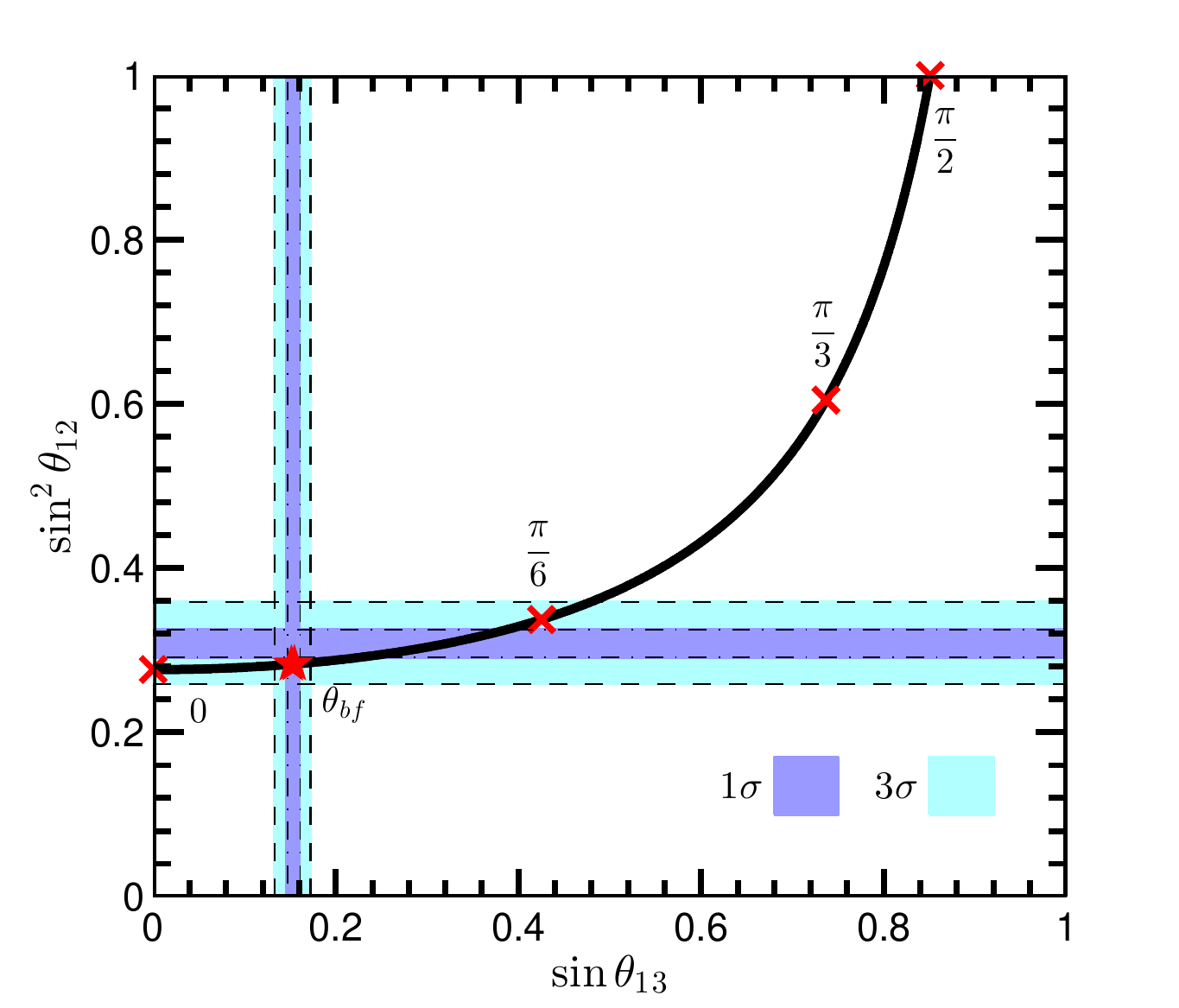}
\hskip-0.30in \includegraphics[width=0.51\textwidth]{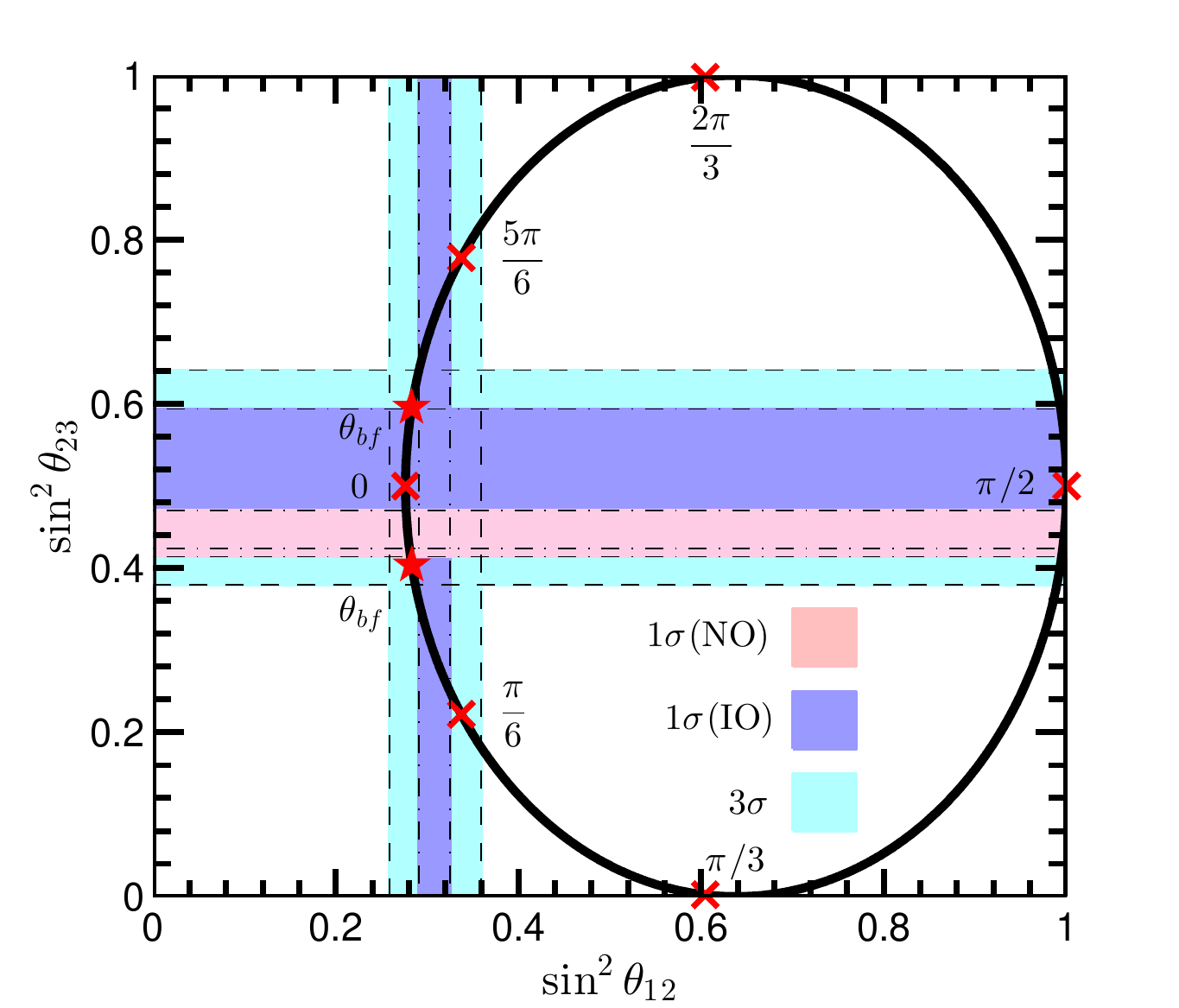} \\
\includegraphics[width=0.51\textwidth]{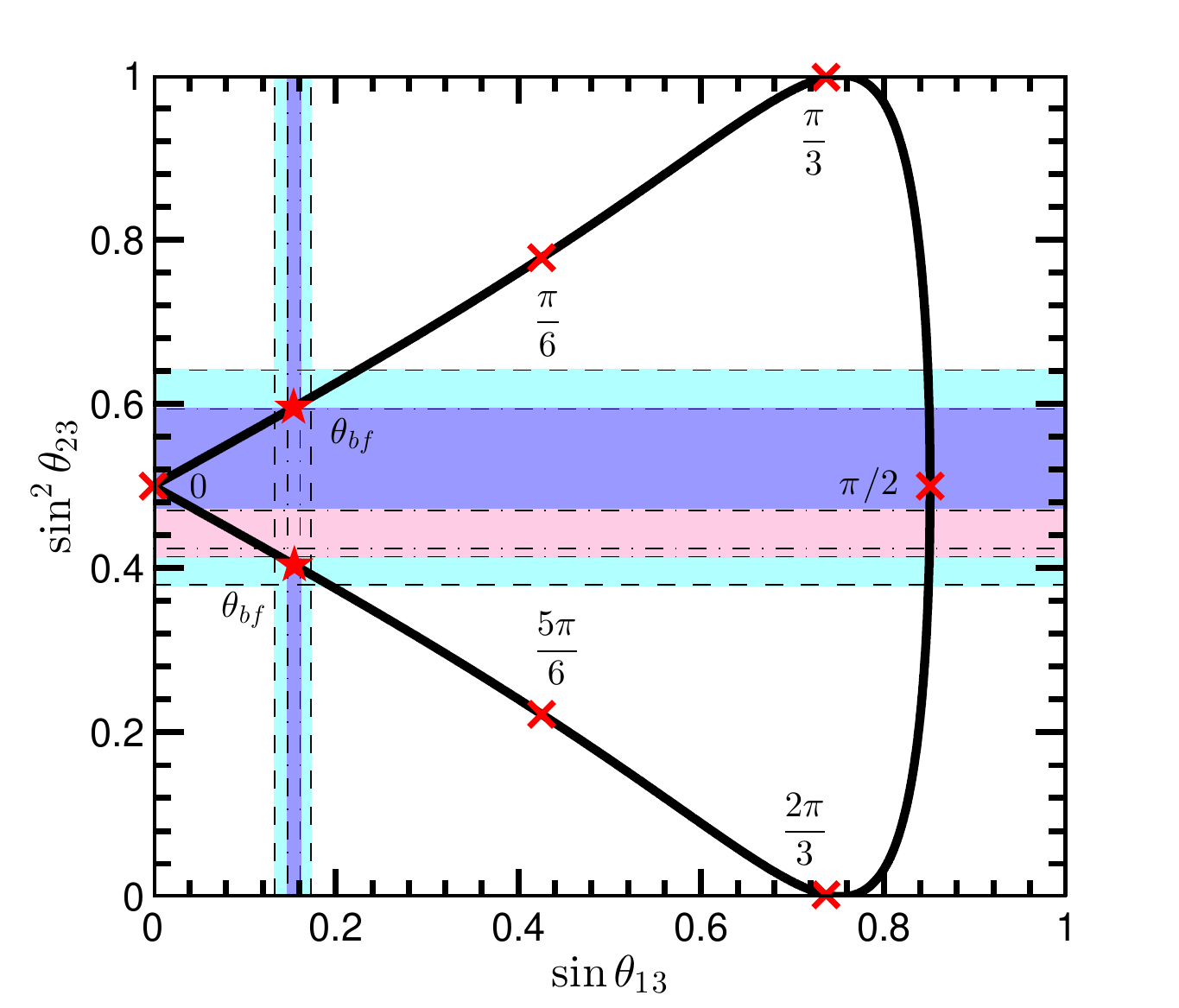}
\hskip-0.30in \includegraphics[width=0.51\textwidth]{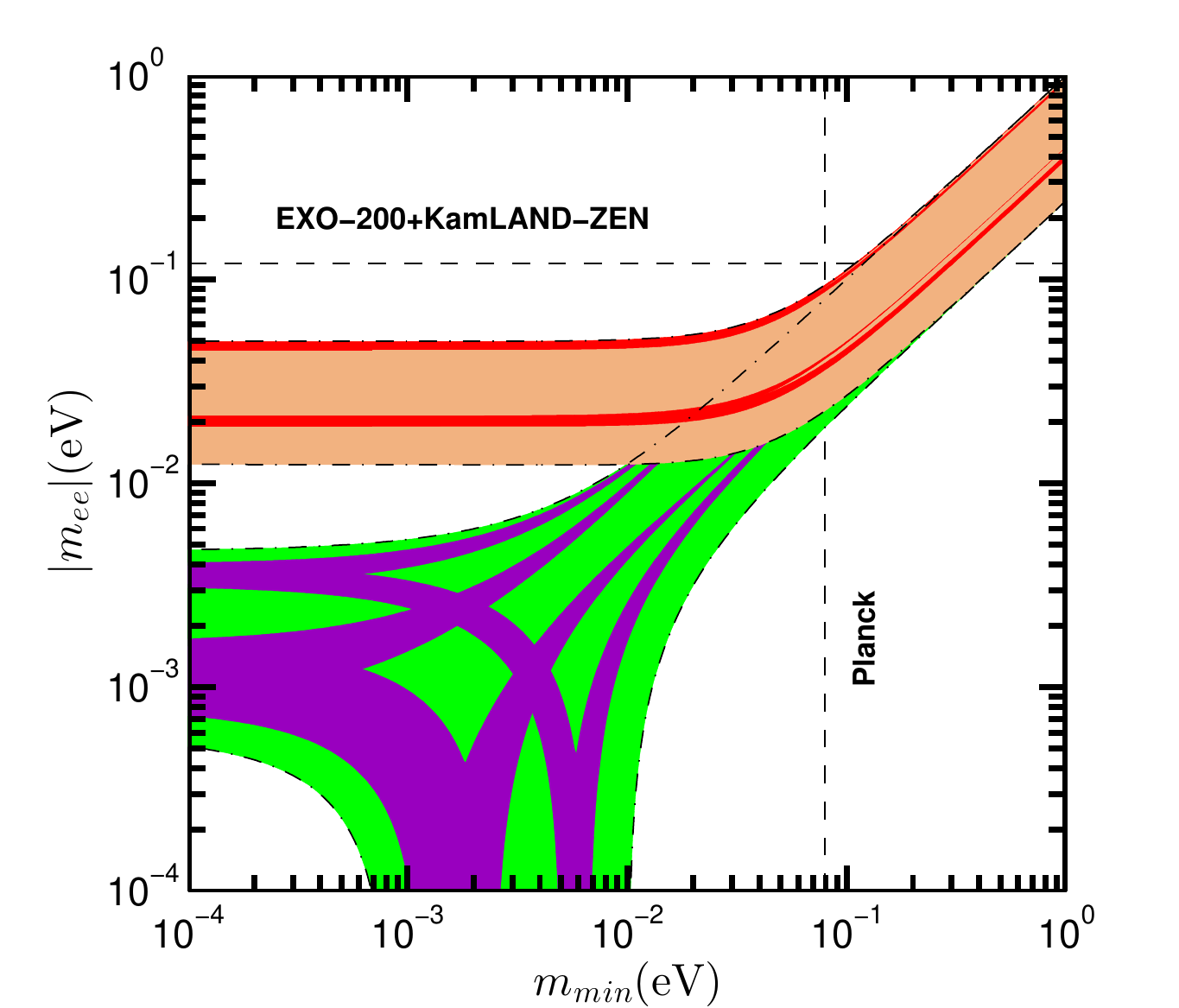}
\caption{\label{fig:general_mixing_caseIII} The correlation among $\sin^{2}\theta_{12}$, $\sin^{2}\theta_{23}$ and $\sin\theta_{13}$ (the former three panels) and the allowed values of the effective mass $|m_{ee}|$ (the last panel) in case III. The global minimum of the $\chi^2$ function is labelled with a red pentagram, and the points for $\theta=0$, $\pi/6$, $\pi/3$, $\pi/2$, $2\pi/3$ and $5\pi/6$ are marked with a cross to guide the eye. The $1\sigma$ and $3\sigma$ ranges of the mixing angles are taken from Ref.~\cite{Capozzi:2013csa}. In the last panel, the orange and green bands denote the $3\sigma$ regions for normal ordering and inverted ordering mass spectrum respectively. The red and purple areas are the predictions for the lepton mixing matrix in Eq.~\eqref{eq:PMNS_caseIII}. The present most strict bound $|m_{ee}|<(0.120-0.250)$ eV from
EXO-200~\cite{Auger:2012ar,Albert:2014awa} combined with KamLAND-ZEN~\cite{Gando:2012zm} is represented
by the horizontal dashed line, and the upper limit on $m_{min}$ from the latest Planck result $m_1+m_2+m_3<0.230$ eV at $95\%$ confidence level~\cite{Ade:2013zuv} is shown by vertical dashed line. }
\end{figure}
Note that the second column vector is $\left( \sqrt{\frac{1}{\sqrt{5}\,\kappa}}, \sqrt{\frac{\kappa}{2\sqrt{5}}}, \sqrt{\frac{\kappa}{2\sqrt{5}}}\right)^{T}$ which coincides with the second column of the GR mixing. The lepton mixing parameters are predicted to be
\begin{eqnarray}\label{eq:mixing_parameters_caseIII}
\nonumber &&\qquad \sin^{2}\theta_{13}=\frac{\kappa}{\sqrt{5}}\sin^2\theta\,, \quad \sin^{2}\theta_{12}=\frac{4-2\kappa}{5-2\kappa+\cos{2\theta}}\,,\\
&&\sin^{2}\theta_{23}=\frac{1}{2}-\frac{\sqrt{3-\kappa}\sin2\theta}{3\kappa-2+\kappa\cos2\theta}\,, \quad \sin\delta_{CP}=\sin\alpha_{21}=\sin\alpha_{31}=0\,.
\end{eqnarray}
We see that $\theta_{23}$ deviates from maximal mixing and all the three CP violating phases are trivial due to a common CP transformation $\rho_{\bf{r}}(1)$ of the charged lepton and neutrino sectors. The mixing angles $\theta_{12}$, $\theta_{13}$ and $\theta_{23}$ only depend on the parameter $\theta$, and they fulfill the following relations,
\begin{equation}
\sin^{2}\theta_{12}\cos^2\theta_{13}=\frac{3-\kappa}{5},\quad \sin^{2}\theta_{23}=\frac{1}{2}\pm(\kappa-1)\tan\theta_{13}\sqrt{1+(\kappa-2)\tan^{2}\theta_{13}}\;,
\end{equation}
which are plotted in Fig.~\ref{fig:general_mixing_caseIII}. Obviously the mixing angles can be very close to the their measured values for certain values of the parameter $\theta$. The global minimum of the $\chi^2$ function is rather small, as shown in Table~\ref{tab:best_fitting}. The predictions for the effective mass $\left|m_{ee}\right|$ are also displayed in Fig.~\ref{fig:general_mixing_caseIII}.

\item[~~(\uppercase\expandafter{\romannumeral4})] $X_{\nu{\bf r}}=\rho_{\bf r}(S),\rho_{\bf r}(T^3ST^2ST^3S)$\\
Invariance of the neutrino mass matrix $m_{\nu}$ under the action of these residual CP transformations implies that $\alpha$, $\beta$, $\gamma$ are real while $\delta$ is pure imaginary. The diagonalization matrix of $m^{\prime}_{\nu}$ is
\begin{equation}
U^{\prime}_{\nu}=\left(\begin{array}{ccc}
   ~i\cos\theta  &~  0  ~&  i\sin\theta  \\
  0  &~ 1 ~&  0 \\
  -\sin\theta  &~  0 ~&  \cos\theta
  \end{array}\right)\,,
\end{equation}
where
\begin{equation}
\tan{2\theta}=-\frac{4 i \sqrt{2+\kappa}~\delta }{3 (\kappa \beta  -2 \gamma)}\,.
\end{equation}
The neutrino masses are given by
\begin{eqnarray}\label{eq:nu_mass_caseIV}
\nonumber && m_{1}=\frac{1}{2}\left|-2\alpha+(3\kappa-2) \beta +2 \gamma+\frac{3 (\kappa\beta  -2 \gamma )}{\cos{2\theta}}\right|\,,\\
\nonumber && m_{2}=\left|\alpha+(3 \kappa-2) \beta  +2\gamma\right|\,, \\
&& m_{3}=\frac{1}{2}\left|-2\alpha+(3\kappa-2) \beta +2 \gamma-\frac{3 (\kappa\beta  -2 \gamma )}{\cos{2\theta}}\right|\,.
\end{eqnarray}
The PMNS matrix is of the form
\begin{equation}\label{eq:PMNS_caseIV}
U_{PMNS}=U_{GR}U^{\prime}_{\nu}=
\left(\begin{array}{ccc}
 -i\sqrt{\frac{\kappa}{\sqrt{5}}} \cos\theta &~ \sqrt{\frac{1}{\sqrt{5}\kappa}} ~& -i\sqrt{\frac{\kappa}{\sqrt{5}}} \sin\theta \\
 \frac{i\cos\theta}{\sqrt{2\sqrt{5}\kappa}}+\frac{\sin\theta}{\sqrt{2}} &~ \sqrt{\frac{\kappa}{2\sqrt{5}}} ~& \frac{i\sin\theta}{\sqrt{2\sqrt{5}\kappa}}-\frac{\cos\theta}{\sqrt{2}} \\
 \frac{i\cos\theta}{\sqrt{2\sqrt{5}\kappa}}-\frac{\sin\theta}{\sqrt{2}} &~ \sqrt{\frac{\kappa}{2\sqrt{5}}} ~& \frac{i\sin\theta}{\sqrt{2\sqrt{5}\kappa}}+\frac{\cos\theta}{\sqrt{2}}
\end{array}\right)\,.
\end{equation}
The second column has the same form as for the GR mixing. The lepton mixing angles and CP phases are determined to be
\begin{eqnarray}\label{eq:mixing_parameters_caseIV}
 \nonumber &&~ \sin^{2}\theta_{13}=\frac{\kappa}{\sqrt{5}}\sin^{2}\theta\,, \quad \sin^{2}\theta_{12}=\frac{4-2\kappa}{5-2\kappa+\cos{2\theta}}\,,\\
 && \sin^{2}\theta_{23}=\frac{1}{2}\,, \quad \left|\sin\delta_{CP}\right|=1\,, \quad \sin\alpha_{21}=\sin\alpha_{31}=0\,.
\end{eqnarray}
We see that both $\theta_{23}$ and $\delta_{CP}$ are maximal and the two Majorana CP phases $\alpha_{21}$ and $\alpha_{31}$ are trivial. Similar to case III, the relation $\sin^2\theta_{12}\cos^2\theta_{13}=(3-\kappa)/5$ is satisfied as well. The best fitting results for the three mixing angles are listed in Table~\ref{tab:best_fitting}. The predictions for the $(\beta\beta)_{0\nu}-$decay effective mass $|m_{ee}|$ are shown in Fig.~\ref{fig:general_mixing_caseIV}.

\begin{figure}[t!]
\centering
\includegraphics[width=0.65\textwidth]{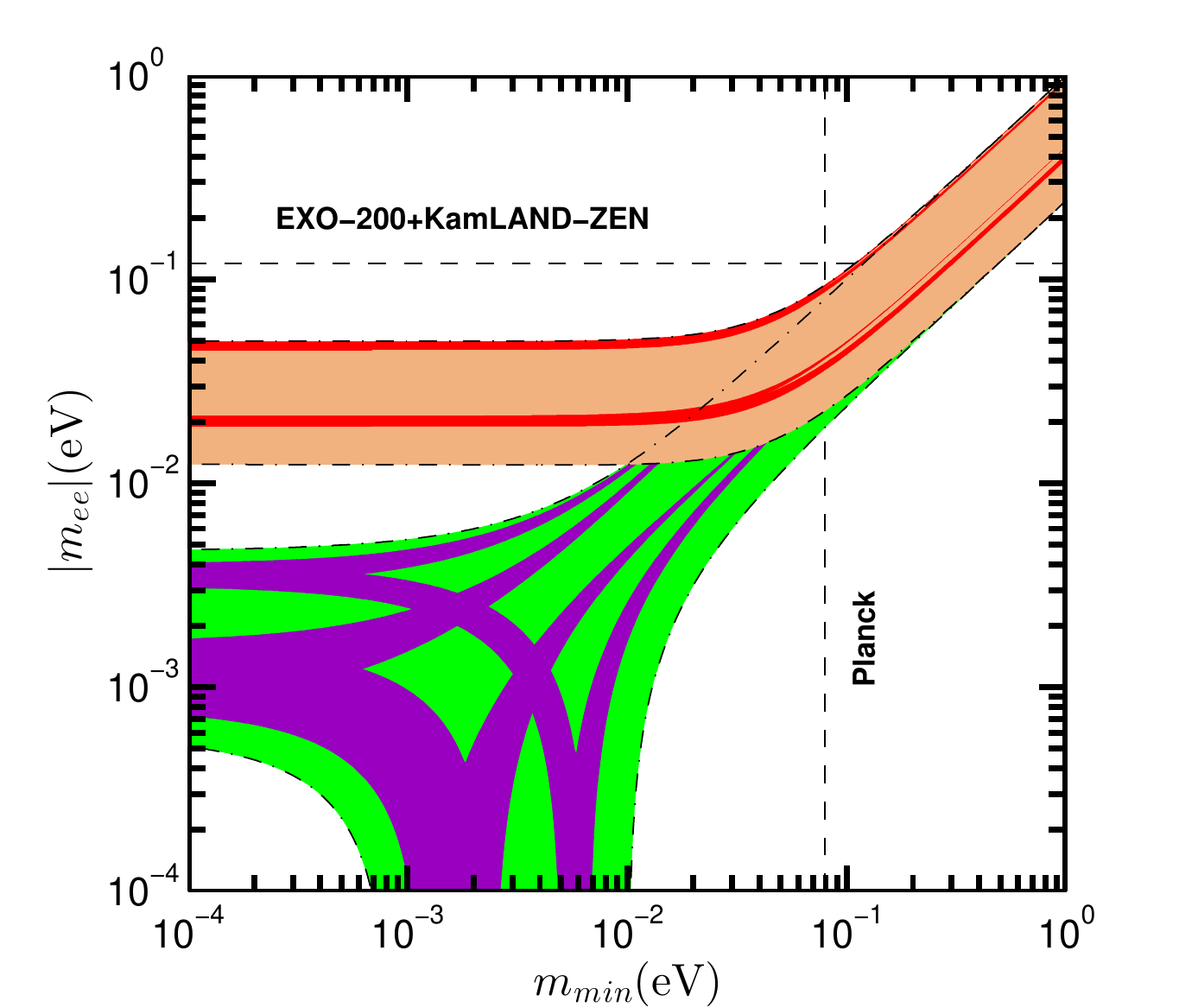}
\caption{\label{fig:general_mixing_caseIV} The $(\beta\beta)_{0\nu}-$decay effective mass $|m_{ee}|$ with respect the lightest neutrino mass $m_{min}$ in case IV. The orange and green bands denote the $3\sigma$ regions for normal ordering and inverted ordering mass spectrum respectively. The red and purple areas are the predictions for the lepton mixing matrix in Eq.~\eqref{eq:PMNS_caseIV}. The present most strict bound $|m_{ee}|<(0.120-0.250)$ eV from EXO-200~\cite{Auger:2012ar,Albert:2014awa} combined with KamLAND-ZEN~\cite{Gando:2012zm} is represented
by the horizontal dashed line, and the upper limit on $m_{min}$ from the latest Planck result $m_1+m_2+m_3<0.230$ eV at $95\%$ confidence level~\cite{Ade:2013zuv} is shown by vertical dashed line. Note that the correlation between $\sin^2\theta_{12}$ and $\sin\theta_{13}$ is the same as that of case III and can be found in Fig.~\ref{fig:general_mixing_caseIII}. }
\end{figure}
\end{description}

\subsection{\label{sec:3.2.3} $G_{l}=Z^{T^3ST^2S}_{3}$, $G_{\nu}=Z^{ST^2ST^3S}_{2}$}

In the charged lepton sector, the remnant CP transformation $H^{l}_{CP}$ is determined by the consistency condition
\begin{equation}
X_{l\mathbf{r}}\rho^{*}_{\mathbf{r}}(T^3ST^2S) X^{-1}_{l\mathbf{r}}=\rho_{\mathbf{r}}(g^{\prime}), \qquad g^{\prime}\in Z^{T^3ST^2S}_{3}\,.
\end{equation}
We find that there are 6 possible solutions for $X_{l\mathbf{r}}$, i.e.,
\begin{equation}\label{eq:CP_transformation_Z31}
 H^{l}_{CP}=\{\rho_{\bf r}(ST^3),\rho_{\bf r}(ST^3S),\rho_{\bf r}(T^3),\rho_{\bf r}(T^3S),\rho_{\bf r}(T^3ST^2ST^3),\rho_{\bf r}(T^3ST^2ST^3S)\}\,.
\end{equation}
The charged lepton mass matrix should respect both the remnant family symmetry $Z^{T^3ST^2S}_{3}$ and the remnant CP symmetry $H^{l}_{CP}$ :
\begin{equation}\label{eq:rem_flavour_ch}
\rho^{\dagger}_{\mathbf{3}}(T^3ST^2S)m^{\dagger}_{l}m_{l}\rho_{\mathbf{3}}(T^3ST^2S)=m^{\dagger}_{l}m_{l}, \quad X^{\dagger}_{l\bf{3}}m^{\dagger}_{l}m_{l}X_{l\bf{3}}=(m^{\dagger}_{l}m_{l})^{*}, \quad  X_{l\bf{3}}\in H^{l}_{CP}\,.
\end{equation}
Notice that the three residual CP transformations $X_{l\bf{r}}=$$\rho_{\bf r}(ST^3)$, $\rho_{\bf r}(T^3S)$, $\rho_{\bf r}(T^3ST^2ST^3S)$ lead to degenerate charged lepton masses since both $\rho_{\bf r}(ST^3)$ and $\rho_{\bf r}(T^3S)$ are not symmetric. For the remaining ones $X_{l\bf{r}}=$$\rho_{\bf r}(ST^3S)$, $\rho_{\bf r}(T^3)$, $\rho_{\bf r}(T^3ST^2ST^3)$, the hermitian combination $m^{\dagger}_{l}m_{l}$ is constrained to take the following form
\begin{eqnarray}
\hskip-0.30in m^{\dagger}_{l}m_{l}=\left(\begin{array}{ccc}
a &~ 2\left(\kappa b+\sqrt{2}( 2\kappa-3)c\right)e^{-\frac{3\pi i}{5}} ~& 2\kappa be^{-\frac{2\pi i}{5}} \\
2\left(\kappa b+\sqrt{2}( 2\kappa-3)c\right)e^{\frac{3\pi i}{5}} &~ a+\frac{\sqrt{2}}{\kappa}b+(8\kappa-14)c ~&  2(\kappa-1)ce^{\frac{\pi i}{5}}  \\
2\kappa be^{\frac{2\pi i}{5}} &~ 2(\kappa-1)ce^{-\frac{\pi i}{5}} ~& a-\frac{\sqrt{2}}{\kappa}(b+\sqrt{2} c)
\end{array}\right),
\end{eqnarray}
where $a$, $b$ and $c$ are real parameters. It can be diagonalized by the unitary matrix
\begin{equation}
U_{l}=\left(\begin{array}{ccc}
  \sqrt{\frac{7-4\kappa}{15}}e^{-\frac{2\pi i}{5}} &~  \sqrt{\frac{2\sqrt{5}\,\kappa}{15} }e^{\frac{3\pi i}{5}} ~&  \sqrt{\frac{2\sqrt{5}\,\kappa}{15}}e^{-\frac{2\pi i}{5}} \\
  \sqrt{\frac{2\sqrt{5}\,\kappa}{15}}e^{-\frac{4\pi i}{5}} &~ \frac{1}{2} \left(1-\sqrt{\frac{7-4\kappa}{15}}\right)e^{\frac{\pi i}{5}} ~& \frac{1}{2} \left(1+\sqrt{\frac{7-4\kappa }{15}}\right)e^{\frac{\pi i}{5}} \\
 \sqrt{\frac{2\sqrt{5}\,\kappa}{15} } &~ \frac{1}{2} \left(1+\sqrt{\frac{7-4\kappa}{15}}\right) ~& \frac{1}{2} \left(1-\sqrt{\frac{7-4\kappa}{15}}\right) \\
\end{array}\right)\,,
\end{equation}
with $U^{\dagger}_{l} m^{\dagger}_{l}m_{l}U_{l}=\textrm{diag}(m^{2}_{e},m^{2}_{\mu},m^{2}_{\tau})$, where the charged lepton masses are
\begin{eqnarray}
\nonumber&&m^{2}_{e}=a-4(\kappa-1)c, \quad m^{2}_{\mu}=a-\sqrt{6(2+\kappa)}~b-\left(8-5\kappa+\sqrt{3(47-29\kappa)}\right)c\,,\\
&&m^{2}_{\tau}=a+\sqrt{6(2+\kappa)}~b+\left(5\kappa-8+\sqrt{3(47-29\kappa)}\right)c\,.
\end{eqnarray}
The symmetry group $A_{5}\rtimes H_{CP}$ is broken into $Z^{ST^2ST^3S}_{2}\times H^{\nu}_{CP}$ in the neutrino sector. By solving the restricted consistency equation of Eq.~\eqref{eq:consistency_remnant_nu}, we find
\begin{eqnarray}\label{eq:CP_transformations_Z21}
H^\nu_{CP}=\left\{\rho_{\bf r}(T^2), \rho_{\bf r}(TST), \rho_{\bf r}(T^3ST^2ST^3S), \rho_{\bf r}((ST^2)^2S)\right\}\,.
\end{eqnarray}
The neutrino mass matrix preserving the remnant family symmetry $G_{\nu}=Z^{ST^2ST^3S}_{2}$ is of the form
{\small
\begin{equation}
m_{\nu}=\alpha
\left(\begin{array}{ccc}
 1 & 0 & 0 \\
 0 & 0 & 1 \\
 0 & 1 & 0 \\
\end{array}\right)
+\beta
\left(\begin{array}{ccc}
 2 & 0 & 0 \\
 0 &~ 3 e^{-\frac{4\pi i}{5}} & -1 \\
 0 & -1 & 3 e^{\frac{4\pi i}{5}} \\
\end{array}\right)
+\gamma
\left(\begin{array}{ccc}
 0 & e^{\frac{3\pi i}{5}} & e^{-\frac{3\pi i}{5}} \\
 e^{\frac{3\pi i}{5}} &  \sqrt{2}e^{\frac{\pi i}{5}} & 0 \\
 e^{-\frac{3\pi i}{5}} & 0 &  \sqrt{2}e^{-\frac{\pi i}{5}} \\
\end{array}\right)
+\delta
\left(\begin{array}{ccc}
 2 \sqrt{2} & e^{-\frac{2\pi i}{5}} & e^{\frac{2\pi i}{5}} \\
 e^{-\frac{2\pi i}{5}} &  \sqrt{2}e^{\frac{\pi i}{5}} & -\sqrt{2} \\
 e^{\frac{2\pi i}{5}} & -\sqrt{2} &  \sqrt{2}e^{-\frac{\pi i}{5}} \\
\end{array}\right)\,,
\end{equation}}
where parameters $\alpha$, $\beta$, $\gamma$ and $\delta$ are generally complex, and they are further constrained to be either real or imaginary by CP symmetry. It is convenient to firstly perform a constant unitary transformation $U_{GRP}$ and yield
\begin{eqnarray}
\nonumber m^{\prime}_{\nu}&=&U^{T}_{GRP}m_{\nu}U_{GRP}\\
&=&\left(\begin{array}{ccc}
 \alpha +2 \beta -\sqrt{2(1+\kappa)}\, \gamma  & 0 & -\sqrt{10}\,\delta  \\
 0 & -\alpha +4 \beta -\sqrt{2}\,\gamma  & 0 \\
 -\sqrt{10}\,\delta  & 0 & \alpha +2 \beta +\sqrt{2(2-\kappa)}\,\gamma
\end{array}\right)\,,
\end{eqnarray}
where
\begin{equation}
U_{GRP}=
\left(\begin{array}{ccc}
 \sqrt{\frac{1}{\sqrt{5}\kappa}} &~ 0 ~& - \sqrt{\frac{\kappa}{\sqrt{5}}} \\
 \sqrt{\frac{\kappa}{2\sqrt{5}}}e^{\frac{2\pi i}{5}} &~ \frac{1}{\sqrt{2}}e^{-\frac{3\pi i}{5}} ~& \frac{1}{\sqrt{2\sqrt{5}\kappa}}e^{\frac{2\pi i}{5}} \\
 \sqrt{\frac{\kappa}{2\sqrt{5}}}e^{-\frac{2\pi i}{5}} &~ \frac{1}{\sqrt{2}}e^{-\frac{2\pi i}{5}} ~& \frac{1}{\sqrt{2\sqrt{5}\kappa}}e^{-\frac{2\pi i}{5}} \\
\end{array}\right)\,.
\end{equation}
Next we discuss the constraints of the residual CP symmetry on the neutrino mass matrix $m_{\nu}$.

\begin{description}[labelindent=-0.7em, leftmargin=0.1em]
\item[~~(\uppercase\expandafter{\romannumeral5})] $X_{\nu{\bf r}}=\rho_{\bf r}(T^2), \rho_{\bf r}(T^3ST^2ST^3S)$\\
In this case, $\alpha$, $\beta$, $\gamma$ and $\delta$ are determined to be real. Then neutrino mass matrix $m^{\prime}_{\nu}$ is a real symmetric matrix, and it can be diagonalized by a rotation matrix $U^{\prime}_{\nu}$ in the (2,3) sector,
\begin{equation}
U^{\prime}_{\nu}=
\left(\begin{array}{ccc}
 \cos \theta  &~ 0 ~& -\sin \theta  \\
 0 &~ 1 ~& 0 \\
 \sin \theta  &~ 0 ~& \cos \theta  \\
\end{array}\right)\,,
\end{equation}
with
\begin{equation}
\tan{2\theta}=2 \delta /\gamma \,.
\end{equation}
The three light neutrino masses are given by
\begin{eqnarray}
\nonumber && m_{1}= \frac{1}{2}\left|2\alpha+4\beta-\sqrt{2}\gamma-\frac{\sqrt{10}\gamma}{\cos{2\theta}}\right| \,, \\
\nonumber && m_{2}=\left|-\alpha +4 \beta -\sqrt{2} \gamma\right|\,, \\
&& m_{3}=\frac{1}{2}\left|2\alpha+4\beta-\sqrt{2}\gamma+\frac{\sqrt{10}\gamma}{\cos{2\theta}}\right| \,.
\end{eqnarray}
The lepton mixing matrix is of the form
\begin{equation}\label{eq:PMNS_caseV}
U_{PMNS}=U^{\dagger}_{l}U_{GRP}U^{\prime}_{\nu}=\frac{1}{\sqrt{3}}
\left(\begin{array}{ccc}
 \cos \theta +\sin \theta &~ 1 ~& \cos \theta -\sin \theta \\
  e^{\frac{2\pi i}{3}} \cos \theta -e^{\frac{\pi i}{3}}\sin \theta &~ 1 ~& e^{\frac{4\pi i}{3}} \cos \theta -e^{\frac{2\pi i}{3}}\sin\theta \\
 e^{\frac{4\pi i}{3}}\cos \theta +e^{\frac{2\pi i}{3}}\sin\theta &~ 1 ~& e^{\frac{2\pi i}{3}}\cos \theta +e^{\frac{\pi i}{3}} \sin \theta \\
\end{array}\right)\,.
\end{equation}
We see that the second column of the PMNS matrix is $\left(1, 1, 1\right)^{T}/\sqrt{3}$, which frequently appears in discrete flavor symmetry models. The leptonic mixing parameters read as\footnote{For $\cos2\theta=0$, we have $\sin\theta_{13}=0$ or $\cos\theta_{12}=0$ so that $\delta_{CP}$ cannot be determined uniquely. }
\begin{eqnarray}
\nonumber && \sin^2\theta_{13}=\frac{1}{3}(1-\sin{2\theta}), \quad \sin^2\theta_{12}=\frac{1}{2+\sin{2\theta}}, \quad \sin^2\theta_{23}=\frac{1}{2}\,, \\
&& \left|\sin\delta_{CP}\right|=1, \quad \sin\alpha_{21}=\sin\alpha_{31}=0\,.
\end{eqnarray}
Both Dirac CP phase and $\theta_{23}$ are maximal while Majorana CP phases are conserved in this case. In common with all trimaximal mixings, $\theta_{12}$ and $\theta_{13}$ are related with each other by
\begin{equation}
3\sin^{2}\theta_{12}\cos^2\theta_{13}=1\,.
\end{equation}
The measured $3\sigma$ range $0.0176\leq\sin^2\theta_{13}\leq0.0295$~\cite{Capozzi:2013csa} gives rise to $0.339\leq\sin^2\theta_{12}\leq0.343$ which can be directly tested by JUNO in near future~\cite{JUNO}. The correlation between $\theta_{12}$ and $\theta_{13}$ and the predictions for the $(\beta\beta)_{0\nu}-$decay are displayed in Fig.~\ref{fig:general_mixing_caseV}. All the three mixing angles can agree within $3\sigma$ with the experimental data for certain values of $\theta$. The best fitting results are listed in Table~\ref{tab:best_fitting}, and the minimum values of the $\chi^{2}$ functions are $3.987$ and $7.480$ for IO and NO, respectively.
\begin{figure}[t!]
\centering
\includegraphics[width=0.51\textwidth]{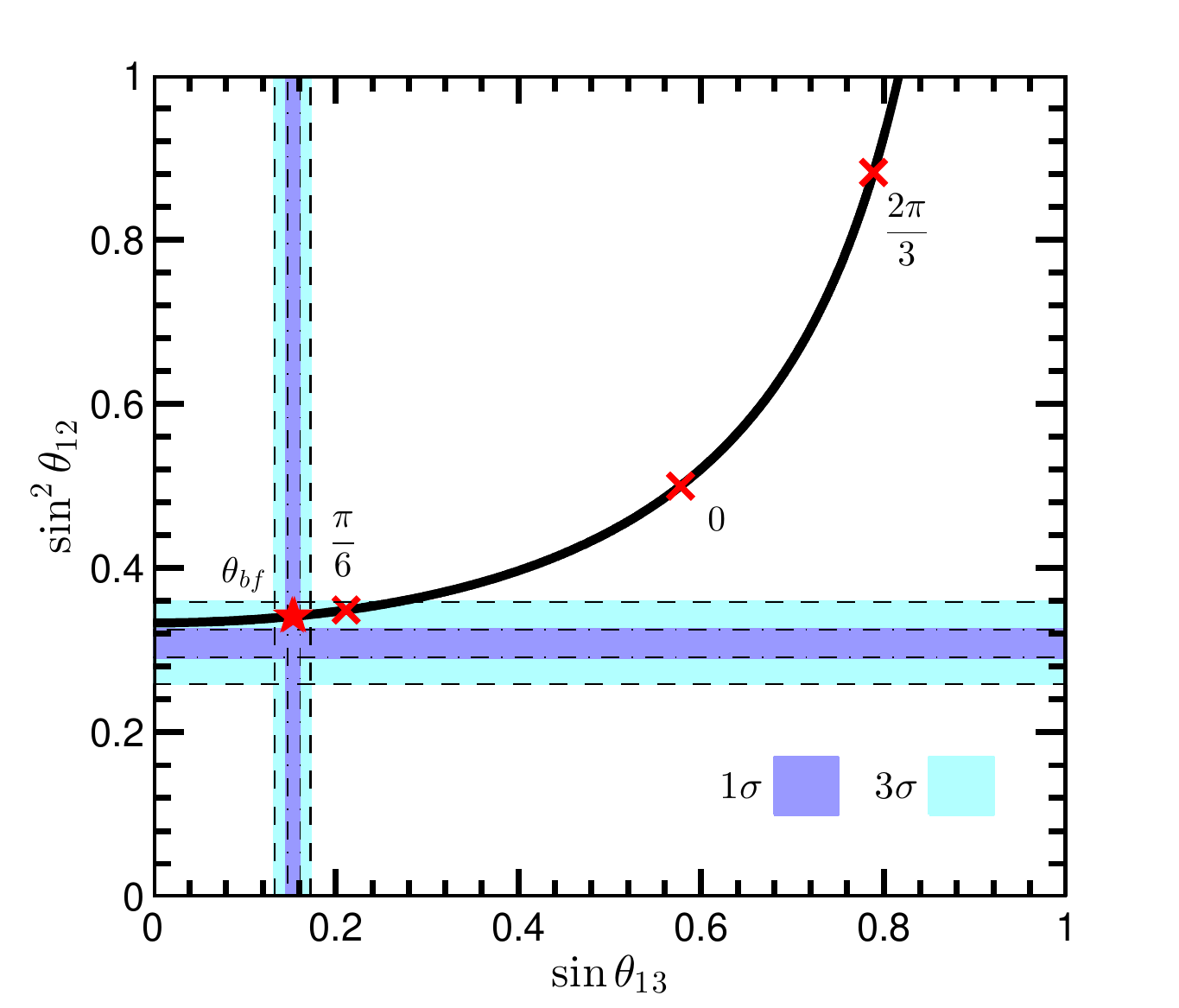}
\hskip-0.30in \includegraphics[width=0.51\textwidth]{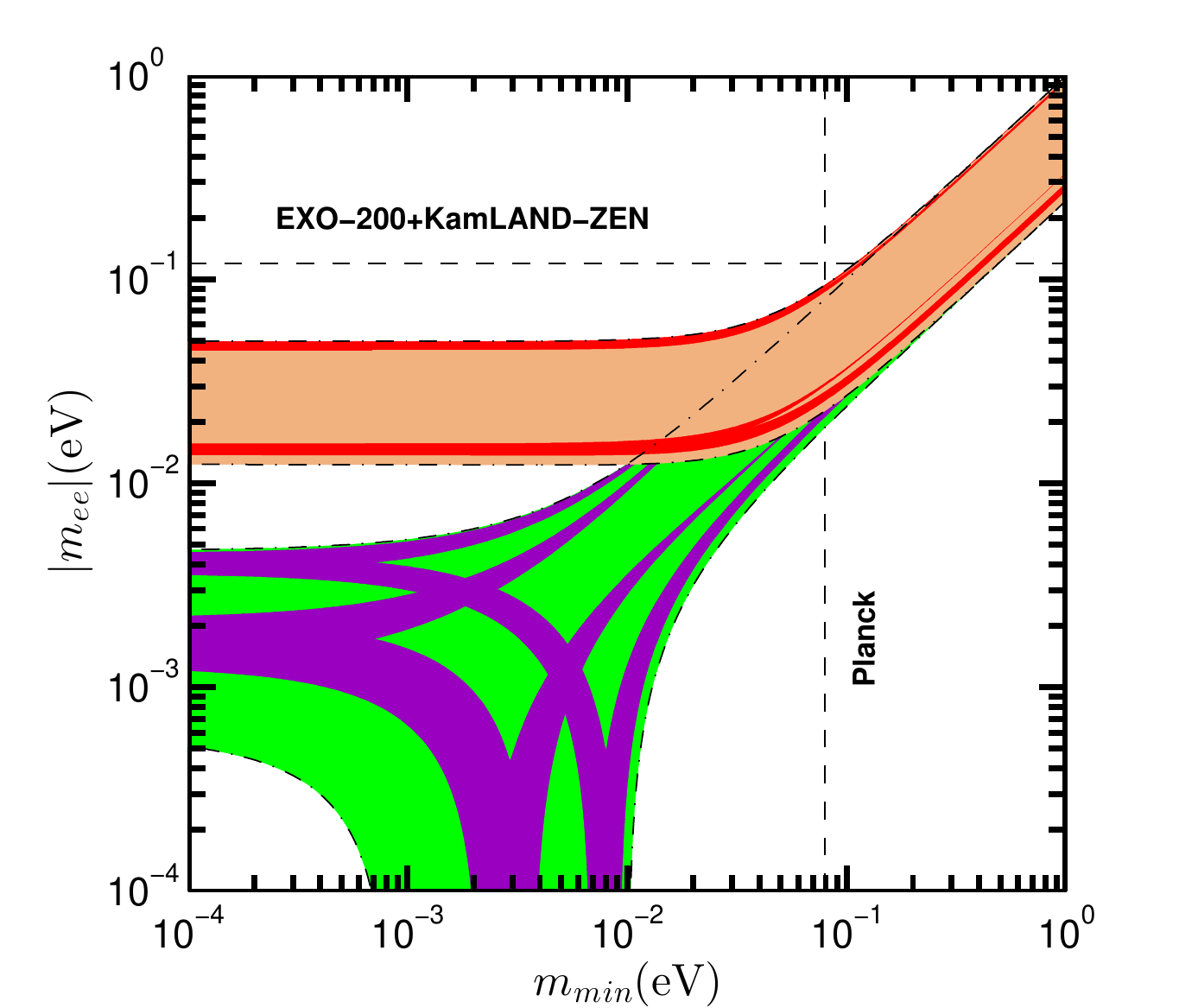}
\caption{\label{fig:general_mixing_caseV} Results for $\sin^{2}\theta_{12}$ and $\sin\theta_{13}$ (left panel) and the allowed values of the effective mass $|m_{ee}|$ (right panel) in case V. On the left panel, the best fitting value is labelled with a red pentagram, and the points for $\theta=0$, $\pi/6$ and $2\pi/3$ are marked with a cross to guide the eye. The $1\sigma$ and $3\sigma$ ranges of the mixing angles are taken from Ref.~\cite{Capozzi:2013csa}. On the right panel, the orange and green bands denote the $3\sigma$ regions for normal ordering and inverted ordering mass spectrum respectively. The red and purple areas are the predictions for the lepton mixing matrix in Eq.~\eqref{eq:PMNS_caseV}. The present most strict bound $|m_{ee}|<(0.120-0.250)$ eV from
EXO-200~\cite{Auger:2012ar,Albert:2014awa} combined with KamLAND-ZEN~\cite{Gando:2012zm} is represented
by the horizontal dashed line, and the upper limit on $m_{min}$ from the latest Planck result $m_1+m_2+m_3<0.230$ eV at $95\%$ confidence level~\cite{Ade:2013zuv} is shown by vertical dashed line.}
\end{figure}

\item[~~(\uppercase\expandafter{\romannumeral6})] $X_{\nu{\bf r}}=\rho_{\bf r}(TST), \rho_{\bf r}((ST^2)^2S)$\\
The requirement of real $\alpha$, $\beta$, $\gamma$ and pure imaginary $\delta$ follows immediately from the remnant CP invariant condition.
In the same way as previous cases, the PMNS mixing matrix is found to be
\begin{equation}
U_{PMNS}=\frac{1}{\sqrt{3}}
\left(\begin{array}{ccc}
 e^{\frac{5\pi i}{6}} \cos \theta +e^{\frac{2\pi i}{3}}\sin \theta  &~ 1 ~& e^{\frac{2\pi i}{3}}\cos \theta -e^{\frac{5\pi i}{6}} \sin \theta  \\
 e^{\frac{\pi i}{6}}\cos \theta -e^{\frac{\pi i}{3}} \sin \theta  &~ 1 ~& e^{\frac{4\pi i}{3}} \cos \theta +e^{\frac{7\pi i}{6}}\sin \theta  \\
 \sin \theta -i \cos \theta  &~ 1 ~& \cos \theta +i \sin \theta  \\
\end{array}\right)\,,
\end{equation}
The expressions for the lepton mixing parameters are as follows,
\begin{eqnarray}
\nonumber && \sin^{2}\theta_{13}=\frac{1}{3}-\frac{\sqrt{3}\sin{2\theta}}{6}, \quad \sin^{2}\theta_{12}=\frac{2}{4+\sqrt{3}\sin{2\theta}}\,, \\ \nonumber &&\sin^{2}\theta_{23}=\frac{2+\sqrt{3}\sin{2\theta}}{4+\sqrt{3}\sin{2\theta}}, \quad
\left|\sin\delta_{CP}\right|=\left|\frac{8\cos{2\theta+\sqrt{3}\sin{4\theta}}}{2(2+\sqrt{3}\sin{2\theta})\sqrt{4-2\sqrt{3}\sin{2\theta}}}\right|\,, \\
&& \left|\sin{\alpha_{21}}\right|=\left|\frac{2 \sin2\theta +\sqrt{3}}{2+\sqrt{3} \sin2 \theta}\right|, \quad \left|\sin{\alpha^{\prime}_{31}}\right|=\left|\frac{4 \sqrt{3} \cos2 \theta}{5+3 \cos4 \theta}\right|\,,
\end{eqnarray}
where $\alpha^{\prime}_{31}=\alpha_{31}-2\delta_{CP}$. It is remarkable that all the three CP violating phases nontrivially depend on the parameter $\theta$. However, we see that in case of $\theta=\pi/4$ the minimum value of $\theta_{13}$ is obtained with $\sin^2\theta_{13}|_{\theta=\pi/4}=(2-\sqrt{3})/6\simeq0.0447$ which is outside the $3\sigma$ range~\cite{Capozzi:2013csa}. Furthermore, we note that the atmospheric angle $\theta_{23}$ is the complementary angle of $\theta_{12}$ or is equal to $\theta_{12}$ if the second and the third rows of the PMNS matrix is interchanged. As a result, this mixing pattern is not compatible with experimental data and consequently we don't included it in Table~\ref{tab:best_fitting}.

\end{description}

\subsection{\label{sec:3.2.4}$G_{l}=K^{(ST^2ST^3S,TST^4)}_{4}$, $G_{\nu}=Z^{S}_{2}$}

In the last case, the residual symmetries are assumed to be  $K^{(ST^2ST^3S,TST^4)}_{4}\rtimes H^{l}_{CP}$ in the charged lepton sector and $Z^{S}_{2}\times H^{\nu}_{CP}$ in the neutrino sector. For the remnant family symmetry $K^{(ST^2ST^3S,TST^4)}_{4}$ to hold, the mass matrix $m^{\dagger}_{l}m_{l}$ has to fulfill
\begin{equation}
\rho^{\dagger}_{\mathbf{3}}(ST^2ST^3S)m^{\dagger}_{l}m_{l}\rho_{\mathbf{3}}(ST^2ST^3S)=m^{\dagger}_{l}m_{l},\qquad
\rho^{\dagger}_{\mathbf{3}}(TST^4)m^{\dagger}_{l}m_{l}\rho_{\mathbf{3}}(TST^4)=m^{\dagger}_{l}m_{l}\,.
\end{equation}
Then $m^{\dagger}_{l}m_{l}$ is constrained to take the form
\begin{equation}
m^{\dagger}_{l}m_{l}=
\left(\begin{array}{ccc}
 a &~ 2\kappa be^{-\frac{2\pi i}{5}} ~& 2\kappa be^{\frac{2\pi i}{5}} \\
 2\kappa be^{\frac{2\pi i}{5}}  &~ a+2\sqrt{2}\kappa b+2(\kappa-1)c ~& 2(\kappa-1)ce^{-\frac{\pi i}{5}}  \\
 2\kappa be^{-\frac{2\pi i}{5}}  &~ 2(\kappa-1)ce^{\frac{\pi i}{5}}   ~& a+2\sqrt{2}\kappa b+2(\kappa-1)c
\end{array}\right)\,,
\end{equation}
where $a$, $b$ and $c$ are real. It is diagonalized by the unitary matrix
\begin{equation}
\label{eq:Uc_caseVII_VIII}U_{l}=
\left(\begin{array}{ccc}
 \sqrt{\frac{\kappa}{\sqrt{5}}} &~ 0 ~& \sqrt{\frac{1}{\sqrt{5}\,\kappa}} \\
 \frac{1}{\sqrt{2\sqrt{5}\kappa}}e^{-\frac{3\pi i}{5}} &~ \frac{1}{\sqrt{2}}e^{-\frac{\pi i}{10}} ~& \sqrt{\frac{\kappa}{2\sqrt{5}}}e^{\frac{2\pi i}{5}} \\
 \frac{1}{\sqrt{2\sqrt{5}\,\kappa}}e^{\frac{3\pi i}{5}} &~ \frac{1}{\sqrt{2}}e^{\frac{\pi i}{10}} ~& \sqrt{\frac{\kappa}{2\sqrt{5}}}e^{-\frac{2\pi i}{5}} \\
\end{array}\right)\,,
\end{equation}
with $U^{\dagger}_{l}m^{\dagger}_{l} m_{l}U_{l}=\textrm{diag}(m^{2}_{e},m^{2}_{\mu},m^{2}_{\tau})$ where
\begin{equation}
m^{2}_{e}=a-2\sqrt{2}\,b, \quad m^{2}_{\mu}=a+2\sqrt{2}\,\kappa b+4(\kappa-1)c, \quad m^{2}_{\tau}=a+2\sqrt{2}\,\kappa^{2}b\,.
\end{equation}
The mass matrix $m^{\dagger}_{l}m_{l}$ is also subject to the constraint of the residual CP symmetry $H^{l}_{CP}$. It is straightforward to determine that $H^{l}_{CP}$ can take the value
\begin{eqnarray}\label{eq:CP_transformation_K41}
\nonumber \hskip-0.17in H^{l}_{CP}=&&\{\rho_{\bf r}(ST^2ST),\rho_{\bf r}((ST^2)^2S),\rho_{\bf r}(ST^3),\rho_{\bf r}(T^2),\rho_{\bf r}((T^2S)^2T^3),\rho_{\bf r}(T^2ST^4),\rho_{\bf r}(T^3S),\\
 && ~\rho_{\bf r}(T^3(ST^2)^2),\rho_{\bf r}(T^3ST^2ST^3S),\rho_{\bf r}(T^4ST^2),\rho_{\bf r}(TST^2S),\rho_{\bf r}(TST)\}\,.
\end{eqnarray}
The twelve CP transformations can be classified into two categories. For
$X_{l\bf{r}}=\rho_{\bf{r}}((ST^2)^2S)$, $\rho_{\bf{r}}(T^2)$, $\rho_{\bf{r}}(T^3ST^{2}ST^3S)$, $\rho_{\bf{r}}(TST)$, the remnant CP invariant condition $X^{\dagger}_{l\bf{3}}m^{\dagger}_{l}m_{l}X_{l\bf{3}}=(m^{\dagger}_{l}m_{l})^{\ast}$ is automatically satisfied, and therefore no additional constraint is produced. Nevertheless, the remaining eight CP transformations $X_{l\bf{r}}=\rho_{\bf r}(ST^2ST)$, $\rho_{\bf r}(ST^3)$, $\rho_{\bf r}((T^2S)^2T^3)$, $\rho_{\bf r}(T^2ST^4)$, $\rho_{\bf r}(T^3S)$, $\rho_{\bf r}(T^3(ST^2)^2)$, $\rho_{\bf r}(T^4ST^2)$ and $\rho_{\bf r}(TST^2S)$ are not viable, as they require $b=c=0$ so that the charged lepton mass spectrum is completely degenerate with $m^{2}_{e}=m^{2}_{\mu}=m^{2}_{\tau}=a$. In neutrino sector, the remnant symmetry $Z^{S}_{2}\times H^{\nu}_{CP}$ and its phenomenological implications have been studied in section~\ref{sec:3.2.1}. The neutrino mass matrix $m_{\nu}$ is found to be given by Eq.~\eqref{eq:nu_general_mass_one}, where the parameters $\alpha$, $\beta$ and $\gamma$ are real while $\delta$ is real or pure imaginary depending on the residual CP transformation $X_{\nu{\bf r}}$.

\begin{description}[labelindent=-0.7em, leftmargin=0.1em]
\item[~~(\uppercase\expandafter{\romannumeral7})] $X_{\nu{\bf r}}=\rho_{\bf r}(T^3ST^2ST^3), \rho_{\bf r}(T^3ST^2ST^3S)$\\
In this case, the neutrino mass matrix is diagonalized by the unitary matrix in Eq.~\eqref{eq:PMNS_caseII}. Combining the unitary transformation $U_{l}$ in Eq.~\eqref{eq:Uc_caseVII_VIII} from the charged lepton sector, we obtain the lepton flavor mixing matrix:
\begin{equation}
\label{eq:PMNS_caseVII}U_{PMNS}=\frac{1}{2}
\left(\begin{array}{ccc}
 \kappa  &~  \cos\theta +(\kappa-1)\sin\theta  ~& (\kappa-1)\cos\theta -\sin\theta  \\
 -1 &~ (\kappa-1)\cos\theta +\kappa\sin\theta  ~&  \kappa\cos\theta -(\kappa-1)\sin\theta  \\
 \kappa-1 &~  \sin\theta -\kappa\cos\theta  ~& \cos\theta +\kappa\sin\theta   \\
\end{array}\right)\,,
\end{equation}
where the parameter $\theta$ is specified by Eq.~\eqref{eq:theta_caseII}. The lepton mixing parameters are predicted to be
\begin{eqnarray}
\nonumber && \sin^{2}\theta_{13}=\frac{(\cos{\theta}-\kappa\sin{\theta})^{2}}{4\kappa^{2}}, \qquad \sin^{2}\theta_{12}=\frac{(\kappa\cos{\theta}+\sin{\theta})^{2}}{4\kappa^{2}-(\cos{\theta}-\kappa\sin{\theta})^{2}}\,, \\
\label{eq:mixing_parameters_caseVII}&&\sin^{2}\theta_{23}=\frac{(\kappa^{2}\cos{\theta}-\sin{\theta})^{2}}{4\kappa^{2}-(\cos{\theta}-\kappa\sin{\theta})^{2}}, \qquad
\sin\delta_{CP}=\sin\alpha_{21}=\sin\alpha_{31}=0\,.
\end{eqnarray}
We find all the three CP violating phases $\delta_{CP}$, $\alpha_{21}$ and $\alpha_{31}$ are conserved, this is because that a common CP transformation $\rho_{\bf r}(T^3ST^2ST^3S)$ is shared by the neutrino and charged lepton sectors. In addition, $\theta_{23}$ deviates from maximal value. After some tedious calculations, we find the following relations between the mixing angles
\begin{eqnarray}
\nonumber && \hskip-0.35in 4\cos^{2}\theta_{12}\cos^{2}\theta_{13}=1+\kappa\,, \\
\label{eq:mixing_angles_correlations}&&\hskip-0.35in  5\sin^{2}\theta_{23}=3-\kappa+(1+2\kappa)\tan^{2}\theta_{13}\pm2\kappa\tan\theta_{13}\sqrt{2+\kappa-(2+3\kappa)\tan^{2}\theta_{13}}\,\,,
\end{eqnarray}
which is plotted in Fig.~\ref{fig:general_mixing_caseVII}. For the
$3\sigma$ interval $0.0176\leq\sin^2\theta_{13}\leq0.0295$~\cite{Capozzi:2013csa}, we have $0.326\leq\sin^2\theta_{12}\leq0.334$ and $0.454\leq\sin^2\theta_{23}\leq0.511$, which are in the experimentally favored ranges~\cite{Capozzi:2013csa}. The global minimum of the $\chi^2$ function is rather small 3.503 (1.626) for NO (IO) neutrino mass spectrum, therefore this mixing pattern can describe the experimental data very well. Moreover, we note that the best fitting value of $\theta_{23}$ is in the first octant with $\sin^{2}\theta_{23}(\theta_{bf})=0.480$ (0.486) for NO (IO) spectrum. Agreement with experimental data can also be achieved if the second and third rows of the PMNS matrix in Eq.~\eqref{eq:PMNS_caseVII} are exchanged. Then the atmospheric mixing angle $\theta_{23}$ changes to
\begin{equation}
\label{eq:atmos_exch_caseVII}\sin^{2}\theta_{23}=\frac{\kappa^{2}(\cos{\theta}+\kappa\sin{\theta})^{2}}{4\kappa^{2}-(\cos{\theta}-\kappa\sin{\theta})^{2}}\,,
\end{equation}
and the predictions for the other mixing parameters remain as Eq.~\eqref{eq:mixing_parameters_caseVII}. The allowed region of $\sin^2\theta_{23}$ becomes $0.489\leq\sin^2\theta_{23}\leq0.546$ with the best fitting value $\sin^{2}\theta_{23}(\theta_{bf})=0.510$ (0.513) for NO (IO) spectrum. Obviously $\theta_{23}(\theta_{bf})$ is in the second octant. Comparing with other mixing patterns shown in Table~\ref{tab:best_fitting}, we see that this case gives rise to the smallest $\chi^2_{min}$ for both NO and IO. The above predictions for solar and atmospheric mixing angles could be tested directly in near future, since the next generation neutrino oscillation experiments are expected to reduce the experimental error on $\theta_{12}$ and $\theta_{23}$ to few degrees. The theoretical results for the $(\beta\beta)_{0\nu}-$decay effective mass $|m_{ee}|$ are displayed in Fig.~\ref{fig:general_mixing_caseVII}. Note that interchanging the second and third rows does't matter since $|m_{ee}|$ is independent of $\theta_{23}$. Again, the predictions for IO neutrino spectrum are within the sensitivity of forthcoming experiments.

\begin{figure}[t!]
\centering
\includegraphics[width=0.51\textwidth]{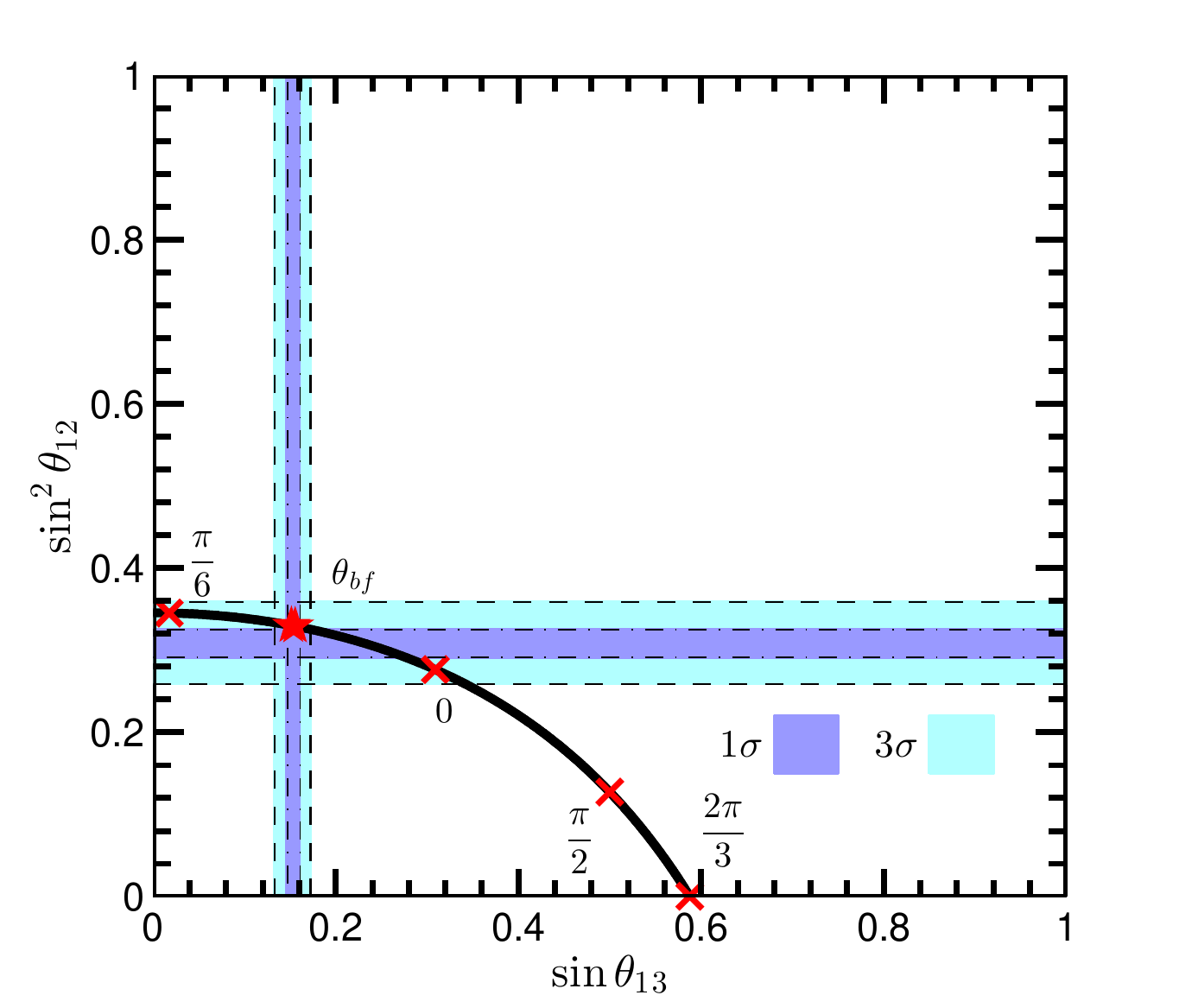}
\hskip-0.30in \includegraphics[width=0.51\textwidth]{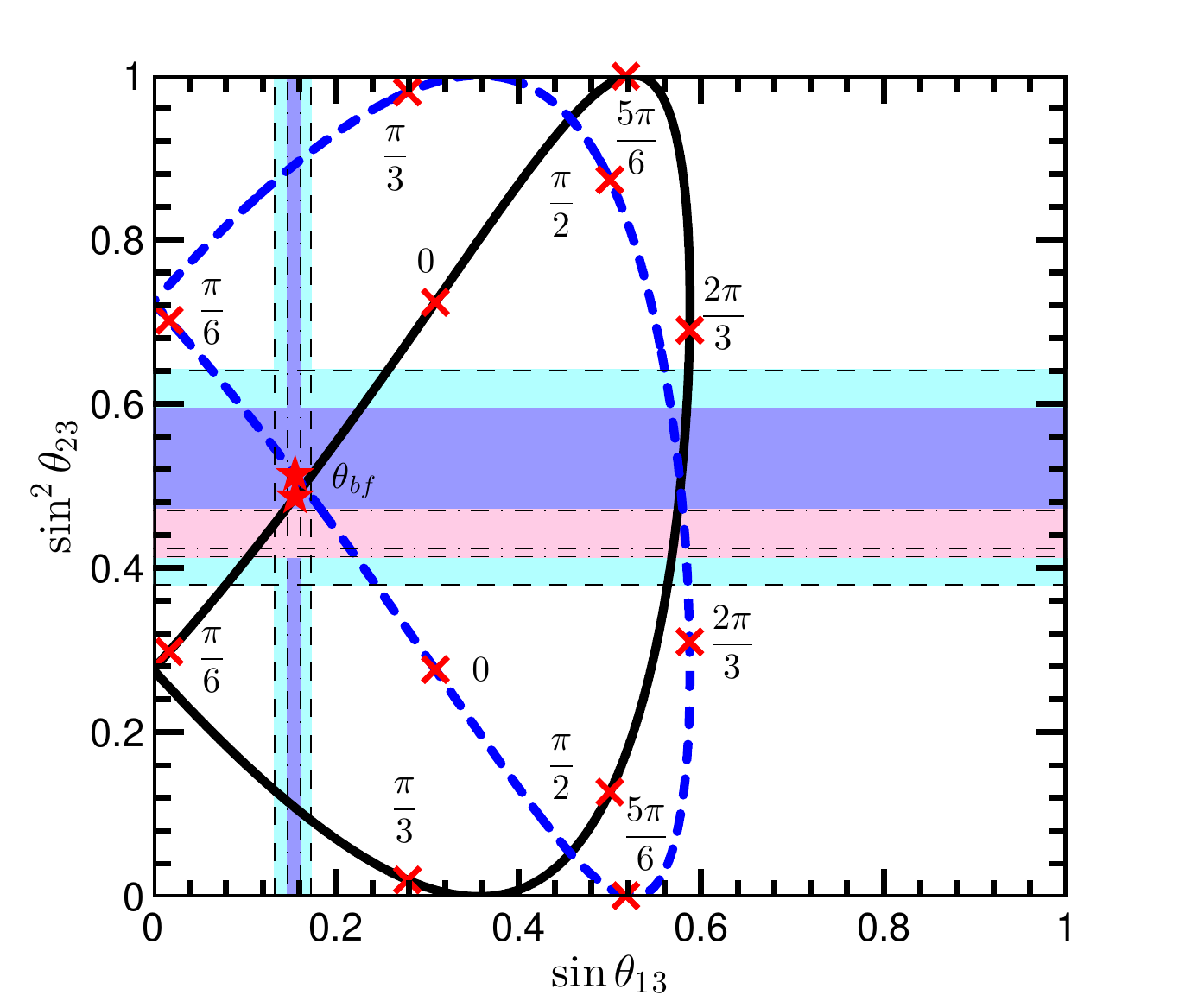} \\
\includegraphics[width=0.51\textwidth]{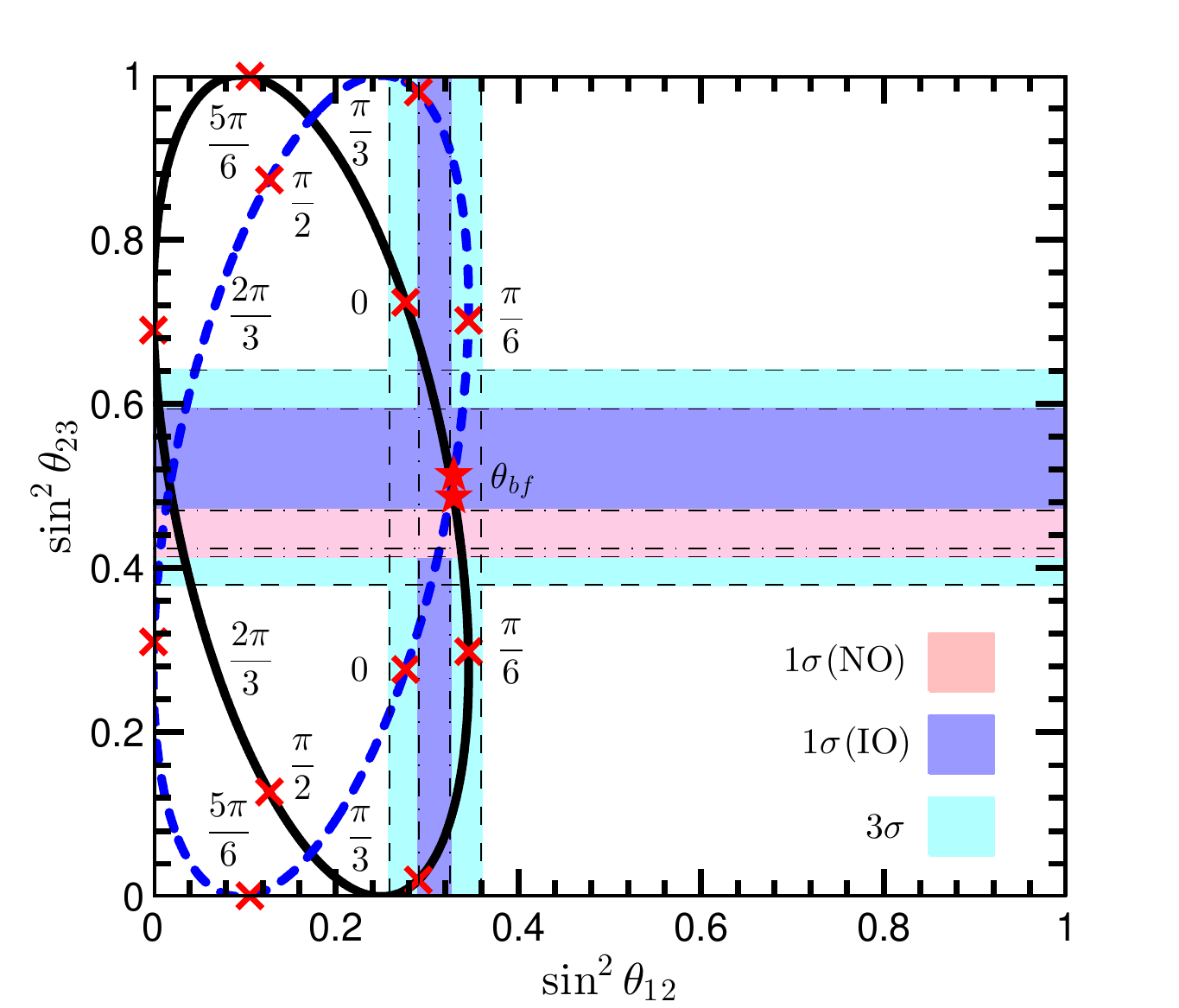}
\hskip-0.30in \includegraphics[width=0.51\textwidth]{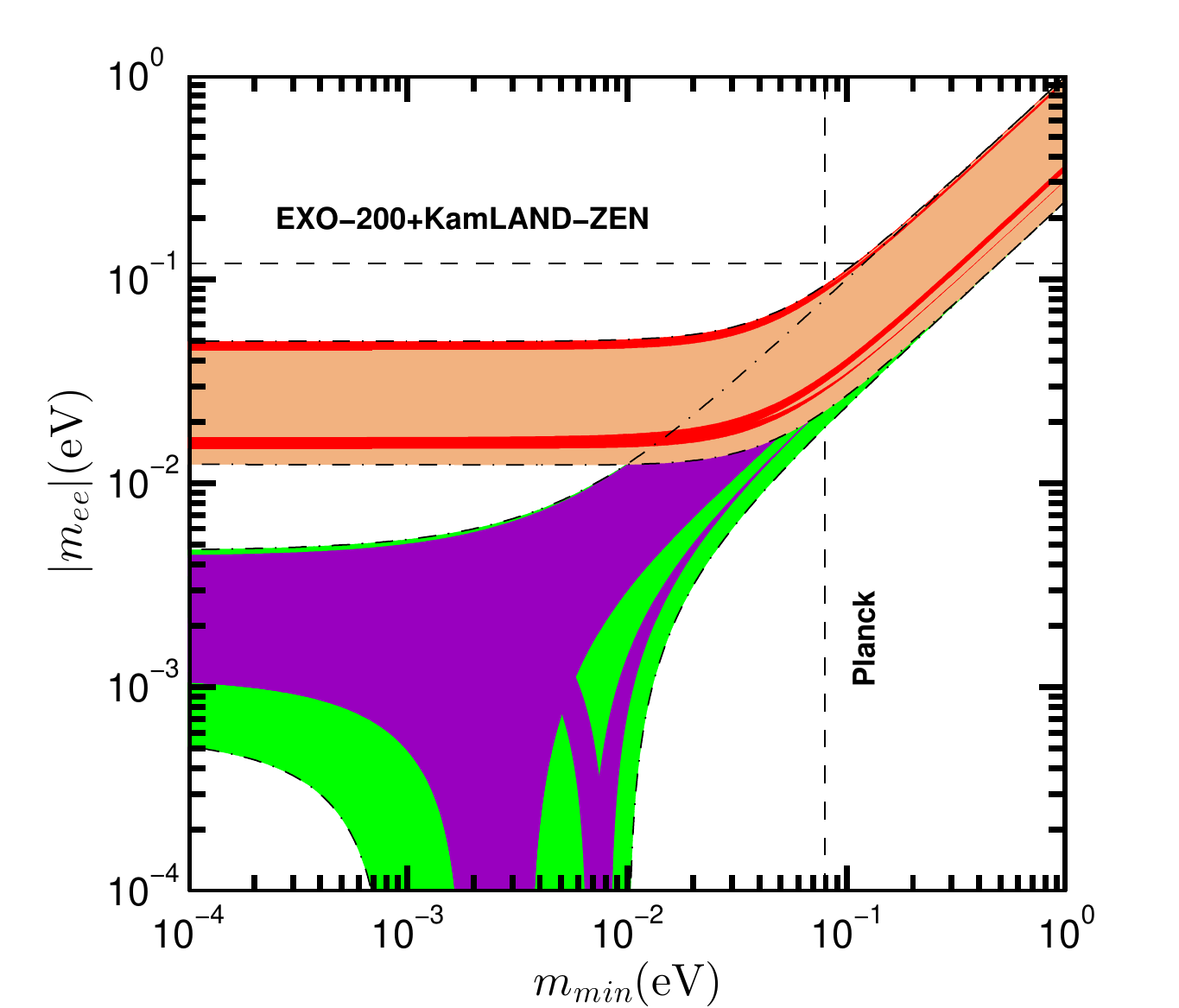}
\caption{\label{fig:general_mixing_caseVII}  The results for $\sin^{2}\theta_{12}$, $\sin^{2}\theta_{23}$ and $\sin\theta_{13}$ (the former three panels) and the allowed values of the effective mass $|m_{ee}|$ (the last panel) in case VII. The global minimum of the $\chi^2$ function is labelled with a red pentagram, and the points for $\theta=0$, $\pi/6$, $\pi/3$, $\pi/2$, $2\pi/3$ and $5\pi/6$ are marked with a cross to guide the eye. The black solid lines and blue dashed lines in the upper-right and lower-left panels represent the two solutions for $\theta_{23}$ shown in Eq.~\eqref{eq:mixing_parameters_caseVII} and Eq.~\eqref{eq:atmos_exch_caseVII} respectively. The corresponding PMNS matrices are related through a exchange of the second and third rows. The $1\sigma$ and $3\sigma$ ranges of the mixing angles are taken from Ref.~\cite{Capozzi:2013csa}. In the last panel, the orange and green bands denote the $3\sigma$ regions for normal ordering and inverted ordering mass spectrum respectively. The red and purple areas are the predictions for the lepton mixing matrix in Eq.~\eqref{eq:PMNS_caseVII}. The present most strict bound $|m_{ee}|<(0.120-0.250)$ eV from
EXO-200~\cite{Auger:2012ar,Albert:2014awa} combined with KamLAND-ZEN~\cite{Gando:2012zm} is are represented
by the horizontal dashed line, while the upper limit on $m_{min}$ from the latest Planck result $m_1+m_2+m_3<0.230$ eV at $95\%$ confidence level~\cite{Ade:2013zuv} is shown by dashed line. }
\end{figure}

\item[~~(\uppercase\expandafter{\romannumeral8})] $X_{\nu{\bf r}}=\rho_{\bf r}(1), \rho_{\bf r}(S)$\\
The neutrino mass matrix is diagonalized by the unitary transformation in Eq.~\eqref{eq:PMNS_caseI}. The PMNS matrix is found to take the following form
\begin{equation}
\label{eq:PMNS_caseVIII}U_{PMNS}=\frac{1}{2}
\left(\begin{array}{ccc}
 \sin \theta -i \kappa  \cos \theta  &~ \cos \theta +i \kappa  \sin \theta  ~& \kappa-1 \\
 i \cos \theta +(\kappa-1) \sin \theta &~ (\kappa-1) \cos\theta -i \sin \theta  ~& \kappa  \\
 i (\kappa-1)\cos\theta+\kappa  \sin\theta &~ \kappa  \cos \theta -i (\kappa-1) \sin\theta ~& -1 \\
\end{array}\right)\,,
\end{equation}
up to permutations of rows and columns. The lepton mixing angles and CP phases can be read off as
\begin{eqnarray}
\nonumber &&\hskip-0.30in \sin^{2}\theta_{13}=\frac{3-\sqrt{5}}{8}\simeq0.0955, \quad \sin^{2}\theta_{12}=\frac{1}{2}-\frac{\sqrt{5}}{10}\cos{2\theta}, \quad \sin^{2}\theta_{23}=\frac{5+\sqrt{5}}{10}\simeq0.724\,, \\
&&\hskip-0.30in \left|\sin{\delta_{CP}}\right|=\left|\frac{\sqrt{10}\sin{2\theta}}{\sqrt{9-\cos{4\theta}}}\right|, \quad \left|\sin{\alpha_{21}}\right|=\left|\frac{8\sin{2\theta}}{9-\cos{4\theta}}\right|, \quad \left|\sin{\alpha^{\prime}_{31}}\right|=\left|\frac{2 \sin2\theta}{\sqrt{5}+ \cos2\theta}\right|\,.
\end{eqnarray}
We see that the solar mixing angle $\theta_{12}$ has a lower bound given by $\sin^2\theta_{12}\geq(5-\sqrt{5})/10\simeq0.276$, and the experimental data on $\theta_{12}$ can be accommodated for particular values of $\theta$. Both $\theta_{13}$ and $\theta_{23}$ are independent of $\theta$, and they are outside the $3\sigma$ ranges~\cite{Capozzi:2013csa}. Furthermore, $6\times6=$36 possible permutations of rows and columns of this mixing pattern are considered. However, none of them can give rise to three mixing angles in the experimentally preferred $3\sigma$ range~\cite{Capozzi:2013csa}.

\end{description}

\section{\label{sec:4}Model building}

\begin{table}[t!]
\renewcommand{\tabcolsep}{1.05mm}
\centering
\begin{tabular}{|c||c|c|c|c|c|c||c|c|c|c|c|c|c|c||c|c|c|c|c|c|c|c|}
\hline \hline
Field &   $l$    &  $\nu^{c}$     &  $e^{c}$     &   $\mu^{c}$    &    $\tau^{c}$  &  $h_{u,d}$ & $\varphi$ &  $\phi$  &  $\psi$ & $\xi$ &  $\zeta$ &  $\chi$ & $\rho$      & $\Delta$  & $\sigma^{0}$ & $\phi^{0}$  & $\psi^{0}$ & $\xi^{0}$ & $\chi^{0}$  &   $\rho^{0}$ & $\Delta^0$  \\ \hline

$A_5$  &   $\mathbf{3}$  &  $\mathbf{3}$  &  $\mathbf{1}$  & $\mathbf{1}$   &  $\mathbf{1}$   & $\mathbf{1}$  & $\mathbf{3}$  & $\mathbf{3}^\prime$ &   $\mathbf{5}$ & $\mathbf{1}$ &   $\mathbf{1}$ &  $\mathbf{3}$  &  $\mathbf{3}^{\prime}$  & $\mathbf{5}$  & $\mathbf{1}$   & $\mathbf{4}$ & $\mathbf{5}$  &  $\mathbf{1}$ & $\mathbf{3}$ & $\mathbf{3}$ & $\mathbf{5}$     \\ \hline

$Z_3$ & $\omega_3$ & $1$ & $\omega^2_3$ & $\omega^2_3$ & $\omega^2_3$ & $1$ & $1$ & $1$ & $1$ & $\omega^2_3$ &$\omega_3$ & $\omega_3$ & $\omega_3$ & $\omega^2_3$ & $1$ & $1$ & $1$ & $\omega_3$  & $1$ & $1$ & $\omega_3$  \\ \hline

$Z_4$ & $-1$ & $1$ & $-1$ & $-1$ & $-1$ & $1$ & $1$ & $1$ & $1$ & $-1$ &$i$ & $i$ & $-i$ & $-1$ & $1$ & $1$ & $1$ & $-1$  & $i$ & $-i$ & $-1$  \\ \hline

$Z_6$ & $1$ & $1$  & $\omega_6$ & $\omega^2_6$ &  $\omega^5_6$ & $1$ & $\omega_6$ & $\omega^2_6$ & $\omega^2_6$ & $1$ & $1$ & $1$ & $1$ & $1$ & $\omega^4_6$  & $\omega^3_6$ & $\omega^4_6$ & $1$ & $1$ &   $1$ & $1$  \\ \hline

$U(1)_R$  &  $1$  & $1$  & $1$  &   $1$  & $1$  &  $0$ &  $0$ &  $0$ & $0$  &  $0$ & $0$  & $0$  & $0$ & $0$ & $2$ & $2$ & $2$ & $2$ &  $2$ & 2 & $2$  \\ \hline \hline
\end{tabular}
\caption{\label{tab:model_fields}The matter fields, flavon fields, driving fields and their transformation properties under the family symmetry $A_{5}\times Z_{3}\times Z_{4}\times Z_{6}$ and $U(1)_R$, where the phase  $\omega_3=e^{\frac{2i\pi}{3}}$ and $\omega_{6}=e^{\frac{i\pi}{3}}$.}
\end{table}

In previous section, we have performed a model-independent analysis of the lepton mixing patterns which can be derived from $A_5\rtimes H_{CP}$. As summarized in Table~\ref{tab:best_fitting}, we find five new mixing patterns which are compatible with current experimental data. In this section, we shall construct a concrete model with both $A_5$ family symmetry and generalized CP symmetry, the symmetry breaking patterns studied in section~\ref{sec:3.2.2} are implemented,
and therefore the lepton flavor mixings given by Eqs.~(\ref{eq:PMNS_caseIII}, \ref{eq:PMNS_caseIV}) in case III and case IV are realized. Note that it would be also interesting to implement other cases such as case VII in a model. In the present model, both the three generations of left-handed lepton doublets $l$ and the three generations of right-handed neutrinos $\nu^{c}$ are assigned to transform as $A_5$ triplet $\bf{3}$, while the right-haned charged leptons $e^c$, $\mu^c$ and $\tau^c$ are all invariant under $A_5$.
In discrete flavor symmetry model building, either cyclic $Z_n$ or continuous $U(1)$ symmetry is frequently introduced to eliminate unwanted operators, to ensure the required vacuum alignment and to reproduce the observed charged lepton mass hierarchies. The auxiliary symmetry is taken to be $Z_3\times Z_4\times Z_6$ in this model. The $A_5$ family symmetry and CP symmetry are broken by some flavons in a proper manner. All the flavon fields are standard model gauge singlets. As anticipated, we formulate our model in the framework of supersymmetry (SUSY). A $U(1)_R$ symmetry related to $R-$parity and the presence of driving fields in the flavon superpotential are common features of supersymmetric formulations. The field content of the model and their classification under the symmetry are listed in Table~\ref{tab:model_fields}. In the following, we first discuss the vacuum alignment of the model, then specify the structure of the model at leading order and next-to-leading order. As we shall show, the lepton mixing is exactly the GR at LO, and a non-vanishing value of the reactor mixing angle $\theta_{13}$ is generated by higher order corrections. Consequently $\theta_{13}$ is naturally of the correct order in our model.

\subsection{\label{sec:4.1}Vacuum alignment}

We utilize the standard supersymmetric driving field mechanism~\cite{Altarelli:2005yx} to solve the vacuum alignment problem.
A global $U(1)_{R}$ continuous symmetry is assumed in this approach, and the usual $R-$parity is a discrete group of this $U(1)_{R}$. The matter fields have $R-$charge equal to one, both flavon fields and Higgs are chargeless and the driving fields carry two units of $R-$charge. At LO the most general driving superpotential $w_d$ invariant under $A_5\times Z_3\times Z_4\times Z_6$ with $R=2$ can be written as
\begin{equation}
w_{d}=w^{l}_{d}+w^{\nu}_{d}\,,
\end{equation}
with
\begin{eqnarray} \label{eq:ch_LO_driving_superpotential}w^{l}_{d}=&&f_{1}\sigma^{0}(\varphi\varphi)_{\bf{1}}+f_{2}(\phi^{0}(\varphi\phi)_{\bf{4}})_{\bf{1}}+f_{3}(\phi^{0}(\varphi\psi)_{\bf{4}})_{\bf{1}}
+M_{\psi}(\psi^{0}\psi)_{\bf{1}}+f_{4}(\psi^{0}(\varphi\varphi)_{\bf{5}})_{\bf{1}}\,,\\
\nonumber w^{\nu}_{d}=&&M_{\xi}\xi^0\xi+g_{1}\xi^0\zeta^2+g_{2}\xi^0(\chi\chi)_{\bf{1}}+g_{3}\xi^0(\rho\rho)_{\bf{1}}+g_{4}\xi(\chi^0\chi)_{\bf{1}}
+g_{5}(\chi^{0}(\chi\Delta)_{\bf{3}})_{\bf{1}}\\
\label{eq:wd_model}&&+g_{6}(\rho^0(\rho\Delta)_{\bf{3}})_{\bf{1}}+M_{\Delta}(\Delta^0\Delta)_{\bf{1}}+g_{7}(\Delta^0(\chi\chi)_{\bf{5}})_{\bf{1}}+g_{8}(\Delta^0(\rho\rho)_{\bf{5}})_{\bf{1}}\,,
\end{eqnarray}
where $(\ldots)_{\bf{R}}$ denotes a contraction into the $A_5$ irreducible representation ${\bf R}$ according to the Clebsch-Gordan coefficients listed in Appendix~\ref{sec:A}. Notice that all the couplings $f_{i}$($i=1,\ldots,4$), $g_{i}$ ($i=1,\ldots,8$) and the mass parameters $M_{\psi}$, $M_{\xi}$, $M_{\Delta}$ are real, since the theory is invariant under the generalized CP defined in Eq.~\eqref{eq:CP_transformations}. In the SUSY limit, the vacuum alignment is achieved via the requirement of vanishing $F-$terms of the driving fields. In the charged lepton sector, the equations for the vanishing of the derivatives of $w^{l}_d$ with respect to each component of the driving fields are:
\begin{eqnarray}
\nonumber&&\frac{\partial w^{l}_{d}}{\partial\sigma^{0}}=f_{1}(\varphi^{2}_{1}+2\varphi_{2}\varphi_{3})=0,\\
\nonumber&&\frac{\partial w^{l}_{d}}{\partial\phi^{0}_{1}}=f_{2}(\varphi_{2}\phi_{3}+\sqrt{2}\varphi_{3}\phi_{1})-f_{3}(2\sqrt{2}\varphi_{1}\psi_{5}+\varphi_{2}\psi_{4}
-\sqrt{6}\varphi_{3}\psi_{1})=0,\\
\nonumber&&\frac{\partial w^{l}_{d}}{\partial\phi^{0}_{2}}=-f_{2}(\sqrt{2}\varphi_{1}\phi_{3}+\varphi_{2}\phi_{2})+f_{3}(\sqrt{2}\varphi_{1}\psi_{4}+3\varphi_{2}\psi_{3}
-2\varphi_{3}\psi_{5})=0,\\
\nonumber&&\frac{\partial w^{l}_{d}}{\partial\phi^{0}_{3}}=-f_{2}(\sqrt{2}\varphi_{1}\phi_{2}+\varphi_{3}\phi_{3})-f_{3}(\sqrt{2}\varphi_{1}\psi_{3}-2\varphi_{2}\psi_{2}
+3\varphi_{3}\psi_{4})=0,\\
\nonumber&&\frac{\partial w^{l}_{d}}{\partial\phi^{0}_{4}}=f_{2}(\sqrt{2}\varphi_{2}\phi_{1}+\varphi_{3}\phi_{2})+f_{3}(2\sqrt{2}\varphi_{1}\psi_{2}-\sqrt{6}\varphi_{2}\psi_{1}
+\varphi_{3}\psi_{3})=0,\\
\nonumber&&\frac{\partial w^{l}_{d}}{\partial\psi^{0}_{1}}=M_{\psi}\psi_{1}+2f_{4}(\varphi^{2}_{1}-\varphi_{2}\varphi_{3})=0,\\
\nonumber&&\frac{\partial w^{l}_{d}}{\partial\psi^{0}_{2}}=M_{\psi}\psi_{5}-2\sqrt{3}f_{4}\varphi_{1}\varphi_{3}=0,\\
\nonumber&&\frac{\partial w^{l}_{d}}{\partial\psi^{0}_{3}}=M_{\psi}\psi_{4}+\sqrt{6}f_{4}\varphi^{2}_{3}=0,\\
\nonumber&&\frac{\partial w^{l}_{d}}{\partial\psi^{0}_{4}}=M_{\psi}\psi_{3}+\sqrt{6}f_{4}\varphi^{2}_{2}=0,\\
&&\frac{\partial w^{l}_{d}}{\partial\psi^{0}_{5}}=M_{\psi}\psi_{2}-2\sqrt{3}f_{4}\varphi_{1}\varphi_{2}=0\,,
\end{eqnarray}
We find one solution to those equations,
\begin{equation}\label{eq:ch_LO_VEVs}
   \langle\varphi\rangle=\left(\begin{array}{c}
    0 \\ 1  \\ 0
  \end{array}\right)v_{\varphi}, \qquad \langle\phi\rangle=\left(\begin{array}{c}
    0 \\ 1  \\  0
  \end{array}\right)v_{\phi}, \qquad \langle\psi\rangle=\left(\begin{array}{c}
  0 \\  0 \\ 1  \\  0 \\ 0
  \end{array}\right)v_{\psi}\,,
\end{equation}
up to $A_5$ family symmetry transformations, where the vacuum expectation values (VEVs) $v_{\varphi}$, $v_{\phi}$ and $v_{\psi}$ are related by
\begin{equation}\label{eq:ch_LO_VEV_relation}
 v_{\phi}=-\frac{3\sqrt{6}f_{3}f_{4}}{M_{\psi}f_{2}}v^{2}_{\varphi}, \qquad v_{\psi}=-\frac{\sqrt{6}f_{4}}{M_{\psi}}v^{2}_{\varphi}\,,
\end{equation}
with $v_{\varphi}$ undetermined. A common order of magnitude for the VEVs (scaled by the cutoff $\Lambda$) is expected. In order to generate the mass hierarchies among the charged lepton, we assume
\begin{equation}
\label{eq:ch_VEV}\frac{v_{\varphi}}{\Lambda}\sim\frac{v_{\phi}}{\Lambda}\sim\frac{v_{\psi}}{\Lambda}\sim\mathcal{O}(\lambda^{2}_c)\,,
\end{equation}
where $\lambda_c\simeq0.23$ is the Cabibbo angle~\cite{Agashe:2014kda}. In the neutrino sector, the minimization equations for the vacuum are
\begin{eqnarray}
\nonumber&&\frac{\partial w^{\nu}_{d}}{\partial\xi^{0}}=M_{\xi}\xi+g_{1}\zeta^2+g_{2}(\chi^2_{1}+2\chi_{2}\chi_{3})+g_{3}(\rho^2_{1}+2\rho_{2}\rho_{3})=0,\\
\nonumber&&\frac{\partial w^{\nu}_{d}}{\partial\chi^{0}_{1}}=g_{4}\xi\chi_{1}-g_{5}(2\chi_{1}\Delta_{1}-\sqrt{3}\chi_{2}\Delta_{5}-\sqrt{3}\chi_{3}\Delta_{2})=0,\\
\nonumber&&\frac{\partial w^{\nu}_{d}}{\partial\chi^{0}_{2}}=g_{4}\xi\chi_{3}+g_{5}(\sqrt{3}\chi_{1}\Delta_{5}-\sqrt{6}\chi_{2}\Delta_{4}+\chi_{3}\Delta_{1})=0,\\
\nonumber&&\frac{\partial w^{\nu}_{d}}{\partial\chi^{0}_{3}}=g_{4}\xi\chi_{2}+g_{5}(\sqrt{3}\chi_{1}\Delta_{2}+\chi_{2}\Delta_{1}-\sqrt{6}\chi_{3}\Delta_{3})=0,\\
\nonumber&&\frac{\partial w^{\nu}_{d}}{\partial\rho^{0}_{1}}=g_{6}(\sqrt{3}\rho_{1}\Delta_{1}+\rho_{2}\Delta_{4}+\rho_{3}\Delta_{3})=0,\\
\nonumber&&\frac{\partial w^{\nu}_{d}}{\partial\rho^{0}_{2}}=g_{6}(\rho_{1}\Delta_{5}-\sqrt{2}\rho_{2}\Delta_{3}-\sqrt{2}\rho_{3}\Delta_{2})=0,\\
\nonumber&&\frac{\partial w^{\nu}_{d}}{\partial\rho^{0}_{3}}=g_{6}(\rho_{1}\Delta_{2}-\sqrt{2}\rho_{2}\Delta_{5}-\sqrt{2}\rho_{3}\Delta_{4})=0,\\
\nonumber&&\frac{\partial w^{\nu}_{d}}{\partial\Delta^{0}_{1}}=M_{\Delta}\Delta_{1}+2g_{7}(\chi^2_{1}-\chi_{2}\chi_{3})+2g_{8}(\rho^2_{1}-\rho_{2}\rho_{3})=0, \\
\nonumber&&\frac{\partial w^{\nu}_{d}}{\partial\Delta^{0}_{2}}=M_{\Delta}\Delta_{5}-2\sqrt{3}g_{7}\chi_{1}\chi_{3}+\sqrt{6}g_{8}\rho^{2}_{2}=0, \\
\nonumber&&\frac{\partial w^{\nu}_{d}}{\partial\Delta^{0}_{3}}=M_{\Delta}\Delta_{4}+\sqrt{6}g_{7}\chi^2_{3}-2\sqrt{3}g_{8}\rho_{1}\rho_{3}=0, \\
\nonumber&&\frac{\partial w^{\nu}_{d}}{\partial\Delta^{0}_{4}}=M_{\Delta}\Delta_{3}+\sqrt{6}g_{7}\chi^2_{2}-2\sqrt{3}g_{8}\rho_{1}\rho_{2}=0, \\
&&\frac{\partial w^{\nu}_{d}}{\partial\Delta^{0}_{5}}=M_{\Delta}\Delta_{2}-2\sqrt{3}g_{7}\chi_{1}\chi_{2}+\sqrt{6}g_{8}\rho^{2}_{3}=0\,.
\end{eqnarray}
A solution to those equations with each flavon acquiring non-zero VEV is given by
\begin{equation}
\label{eq:nu_LO_VEVs}
\langle\xi\rangle=v_{\xi}, ~ \langle\zeta\rangle=v_{\zeta},~
\langle\chi\rangle=\left(\begin{array}{c}
 \frac{\sqrt{2}}{\kappa} \\ 1  \\  1
  \end{array}\right)v_{\chi}, ~  \langle\rho\rangle=\left(\begin{array}{c}
   -\sqrt{2}\kappa \\ 1 \\ 1
  \end{array}\right)v_{\rho}, ~ \langle\Delta\rangle=\left(\begin{array}{c}
  -\sqrt{\frac{2}{3}}\kappa v_{1}  \\ v_{1} \\ -(1+\kappa)v_{1} \\ -(1+\kappa)v_{1} \\ v_{1}
  \end{array}\right)\,.
\end{equation}
These VEVs are related through
\begin{eqnarray}\label{eq:re_nu_VEVs}
 \nonumber && v_{\xi}=\frac{10 (\kappa-3) g_5 g_7 }{g_4 M_{\Delta} }v^{2}_{\chi}, \quad v^2_{\zeta}=\frac{2(\kappa-3)\left[(g_{2}g_{8}+g_{3}g_{7})g_{4}M_{\Delta}-5g_{5}g_{7}g_{8}M_{\xi}\right]}{g_{1}g_{4}g_{8}M_{\Delta}}v^{2}_{\chi}, \\
 \label{eq:VEV_relations_m2} &&  v^{2}_{\rho}=\frac{(2-\kappa)g_{7}}{g_{8}}v^{2}_{\chi}, \quad  v_{1}=\frac{\sqrt{30}(2-\kappa)g_{7}}{M_{\Delta}}v^2_{\chi}\,,
\end{eqnarray}
where $v_{\chi}$ is undetermined. It is easy to check that the VEVs of $\xi$, $\zeta$ and $\Delta$ break the $A_5$ family symmetry down to  $K^{(S,T^3ST^2ST^3)}_{4}$ while the subgroup $Z^{T^3ST^2ST^3}_2$ is preserved by vacuum of $\chi$ and $\rho$. Furthermore, Eq.~\eqref{eq:VEV_relations_m2} implies that $v^2_{\zeta}$, $v^{2}_{\chi}$, $v^{2}_{\rho}$, $v_{\xi}$ and $v_{1}$ have the same phase up to $\pi$, since all couplings are real. In our model, the GR mixing is reproduced exactly and a non-zero reactor mixing angle $\theta_{13}$ is generated after subleading order contributions are included. In order to obtain the correct size of $\theta_{13}$, we could choose
\begin{equation}
\frac{v_{\xi}}{\Lambda}\sim\frac{v_{\zeta}}{\Lambda}\sim\frac{v_{\chi}}{\Lambda}\sim\frac{v_{\rho}}{\Lambda}\sim\frac{v_{1}}{\Lambda}\sim\mathcal{O}(\lambda_c)\,.
\end{equation}

\subsection{\label{sec:LO_model}Leading order results}
The charged lepton mass terms, which are invariant under the imposed family symmetry $A_5\times Z_3\times Z_4\times Z_6$, can be written as
\begin{eqnarray}\label{eq:ch_LO_superpotential}
\nonumber w_{l}=&&\frac{y_{\tau}}{\Lambda}\tau^{c}(l\varphi)_{\bf1}h_{d}+\frac{y_{\mu_{1}}}{\Lambda^{2}}\mu^{c}(l(\phi\psi)_{\bf3})_{\bf1}h_{d}+\frac{y_{\mu_{2}}}{\Lambda^{2}}\mu^{c}
(l(\psi\psi)_{\bf3})_{\bf1}h_{d}+\frac{y_{e_{1}}}{\Lambda^{3}}e^{c}(l\varphi)_{\bf{1}}(\phi\phi)_{\bf1}h_{d}\\
\nonumber&&+\frac{y_{e_{2}}}{\Lambda^{3}}e^{c}((l\varphi)_{\bf{5}}
(\phi\phi)_{\bf5})_{\bf{1}}h_{d}+\frac{y_{e_{3}}}{\Lambda^{3}}e^{c}((l\varphi)_{\bf{3}}(\phi\psi)_{\bf3})_{\bf{1}}h_{d}+\frac{y_{e_{4}}}{\Lambda^{3}}e^{c}((l\varphi)_{\bf{5}}
(\phi\psi)_{\bf5})_{\bf{1}}h_{d}\\
\nonumber&&+\frac{y_{e_{5}}}{\Lambda^{3}}e^{c}(l\varphi)_{\bf{1}}(\psi\psi)_{\bf1}h_{d}+\frac{y_{e_{6}}}{\Lambda^{3}}e^{c}((l\varphi)_{\bf{3}}
(\psi\psi)_{\bf3})_{\bf{1}}h_{d}+\frac{y_{e_{7}}}{\Lambda^{3}}e^{c}((l\varphi)_{\bf{5}}(\psi\psi)_{\bf{5_{1}}})_{\bf{1}}h_{d} \\
&&+\frac{y_{e_{8}}}{\Lambda^{3}}e^{c}((l\varphi)_{\bf{5}}(\psi\psi)_{\bf{5_{2}}})_{\bf{1}}h_{d}+...\,,
\end{eqnarray}
where dots stand for higher dimensional operators which will be discussed later. Note that all couplings here are real due to the generalized CP symmetry. After the electroweak and flavor symmetries breaking by the VEVs shown in Eq.~\eqref{eq:ch_LO_VEVs}, we obtain a diagonal charged lepton mass matrix, and the three charged lepton masses are
\begin{eqnarray}
\nonumber&&m_{e}=\sqrt{2}\left|3y_{e_2}\frac{v^2_{\phi}v_{\varphi}}{\Lambda^3}+(y_{e_3}-\sqrt{3}y_{e_4})\frac{v_{\phi}v_{\varphi}v_{\psi}}{\Lambda^3}+3y_{e_8}\frac{v_{\varphi}v^2_{\psi}}{\Lambda^3}\right|v_{d},\\
&&m_{\mu}=\sqrt{2}\left|y_{\mu_1}\frac{v_{\phi}v_{\psi}}{\Lambda^2}\right|v_{d},\qquad m_{\tau}=\left|y_{\tau}\frac{v_{\varphi}}{\Lambda}\right|v_{d}\,,
\end{eqnarray}
We see that the realistic mass hierarchies $m_{e}:m_{\mu}:m_{\tau}\simeq\lambda^{4}_c:\lambda^{2}_c:1$ is generated for the order of magnitude of the flavon VEVs in Eq.~\eqref{eq:ch_VEV}. Furthermore, as both $m_{l}$ and $\rho_{\bf{3}}(T)$ are diagonal, obviously we have $\rho^{\dagger}_{\bf{3}}(T)m^{\dagger}_{l}m_{l}\rho_{\bf{3}}(T)=m^{\dagger}_{l}m_{l}$, i.e., the residual flavor symmetry of $m^{\dagger}_{l}m_{l}$ is $Z^{T}_{5}$. Next let's discuss the neutrino sector. Neutrino masses are generated by type I see-saw mechanism in this work. The LO superpotential for neutrino masses is
\begin{equation}\label{eq:nu_LO_superpotential}
w_{\nu}=\frac{y_{1}}{\Lambda}\xi(\nu^{c}l)_{\bf1}h_{u}+\frac{y_{2}}{\Lambda}((\nu^{c}l)_{\bf5}\Delta)_{\bf1}h_{u}+M(\nu^{c}\nu^{c})_{\bf1}\,,
\end{equation}
where the coupling constants $y_1$, $y_2$ and the mass $M$ are enforced to be real by the generalized CP symmetry. The Dirac mass matrix is obtained from the first two terms in Eq.~\eqref{eq:nu_LO_superpotential} and it is given by
\begin{equation}
 m_{D}=a\left(\begin{array}{ccc}
  1  &  0  &  0  \\
  0  &  0  &  1  \\
  0  &  1  &  0
\end{array}\right)v_{u}
+b\left(
\begin{array}{ccc}
 -2 \sqrt{2} \kappa  & -3 & -3 \\
 -3 & -3 \sqrt{2} (\kappa +1) &
   \sqrt{2} \kappa  \\
 -3 & \sqrt{2} \kappa  & -3 \sqrt{2}
   (\kappa +1) \\
\end{array}
\right)v_{u}\,,
\end{equation}
where $v_{u}=\langle h_{u}\rangle$, and the parameters $a$, $b$ are
\begin{equation}
a=y_{1}\frac{v_{\xi}}{\Lambda}, \qquad b=y_{2}\frac{v_{1}}{\sqrt{3}\Lambda}\,.
\end{equation}
The common phase of $a$ and $b$ can be absorbed by field redefinition, consequently both $a$ and $b$ can considered as real.
The last term of Eq.~\eqref{eq:nu_LO_superpotential} leads to the Majorana mass matrix:
\begin{equation}
m_{M}=M
\left(\begin{array}{ccc}
 1 & 0 & 0 \\
 0 & 0 & 1 \\
 0 & 1 & 0
\end{array}\right)\,.
\end{equation}
Therefore the three right-handed neutrinos are completely degenerate with mass equal to $M$. The light neutrino mass matrix is then given by the see-saw relation:
\begin{equation}
\label{eq:nu_LO_matrix} m_{\nu}=-m^{T}_{D}m^{-1}_{M}m_{D}=\alpha\left(\begin{array}{ccc}
  1  &  0  &  0  \\
  0  &  0  &  1  \\
  0  &  1  &  0
\end{array}\right)
+\frac{\beta}{\sqrt{2}}\left(\begin{array}{ccc}
  -2\sqrt{2}  &  3  &  3  \\
  3  &  0  &  \sqrt{2}  \\
  3  &  \sqrt{2}  &  0
\end{array}\right)
+\gamma\left(\begin{array}{ccc}
  2  &  0  &  0  \\
  0  &  3 &  -1  \\
  0  &  -1  &  3
\end{array}\right)\,,
\end{equation}
where
\begin{eqnarray}
\nonumber\alpha&=&-\left[a^2+40(1+\kappa)b^2\right]\frac{v^2_{u}}{M}, \\ \nonumber\beta&=&2\left[ \sqrt{2}ab-(3+4\kappa)b^{2}\right]\frac{v^2_{u}}{M}, \\ \gamma&=&\left[2\sqrt{2}(1+\kappa)ab+(1+8\kappa)b^{2}\right]\frac{v^2_{u}}{M}\,.
\end{eqnarray}
We find that the neutrino mass matrix $m_{\nu}$ in Eq.~\eqref{eq:nu_LO_matrix} is of the same form as the general mass matrix in Eq.~\eqref{eq:nu_general_mass_two} with $\delta=0$. Therefore $m_{\nu}$ is exactly diagonalized by the GR mixing pattern, i.e.,
\begin{equation}
U^{T}_{GR}m_{\nu}U_{GR}=\mathrm{diag}(m_{1},m_{2},m_{3})\,,
\end{equation}
where the phase matrix $K_{\nu}$ which encodes the CP parity
of the neutrino state, has been omitted. The mass eigenvalues $m_{1,2,3}$ are
\begin{eqnarray}
\nonumber&&m_1=\left|a^2-2\sqrt{2}(3-\kappa)ab+10(2-\kappa)b^2\right|\frac{v^2_{u}}{M},\\
\nonumber&&m_2=\left|a^2-10\sqrt{2}\kappa ab+50(1+\kappa)b^2\right|\frac{v^2_{u}}{M},\\
&&m_3=\left|a^2+2\sqrt{2}(3+4\kappa)ab+10(5+8\kappa)b^2\right|\frac{v^2_{u}}{M}\,.
\end{eqnarray}
Since the charged lepton mass matrix is diagonal in LO, the lepton mixing is exactly the GR mixing pattern. Here the reason why the GR mixing is produced is because that the flavor symmetry $A_{5}$ is broken to $K^{(S, T^2ST^2ST^3)}_4$ subgroup by the VEVs of $\xi$ and $\Delta$. Furthermore, we see that three neutrino masses $m_{1,2,3}$ only depend on two real parameters $a$ and $b$ which can be fixed by the measured values of the mass-squared difference $\delta m^{2}\equiv m^{2}_{2}-m^{2}_{1}$ and $\Delta m^{2}\equiv m^{2}_{3}-(m^{2}_{1}+m^{2}_{2})$. For the best fitting values $\delta m^{2}=7.54\times 10^{-5}\text{eV}^{2}$ and $\Delta m^{2}=2.43\times 10^{-3}\text{eV}^{2}$~\cite{Capozzi:2013csa}, we find the neutrino mass spectrum can only be NO, and the absolute values of the neutrino masses are $m_{1}=4.81\times 10^{-4}\text{eV}$, $m_{2}=8.70\times 10^{-3}\text{eV}$ and $m_{3}=0.0497\text{eV}$.

\subsection{\label{subsec:NLO_model2}Next-to-leading-order corrections}

At LO our model gives rise to the GR mixing pattern $U_{GR}$ which predicts a vanishing reactor mixing angle ($\theta_{13}=0^{\circ}$). Hence
substantial next-to-leading-order corrections are needed to bring the model to agree with the experimental data on $\theta_{13}$. We will demonstrate in the following that a non-zero $\theta_{13}$ can be obtained after the NLO contributions are included. Moreover, the LO remnant symmetry $K^{(S,T^3ST^2ST^3)}_{4}$ of neutrino sector is further broken down to $Z^{T^3ST^2ST^3}_{2}$ such that the mixing patterns of case III and case IV discussed in section~\ref{sec:3.2.2} are realized. Firstly we consider the corrections to the flavon superpotential $w^{l}_{d}$ in Eq.~\eqref{eq:ch_LO_driving_superpotential} which determines the vacuum alignment of the charged lepton sector. The symmetry allowed NLO operators are of the following form
\begin{eqnarray}
\delta w^{l}_{d}=((\phi^0\varphi)_{\bf5}(\varphi\varphi)_{\bf5})_{\bf1}/\Lambda+(\Psi^0_{l}\Psi_{l}\Psi^{2}_{\nu}\Psi^{\prime}_{\nu}\rho)_{\bf1}/\Lambda^3\,,
\end{eqnarray}
where all possible $A_5$ contractions should be considered, and all dimensionless coupling constants are omitted with $\Psi^0_{l}\equiv\{\sigma^0,\psi^0\}$, $\Psi_{l}\equiv\{\phi,\psi\}$, $\Psi_{\nu}\equiv\{\xi,\Delta\}$ and $\Psi^{\prime}_{\nu}\equiv\{\zeta,\chi\}$. Note that $\delta w^{l}_{d}$ is suppressed by $\lambda^2_{c}$ with respect to the LO superpotential $w^{l}_{d}$ in Eq.~\eqref{eq:ch_LO_driving_superpotential}. The NLO vacuum configuration is determined by searching for the zeros of the $F-$terms of $w^{l}_{d}+\delta w^{l}_d$ with respect to the driving fields $\sigma^{0}$, $\phi^{0}$ and $\psi^{0}$. We find that the NLO vacuum of $\varphi$, $\phi$ and $\psi$ are given by
\begin{equation}
\label{eq:correct_vacuum_five} \langle\varphi\rangle =\left(\begin{array}{c}
   \epsilon_{1}\lambda^2_c \\  1  \\   \epsilon_{2}\lambda^2_c
  \end{array}\right)v_{\varphi}, \qquad
    \langle\phi\rangle =\left(\begin{array}{c}
    \epsilon_{3}\lambda^2_c \\  1+ \epsilon_{4}\lambda^2_c  \\   \epsilon_{5}\lambda^2_c
  \end{array}\right)v_{\phi}, \qquad
   \langle\psi\rangle =\left(\begin{array}{c}
    \epsilon_{6}\lambda^2_c \\   \epsilon_{7}\lambda^2_c \\ 1+ \epsilon_{8}\lambda^2_c  \\   \epsilon_{9}\lambda^2_c \\   \epsilon_{10}\lambda^2_c \\
  \end{array}\right)v_{\psi}\,,
\end{equation}
where $\epsilon_{i}$ ($i=1,\ldots,10$) are general complex numbers with absolute values of order one. The higher dimensional operators contributing to the charged lepton masses are:
\begin{eqnarray}\label{eq:charged_superpotential_four}
\delta w_{l}=\mu^c(l\varphi^2\Psi_{l})h_d/\Lambda^3+e^c(l\varphi^3\Psi_l)/\Lambda^4\,.
\end{eqnarray}
The charged lepton mass matrix can be obtained by inserting the NLO VEVs of Eq.~\eqref{eq:correct_vacuum_five} into the LO mass terms  plus the contribution of $\delta w_{l}$ evaluated with the LO VEVs of Eq.~\eqref{eq:ch_LO_VEVs}. We find that the NLO charged lepton mass matrix is of the following form:
\begin{eqnarray}
m_{l}\simeq\left(\begin{array}{ccc}
m_{e}  & \lambda^2_c m_{e}  & \lambda^2_c m_{e} \\
\lambda^2_cm_{\mu}   & m_{\mu} &  \lambda^2_cm_{\mu}  \\
\lambda^2_cm_{\tau}   & \lambda^2_cm_{\tau}    &  m_{\tau}
\end{array}\right)\,.
\end{eqnarray}
Therefore the contributions of charged lepton sector to the lepton mixing angles is of order $\lambda^2_c$ and can be neglected.

We proceed to discuss the subleading corrections in the neutrino sector. The higher order corrections to the flavon superpotential of $\xi$, $\zeta$, $\chi$, $\rho$ and $\Delta$ read
\begin{eqnarray}
\nonumber \delta w^{\nu}_{d}&=&\frac{g_{9}}{\Lambda}\zeta^2(\chi^0\chi)_{\bf1}+\frac{g_{10}}{\Lambda}\zeta(\chi^0(\chi\chi)_{\bf{3}})_{\bf1}
+\frac{g_{11}}{\Lambda}(\chi^0\chi)_{\bf1}(\chi\chi)_{\bf1}+\frac{g_{12}}{\Lambda}((\chi^0\chi)_{\bf3}(\chi\chi)_{\bf3})_{\bf1} \\
\nonumber &&+\frac{g_{13}}{\Lambda}((\chi^0\chi)_{\bf5}(\chi\chi)_{\bf{5}})_{\bf1}+\frac{g_{14}}{\Lambda}(\chi^0\chi)_{\bf1}(\rho\rho)_{\bf1}
 +\frac{g_{15}}{\Lambda}((\chi^0\chi)_{\bf5}(\rho\rho)_{\bf{5}})_{\bf1}\\
&&+\frac{g_{16}}{\Lambda}((\rho^0\rho)_{\bf5}(\chi\chi)_{\bf5})_{\bf1}+\frac{g_{17}}{\Lambda}((\rho^0\rho)_{\bf5}(\rho\rho)_{\bf5})_{\bf1}\,,
\end{eqnarray}
where all couplings $g_{i}$ ($i=9,\ldots,17$) are real because of the generalized CP symmetry. The resulting contributions to the $F-$terms of the driving fields $\sigma^{0}$, $\rho^0$, $\chi^0$ and $\Delta^{0}$ are suppressed by $\langle\Psi\rangle/\Lambda\sim\lambda_c$ ($\Psi\equiv\{\zeta,\chi,\rho\}$) compared to the contribution from the LO terms in Eq.~\eqref{eq:wd_model}. Hence they induce shifts in the VEVs of $\xi$, $\zeta$, $\chi$, $\rho$ and $\Delta$ at relative order $\lambda_c$ with respect to the LO results. After some straightforward algebra, the new vacuum configuration can be written as
\begin{eqnarray}\label{eq:nu_NLO_VEVs}
\nonumber&&\langle\xi\rangle=v_{\xi}+\delta v_{\xi}, \quad\langle\zeta\rangle=v_{\zeta}+\delta v_{\zeta}, \quad \langle\chi\rangle=\left(\begin{array}{c}
    \frac{\sqrt{2}}{\kappa}v_{\chi}  \\  v_{\chi} \\v_{\chi}
  \end{array}\right), \\
 \label{eq:correct_vacuum_seven} && \langle\rho\rangle=\left(\begin{array}{c}
    -\sqrt{2}\kappa(v_{\rho}+\delta v_{\rho})  \\  v_{\rho}+\delta v_{\rho}  \\v_{\rho}+\delta v_{\rho}
  \end{array}\right),\quad  \langle\Delta\rangle=\left(\begin{array}{c}
    \sqrt{\frac{2}{3}}\left(-\kappa v_{1}+\left(1+2\kappa\right)\delta v_{\Delta}\right) \\ v_{1}+\delta v_{\Delta}  \\ -\left(1+\kappa\right)v_{1}+2\kappa\delta v_{\Delta} \\ -\left(1+\kappa\right)v_{1}+2\kappa\delta v_{\Delta} \\v_{1}+\delta v_{\Delta}
  \end{array}\right) \,,
\end{eqnarray}
where
\begin{eqnarray}
\nonumber && \delta v_{\xi}=-\frac{X_{1}+g_{5}X_{2}}{g_{4}g_{6}\Lambda}, \quad \delta v_{\zeta}=\frac{g_{8}M_{\xi}X_{1}+(g_{5}g_{8}M_{\xi}-g_{3}g_{4}M_{\Delta})X_{2}}{2g_{1}g_{4}g_{6}g_{8}\Lambda v_{\zeta}}\,,   \\
&& \delta v_{\rho}=\frac{2(\kappa-2)g_{16}M_{\Delta}v^{2}_{\chi}+2g_{17}M_{\Delta}v^{2}_{\rho}}{4g_{6}g_{8}\Lambda v_{\rho}}, \quad \delta v_{\Delta}=\frac{2\sqrt{6}((\kappa-1)g_{16}v^{2}_{\chi}-\kappa g_{17}v^{2}_{\rho})}{g_{6}\Lambda}\,,
\end{eqnarray}
with
\begin{eqnarray}
\nonumber && X_{1}=g_{6}\left(g_{9}v^{2}_{\zeta}+2(3-\kappa)(g_{11}+4g_{13})v^{2}_{\chi}+2\sqrt{5}\kappa(g_{14}+g_{15})v^{2}_{\rho}\right) \,, \\
&& X_{2}=2(\kappa-3)g_{16}v^{2}_{\chi}+2\sqrt{5}\kappa g_{17}v^{2}_{\rho}\,.
\end{eqnarray}
Obviously the vacuum of $\chi$ is kept intact, $\langle\rho\rangle$ acquires $\mathcal{O}(\lambda_c)$ corrections in the same direction, while the alignment of $\Delta$ is tilted. Moreover, from the relations in Eq.~\eqref{eq:re_nu_VEVs}, we see that the shifts $\delta v_{\xi}$, $\delta v_{\zeta}$, $\delta v_{\rho}$ and $\delta v_{\Delta}$ carry the same phase as $v_{\xi}$, $v_{\zeta}$, $v_{\rho}$ and $v_{1}$ up to $\pi$, respectively.

The light neutrino mass matrix receives corrections from both the modified vacuum and the higher dimensional operators in the superpotential $w_{\nu}$. It is easy to check that the NLO corrections to the Majorana mass terms are suppressed by $1/\Lambda^{4}$ which can be safely neglected. The subleading operators contributing to the neutrino Dirac masses are as follows
\begin{eqnarray}
\nonumber \delta w_{\nu}&=&\frac{y_{3}}{\Lambda^2}\zeta^2(l\nu^{c})_{\bf1}h_{u}+\frac{y_{4}}{\Lambda^2}\zeta((l\nu^{c})_{\bf{3}}\chi)_{\bf1}h_{u}+\frac{y_{5}}{\Lambda^2}(l\nu^{c})_{\bf1}(\chi\chi)_{\bf1}h_{u}
+\frac{y_{6}}{\Lambda^2}((l\nu^{c})_{\bf3}(\chi\chi)_{\bf3})_{\bf1}h_{u}  \\
&& +\frac{y_{7}}{\Lambda^2}((l\nu^{c})_{\bf5}(\chi\chi)_{\bf5})_{\bf1}h_{u}+\frac{y_{8}}{\Lambda^2}(l\nu^{c})_{\bf1}(\rho\rho)_{\bf1}h_{u}
+\frac{y_{9}}{\Lambda^2}((l\nu^{c})_{\bf5}(\rho\rho)_{\bf5})_{\bf1}h_{u} \,.
\end{eqnarray}
As a consequence, the corrected Dirac mass matrix becomes
\begin{eqnarray}
\nonumber m_{D}&=&a
\left(\begin{array}{ccc}
 1 & 0 & 0 \\
 0 & 0 & 1 \\
 0 & 1 & 0
\end{array}\right)v_{u}
+b
\left(\begin{array}{ccc}
 2 \sqrt{2} & -3 & -3 \\
 -3 & 0 & -\sqrt{2} \\
 -3 & -\sqrt{2} & 0
\end{array}\right)v_{u}
+c
\left(\begin{array}{ccc}
 2\sqrt{2} & ~0~ & 0 \\
 0 & ~3\sqrt{2}~ & -\sqrt{2} \\
 0 & ~-\sqrt{2}~ & 3\sqrt{2}
\end{array}\right)v_{u}\\
\label{eq:MD_NLO}&&+d
\left(\begin{array}{ccc}
 0 & 1 ~& -1 \\
 -1 & 0 ~& \frac{\sqrt{5}-1}{\sqrt{2}} \\
 1 & \frac{1-\sqrt{5}}{\sqrt{2}} ~& 0
\end{array}\right)v_{u}\,,
\end{eqnarray}
where the four parameters $a$, $b$, $c$ and $d$ are
\begin{eqnarray}
\nonumber && a=y_{1}\frac{v_{\xi}+\delta v_{\xi}}{\Lambda}+y_{3}\frac{v^{2}_{\zeta}}{\Lambda^2}+2(3-\kappa)y_{5}\frac{v^{2}_{\chi}}{\Lambda^2}+2\sqrt{5}\kappa y_{8}\frac{v^{2}_{\rho}}{\Lambda^2}, \\
\nonumber&&b=\frac{y_{2}}{\sqrt{3}}\frac{v_{1}+\delta v_{\Delta}}{\Lambda}-2\sqrt{2}(\kappa-1)y_{7}\frac{v^{2}_{\chi}}{\Lambda^2}+\sqrt{2}y_{9}\frac{v^{2}_{\rho}}{\Lambda^2}\,, \\
&&c=-\frac{y_{2}}{\sqrt{3}}\frac{\left(1+\kappa\right)v_{1}-2\kappa\delta v_{\Delta}}{\Lambda}+\sqrt{2}y_{7}\frac{v^{2}_{\chi}}{\Lambda^2}+2\sqrt{2}\kappa y_{9}\frac{v^{2}_{\rho}}{\Lambda^2},  \qquad d=y_{4}\frac{v_{\zeta}v_{\chi}}{\Lambda^2}\,.
\end{eqnarray}
Notice that the three parameters $a$, $b$ and $c$ have the same phase with $v^{2}_{\chi}$ up to $\pi$, while the phase difference between $d$ and $v^{2}_{\chi}$ is $0$, $\pi$ or $\pm\frac{\pi}{2}$ depending on the product $g_{1}M_{\Delta}\big[(g_{2}g_{8}+g_{3}g_{7})g^2_{4}g_{8}M_{\Delta}-5g_{4}g_{5}g_{7}g^2_{8}M_{\xi}\big]$ being positive or negative. Since the phase of $v_{\chi}$ can be factorized out as an overall phase of the neutrino mass matrix $m_{\nu}$, the VEV $v_{\chi}$ can be taken to be real without loss of generality. As a result, $a$, $b$ and $c$ are all real and the parameter $d$ is real for $g_{1}M_{\Delta}\big[(g_{2}g_{8}+g_{3}g_{7})g^2_{4}g_{8}M_{\Delta}-5g_{4}g_{5}g_{7}g^2_{8}M_{\xi}\big]<0$  or pure imaginary for $g_{1}M_{\Delta}\big[(g_{2}g_{8}+g_{3}g_{7})g^2_{4}g_{8}M_{\Delta}-5g_{4}g_{5}g_{7}g^2_{8}M_{\xi}\big]>0$. In addition, we see that $d$ are suppressed by $\lambda_c$ with respect to $a$, $b$ and $c$, i.e.,
\begin{equation}
a\sim b\sim c\sim\mathcal{O}(\lambda_c),\qquad d\sim\mathcal{O}(\lambda^2_c)\,.
\end{equation}
Utilizing the see-saw formula, we find the light neutrino mass matrix $m_{\nu}$ is of the same form as Eq.~\eqref{eq:nu_general_mass_two} with
\begin{eqnarray}
\nonumber &&\alpha=-\left[3 a^2+24( 2b^2+ b c+2 c^2)-4(3-\kappa) d^2\right]\frac{v^{2}_{u}}{3 M},  \\ \nonumber&&\beta=\left[6b(\sqrt{2}a+b+4c)-2(\kappa-1)d^{2}\right]\frac{v^2_{u}}{3 M}, \\
\nonumber&& \gamma=-\left[6\sqrt{2}ac+3 (3b+2c)(b-2c)+d^2\right]\frac{v^2_{u}}{3M}, \\
&&\delta=-3 d\left[b+2(\kappa-1)c\right]\frac{v^2_{u}}{M}\,.
\end{eqnarray}
Note that the term proportional to $\delta$ spoils the LO GR mixing, and it is of relative order $\lambda_c$ compared with $\alpha$, $\beta$ and $\gamma$ since it is induced by the NLO corrections. Therefore the correct size of the reactor mixing angle $\theta_{13}$ can be naturally achieved in our model. After extracting the overall phase of $v_{\chi}$, the parameters $\alpha$, $\beta$ and $\gamma$ are real while $\delta$ is real or pure imaginary. In the case of $g_{1}M_{\Delta}\big[(g_{2}g_{8}+g_{3}g_{7})g^2_{4}g_{8}M_{\Delta}-5g_{4}g_{5}g_{7}g^2_{8}M_{\xi}\big]<0$, $\delta$ is real such that the neutrino mass matrix $m_{\nu}$ has the most general form compatible with the preservation of the remnant symmetry $Z^{T^{3}ST^{2}ST^{3}}_{2}\times H^{\nu}_{CP}$ with $H^{\nu}_{CP}=\{\rho_{\bf{r}}(1),\rho_{\bf{r}}(T^{3}ST^{2}ST^{3})\}$. This is the case III investigated in the model independent analysis of section~\ref{sec:3.2.2}. The lepton mixing matrix $U_{PMNS}$ and the corresponding preditions for the lepton mixing parameters are given by Eq.~\eqref{eq:PMNS_caseIII} and Eq.~\eqref{eq:mixing_parameters_caseIII} respectively. There is no CP violation in this case.

In the case of $g_{1}M_{\Delta}\big[(g_{2}g_{8}+g_{3}g_{7})g^2_{4}g_{8}M_{\Delta}-5g_{4}g_{5}g_{7}g^2_{8}M_{\xi}\big]>0$, the parameter $\delta$ becomes imaginary. The origin symmetry $A_{5}\rtimes H_{CP}$ is broken down to $Z^{T^{3}ST^{2}ST^{3}}_{2}\times H^{\nu}_{CP}$ with $H^{\nu}_{CP}=\{\rho_{\bf{r}}(S),\rho_{\bf{r}}(T^{3}ST^{2}ST^{3}S)\}$ in the neutrino sector. The neutrino mass matrix $m_{\nu}$ has the same form as that of case IV discussed in section~\ref{sec:3.2.2}. Both atmospheric
mixing angle and Dirac CP phase are predicted to be maximal while Majorana CP phases are conserved, as shown in Eq.~\eqref{eq:mixing_parameters_caseIV}. In short, our model reproduces the GR mixing at LO, and realistic value of $\theta_{13}$ is obtained after
higher order contributions are taken into account. Depending on the overall sign of the product $g_{1}M_{\Delta}\big[(g_{2}g_{8}+g_{3}g_{7})g^2_{4}g_{8}M_{\Delta}-5g_{4}g_{5}g_{7}g^2_{8}M_{\xi}\big]$,
either case III or case IV can be realized.

\section{\label{sec:Conclusion}Conclusions}

Combining a discrete flavor symmetry with a CP symmetry is a very promising approach of predicting both lepton mixing angles and CP phases. In this work we have performed a comprehensive analysis of the $A_5$ family symmetry and CP symmetry. Since the inverse of each conjugacy class of $A_5$ is equal to itself, all the inner automorphisms of $A_5$ are class-inverting while the unique nontrivial outer automorphism of $A_5$ is not. As a result, the physical CP transformations are defined by the inner automorphisms of $A_5$. In our working basis, the CP transformations are found to be of the same form as the flavor symmetry transformations.

Assuming neutrinos are Majorana particles, we have analyzed the possible symmetry breaking patterns of $A_5\rtimes H_{CP}$ and the corresponding predictions for the PMNS matrix as well as the lepton mixing parameters in a model independent way. We find five phenomenologically interesting mixing patterns summarized in Table~\ref{tab:best_fitting}, and one column of the PMNS matrix is fixed to be $(-\sqrt{\frac{\kappa}{\sqrt{5}}}, \frac{1}{\sqrt{2\sqrt{5}\,\kappa}}, \frac{1}{\sqrt{2\sqrt{5}\,\kappa}})^T$, $(\sqrt{\frac{1}{\sqrt{5}\,\kappa}},\sqrt{\frac{\kappa}{2\sqrt{5}}},\sqrt{\frac{\kappa}{2\sqrt{5}}})^T$, $(\frac{1}{\sqrt{3}},\frac{1}{\sqrt{3}},\frac{1}{\sqrt{3}})^T$ or $(\frac{\kappa}{2}, -\frac{1}{2},\frac{\kappa-1}{2})^T$, where $\kappa=(1+\sqrt{5})/2$ is the golden ratio. All the three mixing angles are determined in terms of a single real parameter $\theta$, and their measured values can be accommodated for certain values of $\theta$. In particular, the Dirac CP violating phase $\delta_{CP}$ is predicted to be trivial or maximal while the Majorana phases are trivial. In contrast, $\delta_{CP}$ is quite weakly constrained and Majorana phases can not be predicted if CP symmetry is not considered, as shown in Appendix~\ref{sec:B}. Our theoretical predictions can be tested by forthcoming long-baseline neutrino oscillation experiments such as LBNE, LBNO and HyperKamiokande. The predicted mixing patterns would be ruled out, if significant deviations of $\delta_{CP}$ from trivial and maximal values were detected. Furthermore, the phenomenological predictions for the $(\beta\beta)_{0\nu}-$decay are investigated. The present experimental bounds are saturated, and the effective mass $|m_{ee}|$ is found to be within the sensitivity of future $(\beta\beta)_{0\nu}-$decay experiments for inverted ordering neutrino mass spectrum.

Guided by above model independent analysis, we construct a flavor model with both $A_5$ flavor symmetry and generalized CP symmetry. The lepton mixing is exactly the GR pattern at LO, the observed mass hierarchies among charged lepton are generated, and the three light neutrino masses effectively depend on two real parameters which can be fixed by the measured values of the mass-squared splittings. Therefore the neutrino mass spectrum can only be normal ordering and the absolute neutrino masses are predicted. The model is built in such a way that the GR mixing is modified by NLO contributions and only the second column of GR mixing matrix is kept. A non-vanishing value of $\theta_{13}$ is generated at NLO and it is naturally of the correct order $\lambda_c$ in our model. In case of $g_{1}M_{\Delta}\big[(g_{2}g_{8}+g_{3}g_{7})g^2_{4}g_{8}M_{\Delta}-5g_{4}g_{5}g_{7}g^2_{8}M_{\xi}\big]<0$, Dirac CP phase $\delta_{CP}$ is 0 or $\pi$, consequently the mixing pattern of case III of general analysis in section~\ref{sec:3.2.2} is reproduced exactly. In case of $g_{1}M_{\Delta}\big[(g_{2}g_{8}+g_{3}g_{7})g^2_{4}g_{8}M_{\Delta}-5g_{4}g_{5}g_{7}g^2_{8}M_{\xi}\big]>0$, Dirac CP phase $\delta_{CP}$ is maximal with $\delta_{CP}=\pm\pi/2$, the mixing pattern of case IV is generated. In other words, our model provides an explicit dynamical realization of the assumed symmetry breaking pattern in section~\ref{sec:3.2.2}.

It is interesting to implement any of the remaining cases II, V and VII in Table~\ref{tab:best_fitting} in a concrete model. Moreover, the group $\mathcal{I}^{\prime}$, which is the double cover of $A_5$, may deserve to be studied in a similar fashion. Since $\mathcal{I}^{\prime}$ has doublet representations~\cite{Everett:2010rd}, quark masses and mixing should be easily reproduced.

\section*{Acknowledgements}

One of the author (G.J.D.) is grateful to Stephen F. King and Alexander J. Stuart for stimulating discussions on generalized CP symmetry. The idea of combining $A_5$ family symmetry with generalized CP was initiated during visiting the School of Physics and Astronomy at the University of Southampton. We would like to thank Luis Lavoura for email correspondence. We acknowledge Chang-Yuan Yao for his help on plotting the figures. This work is supported by the National Natural Science Foundation of China under Grant Nos. 11275188 and 11179007.

\newpage

\section*{Appendix}

\begin{appendix}

\section{\label{sec:A}Group Theory of $A_{5}$}

$A_{5}$ is the group of even permutations of five objects, and it has $5!/2=60$ elements. Geometrically it is the symmetry group of a regular icosahedron.
$A_{5}$ group can be generated by two generators $S$ and $T$ which satisfy the multiplication rules~\cite{Ding:2011cm}:
\begin{equation}
  S^{2}=T^{5}=(ST)^{3}=1\,.
\end{equation}
The 60 element of $A_{5}$ group are divided into 5 conjugacy classes:
\begin{eqnarray}
\nonumber 1C_{1} :&&1\\
\nonumber15C_{2} :&& ST^2ST^3S, TST^4, T^4(ST^2)^2, T^2ST^3, (T^2S)^2T^3S, ST^2ST, S, T^3ST^2ST^3,\\
\nonumber && ~T^3ST^2ST^3S, T^3ST^2, T^4ST^2ST^3S, TST^2S, ST^3ST^2S, T^4ST, (T^2S)^2T^4\\
\nonumber20C_{3} : && ST, TS, ST^4, T^4S, TST^3, T^2ST^2, T^2ST^4, T^3ST, T^3ST^3, T^4ST^2, TST^3S, T^2ST^3S, \\
\nonumber&&~T^3ST^2S, ST^2ST^3, ST^3ST, ST^3ST^2, (T^2S)^2T^2, T^2(T^2S)^2, (ST^2)^2S, (ST^2)^2T^2\\
\nonumber12C_{5}: && T, T^4, ST^2, T^2S, ST^3, T^3S, STS, TST, TST^2, T^2ST, T^3ST^4, T^4ST^3\\
\nonumber12C^{\prime}_5: &&T^2, T^3, ST^2S, ST^3S,(ST^2)^2, (T^2S)^2, (ST^3)^2, (T^3S)^2, (T^2S)^2T^3,\\
&&~T^3(ST^2)^2, T^3ST^2ST^4, T^4ST^2ST^3\,,
\end{eqnarray}
where $nC_k$ denotes a class with $n$ elements which have order $k$.
The group structure of $A_5$ has been elaborately analyzed in Ref.~\cite{Ding:2011cm}. Following the convention of Ref.~\cite{Ding:2011cm}, we find that $A_{5}$ group has thirty-six abelian subgroups in total: fifteen $Z_{2}$ subgroups, ten $Z_{3}$ subgroups, five $K_{4}$ subgroups and six $Z_{5}$ subgroups.
In terms of the generators $S$ and $T$, the concrete forms of these abelian subgroups are as follows:
\begin{itemize}[leftmargin=1.5em]
\item{$Z_{2}$ subgroups}
\begin{eqnarray}
\nonumber&& Z^{ST^{2}ST^{3}S}_{2}=\{1,ST^{2}ST^{3}S\},\quad  Z^{TST^{4}}_{2}=\{1,TST^{4}\}, \quad Z^{T^{4}(ST^{2})^{2}}_{2}=\{1,T^{4}(ST^{2})^{2}\},\\
\nonumber&& Z^{T^{2}ST^{3}}_{2}=\{1,T^{2}ST^{3}\}, \quad  Z^{(T^{2}S)^{2}T^{3}S}_{2}=\{1,(T^{2}S)^{2}T^{3}S\}, \quad  Z^{ST^{2}ST}_{2}=\{1, ST^{2}ST\},\\
\nonumber&&Z^{S}_{2}=\{1,S\}, \quad   Z^{T^{3}ST^{2}ST^{3}}_{2}=\{1,T^{3}ST^{2}ST^{3}\}, \quad  Z^{T^{3}ST^{2}ST^{3}S}_{2}=\{1,T^{3}ST^{2}ST^{3}S\}, \\
\nonumber && Z^{T^{3}ST^{2}}_{2}=\{1,T^{3}ST^{2}\}, \quad  Z^{T^{4}ST^{2}ST^{3}S}_{2}=\{1,T^{4}ST^{2}ST^{3}S\}, \quad  Z^{TST^{2}S}_{2}=\{1,TST^2S\},\\
\nonumber&& Z^{ST^{3}ST^{2}S}_{2}=\{1, ST^{3}ST^{2}S\}, \quad  Z^{T^{4}ST}_{2}=\{1, T^{4}ST\}, \quad  Z^{(T^{2}S)^{2}T^{4}}_{2}=\{1,(T^{2}S)^{2}T^{4}\}.
\end{eqnarray}
All the above fifteen $Z_{2}$ subgroups are conjugate to each other.
\item{$Z_{3}$ subgroups}
\begin{eqnarray}
\nonumber&&Z^{T^{3}ST^{2}S}_{3}=\{1, T^{3}ST^{2}S,  ST^{3}ST^{2}\},\quad  Z^{TST^{3}S}_3=\{1, TST^{3}S, (ST^{2})^{2}T^{2}\}, \\
\nonumber&& Z^{T^{3}ST}_{3}=\{1, T^{3}ST, T^{4}ST^{2}\}, \quad  Z^{ST}_3=\{1, ST, T^{4}S\}, \\
\nonumber &&  Z^{(T^{2}S)^{2}T^{2}}_{3}=\{1, (T^{2}S)^{2}T^{2}, (ST^{2})^{2}S\}, \quad Z^{TST^{3}}_3=\{1, TST^{3}, T^{2}ST^{4}\},\\
\nonumber&&Z^{T^{2}ST^{2}}_{3}=\{1, T^{2}ST^{2}, T^{3}ST^{3}\}, \quad  Z^{TS}_3=\{1, TS,ST^4\},\\
\nonumber&& Z^{ST^{3}ST}_{3}=\{1, ST^{3}ST, T^{2}(T^{2}S)^{2}\}, \quad  Z^{ST^{2}ST^{3}}_3=\{1, ST^{2}ST^{3}, T^{2}ST^{3}S\}.
\end{eqnarray}
The ten $Z_{3}$ subgroups are related with each other by group conjugation.
\item{$K_{4}$ subgroups}
\begin{eqnarray}
\nonumber && K^{(ST^{2}ST^{3}S, TST^{4})}_{4}\equiv Z^{ST^{2}ST^{3}S}_{2}\times Z^{TST^{4}}_{2}=\{1, ST^{2}ST^{3}S, TST^{4}, T^{4}(ST^{2})^{2}\}, \\
\nonumber && K^{(T^{2}ST^{3}, ST^{2}ST)}_{4}\equiv Z^{T^{2}ST^{3}}_{2}\times Z^{ST^{2}ST}_{2}=\{1, T^{2}ST^{3}, (T^{2}S)^{2}T^{3}S, ST^{2}ST\},\\
\nonumber && K^{(S, T^{3}ST^{2}ST^{3})}_{4}\equiv Z^{S}_{2}\times Z^{T^{3}ST^{2}ST^{3}}_{2}=\{1, S, T^{3}ST^{2}ST^{3}, T^{3}ST^{2}ST^{3}S\}, \\
\nonumber && K^{(T^{3}ST^{2}, TST^{2}S)}_{4}\equiv Z^{T^{3}ST^{2}}_{2}\times Z^{TST^{2}S}_{2}=\{1, T^{3}ST^{2}, T^{4}ST^{2}ST^{3}S, TST^{2}S\},\\
\nonumber && K^{(ST^{3}ST^{2}S, T^{4}ST)}_{4}\equiv Z^{ST^{3}ST^{2}S}_{2}\times Z^{T^{4}ST}_{2}=\{1,ST^{3}ST^{2}S, T^{4}ST, (T^{2}S)^{2}T^{4}\}.
\end{eqnarray}
All the five $K_{4}$ subgroups are conjugate as well.
\item{$Z_{5}$ subgroups}
\begin{eqnarray}
\nonumber &&\hskip-0.35in Z^{STS}_5=\{1, STS, ST^{2}S, ST^{3}S, TST\}, \quad Z^{ST^{3}}_5=\{1,ST^{3}, T^{2}S, (ST^{3})^{2}, (T^{2}S)^{2}\},\\
\nonumber &&\hskip-0.35in Z^{T^{2}ST}_5=\{1, T^{2}ST, T^{4}ST^{3},  T^{3}(ST^{2})^{2},T^{4}ST^{2}ST^{3}\}, \quad  Z^{T}_5=\{1, T, T^{2}, T^{3}, T^{4}\},\\
\nonumber &&\hskip-0.35in Z^{TST^{2}}_5=\{1, TST^{2}, T^{3}ST^{4}, (T^{2}S)^{2}T^{3},T^{3}ST^{2}ST^{4}\},~ Z^{ST^{2}}_5=\{1, ST^{2}, T^{3}S, (ST^{2})^{2}, (T^{3}S)^{2}\}.
\end{eqnarray}
All the six $Z_{5}$ subgroups are related to each other under group conjugation.
\end{itemize}
Here the superscript of a subgroup denotes its generator (generators). The $A_5$ group has five irreducible representations: one singlet representation $\bf{1}$, two three-dimensional representations $\bf{3}$ and $\bf{3^\prime}$, one four-dimensional representation $\mathbf{4}$ and one five-dimensional representation $\mathbf{5}$. In the present work, we choose the same basis as that of Ref.~\cite{Ding:2011cm}. The explicit forms of the generators $S$ and $T$ in the five irreducible representations are as follows
\begin{eqnarray}
  \begin{array}{ccc}
    \bf{1:} &   ~S=1\,,~ &  T=1 \,, ~\\[-14pt] \\[4pt]
    \bf{3:} &   ~S=\frac{1}{\sqrt{5}}
\begin{pmatrix}
 1 &~ -\sqrt{2} &~ -\sqrt{2} \\
 -\sqrt{2} &~ -\kappa  &~ \kappa-1 \\
 -\sqrt{2} &~ \kappa-1 &~ -\kappa
\end{pmatrix}\,,~
& T=
\begin{pmatrix}
 1 &~ 0 &~ 0 \\
 0 &~ \omega_{5}  &~ 0 \\
 0 &~ 0 &~ \omega_{5} ^4
\end{pmatrix}
\,,~~\\[-14pt] \\[4pt]
\bf{3^{\prime}:} &  ~S=\frac{1}{\sqrt{5}}
\begin{pmatrix}
 -1 &~ \sqrt{2} &~ \sqrt{2} \\
 \sqrt{2} &~ 1-\kappa &~ \kappa  \\
 \sqrt{2} &~ \kappa  &~ 1-\kappa
\end{pmatrix}\,, ~
&T=
\begin{pmatrix}
 1 &~ 0 &~ 0 \\
 0 &~ \omega_{5} ^2 &~ 0 \\
 0 &~ 0 &~ \omega_{5} ^3
\end{pmatrix}
\,,~~\\[-14pt] \\[4pt]
\bf{4:} &  ~S=\frac{1}{\sqrt{5}}
\begin{pmatrix}
 1 &~ \kappa-1 &~ \kappa  &~ -1 \\
\kappa-1 &~ -1 &~ 1 &~ \kappa  \\
 \kappa  &~ 1 &~ -1 &~ \kappa-1 \\
 -1 &~ \kappa  &~ \kappa-1 &~ 1
\end{pmatrix}\,,~
&T=
\begin{pmatrix}
 \omega_{5}  &~ 0 &~ 0 &~ 0 \\
 0 &~ \omega_{5} ^2 &~ 0 &~ 0 \\
 0 &~ 0 &~ \omega_{5}^3 &~ 0 \\
 0 &~ 0 &~ 0 &~ \omega_{5}^4
\end{pmatrix}\,,~~\\[-14pt] \\[4pt]
\bf{5:} &  ~S=\frac{1}{5}
\begin{pmatrix}
 -1 &~ \sqrt{6} &~ \sqrt{6} &~ \sqrt{6} &~ \sqrt{6} \\
 \sqrt{6} &~ (\kappa-1)^{2} &~ -2 \kappa  &~ 2(\kappa-1) &~ \kappa ^2 \\
 \sqrt{6} &~ -2\kappa  &~ \kappa^2&~ (\kappa-1)^{2} &~ 2(\kappa-1) \\
 \sqrt{6} &~ 2(\kappa-1) &~ (\kappa-1)^{2} &~ \kappa ^2 &~ -2 \kappa  \\
 \sqrt{6} &~ \kappa^2 &~ 2(\kappa-1) &~ -2 \kappa  &~ (\kappa-1)^{2}
\end{pmatrix}\,,~
&T=
\begin{pmatrix}
 1 &~ 0 &~ 0 &~ 0 &~ 0 \\
 0 &~ \omega_{5}  &~ 0 &~ 0 &~ 0 \\
 0 &~ 0 &~ \omega_{5} ^2 &~ 0 &~ 0 \\
 0 &~ 0 &~ 0 &~ \omega_{5} ^3 &~ 0 \\
 0 &~ 0 &~ 0 &~ 0 &~ \omega_{5} ^4
\end{pmatrix}\,,~~
  \end{array}
\end{eqnarray}
where $\omega_{5}=e^{\frac{2\pi i}{5}}$. The character table of $A_{5}$ group is reported in Table \ref{tab:character}. We can straightforwardly obtain the Kronecker products between various representations:
\begin{table}[t]
\begin{center}
\begin{tabular}{|c|c|c|c|c|c|}\hline\hline
   &\multicolumn{5}{c|}{Conjugacy Classes}\\\cline{2-6}
   &$1C_{1}$&$15C_{2}$&$20C_{3}$&$12C_{5}$&$12C^{\prime}_{5}$ \\ \hline
$\bf{1}$&1&1&1&1&1\\\hline

$\bf{3}$&3&$-$1&0& $\kappa$ & $1-\kappa$
\\\hline

$\bf{3}'$&3&$-1$&0&$1-\kappa$& $\kappa$ \\\hline

$\bf{4}$&4&0&1&$-1$ &$-1$ \\\hline

$\bf{5}$&5&1&$-1$&0  & 0 \\\hline\hline

\end{tabular}
\caption{\label{tab:character} The character table of the $A_5$ group, where $\kappa=\frac{1+\sqrt{5}}{2}$.}
\end{center}
\end{table}
\begin{eqnarray}
\nonumber&&\bf{1}\otimes \bf{R}=\bf{R}\otimes\bf{1}=\bf{R},~~~\bf{3}\otimes\bf{3}=\bf{1}\oplus\bf{3}\oplus\bf{5},~~~\bf{3}'\otimes\bf{3}'=\bf{1}\oplus\bf{3}'\oplus\bf{5}, ~~~\mathbf{3}\times\bf{3}'=\mathbf{4}\oplus\bf{5},\\
\nonumber&&\bf{3}\otimes\bf{4}=\bf{3}'\oplus\bf{4}\oplus\bf{5},~~~\bf{3}'\otimes\bf{4}=\bf{3}\oplus\bf{4}\oplus\bf{5},~~~\bf{3}\otimes\bf{5}
=\bf{3}\oplus\bf{3}'\oplus\bf{4}\oplus\bf{5},\\
\nonumber&&\bf{3}'\otimes\bf{5}=\bf{3}\oplus\bf{3}'\oplus\bf{4}\oplus\bf{5},~~~\bf{4}\otimes\bf{4}=\bf{1}\oplus\bf{3}\oplus\bf{3}'\oplus\bf{4}\oplus\bf{5},
~~~\bf{4}\otimes\bf{5}=\bf{3}\oplus\bf{3}'\oplus\bf{4}\oplus\bf{5_{1}}\oplus\bf{5_{2}},\\ \label{eq:Kronecker}&&\mathbf{5}\otimes\bf{5}=\bf{1}\oplus\bf{3}\oplus\bf{3}'\oplus\bf{4_{1}}\oplus\bf{4_{2}}\oplus\bf{5_{1}}\oplus\bf{5_{2}}.
\end{eqnarray}
where $\bf{R}$ represents any irreducible representation of $A_{5}$, and $\bf{4_{1}}$, $\bf{4_{2}}$, $\bf{5_{1}}$ and $\bf{5_{2}}$ stand for the two $\bf{4}$ and two $\bf{5}$ representations that appear in the Kronecker products.

We now list the Clebsch-Gordan coefficients for our basis. We use the notation $\alpha_{i}$ ($\beta_{i}$) to denote the elements of the first (second) representation. The subscript "$\bf{S}$" ("$\bf{A}$") refers to symmetric (antisymmetric) combinations.
\begin{eqnarray*}
\begin{array}{|c|c|c|}\hline\hline
~\bf{3}\otimes\bf{3}=\bf{1_{S}}\oplus\bf{3_{A}} \oplus\bf{5_{S}}~ & ~\bf{3^\prime}\otimes\bf{3^\prime}=\bf{1_{S}}\oplus\bf{3^\prime_{A}} \oplus\bf{5_{S}}~ &  ~\bf{3}\otimes\bf{3^\prime}=\bf{4}\oplus\bf{5} \\ \hline
 & & \\[-0.18in]
~\bf{1_S}:\alpha_{1} \beta_{1}+\alpha_{2} \beta_{3}+\alpha_{3} \beta_{2}~  & ~\bf{1_{S}}: \alpha_{1} \beta_{1}+\alpha_{2} \beta_{3}+\alpha_{3} \beta_{2}~ & ~  \\ & & \\[-0.28in]
~\bf{3}_A:\begin{pmatrix}
 \alpha_{2} \beta_{3}-\alpha_{3} \beta_{2} \\
 \alpha_{1} \beta_{2}-\alpha_{2} \beta_{1} \\
 \alpha_{3} \beta_{1}-\alpha_{1} \beta_{3}
\end{pmatrix}
~ &
~\bf{3^\prime_{A}}:
\begin{pmatrix}
 \alpha_{2} \beta_{3}-\alpha_{3} \beta_{2} \\
 \alpha_{1} \beta_{2}-\alpha_{2} \beta_{1} \\
 \alpha_{3} \beta_{1}-\alpha_{1} \beta_{3}
\end{pmatrix}
~ &
~\bf{4}:
\begin{pmatrix}
 \sqrt{2} \alpha_{2} \beta_{1}+\alpha_{3} \beta_{2} \\
 -\sqrt{2} \alpha_{1} \beta_{2}-\alpha_{3} \beta_{3} \\
 -\sqrt{2} \alpha_{1} \beta_{3}-\alpha_{2} \beta_{2} \\
 \sqrt{2} \alpha_{3} \beta_{1}+\alpha_{2} \beta_{3}
\end{pmatrix} \\
& & \\[-0.20in]
~\bf{5_S}:
\begin{pmatrix}
 2 \alpha_{1} \beta_{1}-\alpha_{2} \beta_{3}-\alpha_{3} \beta_{2} \\
 -\sqrt{3} (\alpha_{1} \beta_{2}+ \alpha_{2} \beta_{1}) \\
 \sqrt{6} \alpha_{2} \beta_{2} \\
 \sqrt{6} \alpha_{3} \beta_{3} \\
-\sqrt{3} (\alpha_{1} \beta_{3}+ \alpha_{3} \beta_{1})
\end{pmatrix}
~  &
~\bf{5_{S}}:
\begin{pmatrix}
 2 \alpha_{1} \beta_{1}-\alpha_{2} \beta_{3} -\alpha_{3} \beta_{2}\\
 \sqrt{6} \alpha_{3} \beta_{3} \\
 -\sqrt{3} (\alpha_{1} \beta_{2} + \alpha_{2} \beta_{1})\\
 -\sqrt{3} (\alpha_{1} \beta_{3}+ \alpha_{3} \beta_{1}) \\
 \sqrt{6} \alpha_{2} \beta_{2}
\end{pmatrix}
~ &
~\bf{5}:
\begin{pmatrix}
 \sqrt{3} \alpha_{1} \beta_{1} \\
 \alpha_{2} \beta_{1}-\sqrt{2} \alpha_{3} \beta_{2} \\
 \alpha_{1} \beta_{2}-\sqrt{2} \alpha_{3} \beta_{3} \\
 \alpha_{1} \beta_{3}-\sqrt{2} \alpha_{2} \beta_{2} \\
 \alpha_{3} \beta_{1}-\sqrt{2} \alpha_{2} \beta_{3}
\end{pmatrix}
 \\ \hline \hline
\end{array}
\end{eqnarray*}

\begin{eqnarray*}
\begin{array}{|c|c|} \hline\hline
~\bf{3}\otimes\bf{4}=\bf{3^\prime}\oplus\bf{4}\oplus\bf{5}~~ & ~\bf{3^\prime}\otimes\bf{4}=\bf{3}\oplus\bf{4}\oplus\bf{5}~ \\ \hline
  & \\[-0.16in]
~\bf{3^\prime}:
\begin{pmatrix}
 -\sqrt{2} (\alpha_{2} \beta_{4}+\alpha_{3} \beta_{1}) \\
 \sqrt{2} \alpha_{1} \beta_{2}-\alpha_{2} \beta_{1}+\alpha_{3} \beta_{3} \\
 \sqrt{2} \alpha_{1} \beta_{3}+\alpha_{2} \beta_{2}-\alpha_{3} \beta_{4}
\end{pmatrix}
~~ &  ~\bf{3}:
\begin{pmatrix}
 -\sqrt{2} (\alpha_{2} \beta_{3}+ \alpha_{3} \beta_{2}) \\
 \sqrt{2} \alpha_{1} \beta_{1}+\alpha_{2} \beta_{4}-\alpha_{3} \beta_{3} \\
  \sqrt{2} \alpha_{1} \beta_{4} -\alpha_{2} \beta_{2}+\alpha_{3} \beta_{1}
\end{pmatrix}\\
  & \\[-0.16in]
~\bf{4}:
\begin{pmatrix}
  \alpha_{1} \beta_{1}-\sqrt{2}\alpha_{3} \beta_{2} \\
 -\alpha_{1} \beta_{2}-\sqrt{2}\alpha_{2} \beta_{1} \\
  \alpha_{1} \beta_{3}+\sqrt{2}\alpha_{3} \beta_{4} \\
  -\alpha_{1} \beta_{4}+\sqrt{2}\alpha_{2} \beta_{3}
\end{pmatrix}
~~ &
~\bf{4}:
\begin{pmatrix}
 \alpha_{1} \beta_{1}+\sqrt{2}\alpha_{3} \beta_{3} \\
  \alpha_{1} \beta_{2}-\sqrt{2}\alpha_{3} \beta_{4} \\
 -\alpha_{1} \beta_{3}+\sqrt{2}\alpha_{2} \beta_{1} \\
   -\alpha_{1} \beta_{4}-\sqrt{2}\alpha_{2} \beta_{2}
\end{pmatrix}
~  \\
 &  \\[-0.16in]
~\bf{5}:
\begin{pmatrix}
 \sqrt{6} (\alpha_{2}\beta_{4}- \alpha_{3} \beta_{1}) \\
 2\sqrt{2} \alpha_{1} \beta_{1}+2 \alpha_{3} \beta_{2}\\
 -\sqrt{2} \alpha_{1} \beta_{2}+\alpha_{2} \beta_{1}+3\alpha_{3} \beta_{3} \\
\sqrt{2} \alpha_{1} \beta_{3}-3\alpha_{2} \beta_{2}-\alpha_{3} \beta_{4}\\
 -2\sqrt{2} \alpha_{1} \beta_{4}-2 \alpha_{2} \beta_{3}
\end{pmatrix}
~~ &
~\bf{5}:
\begin{pmatrix}
 \sqrt{6} (\alpha_{2} \beta_{3}-\alpha_{3}\beta_{2}) \\
   \sqrt{2} \alpha_{1} \beta_{1}-3\alpha_{2} \beta_{4}-\alpha_{3} \beta_{3} \\
   2\sqrt{2} \alpha_{1} \beta_{2}+2 \alpha_{3} \beta_{4} \\
   -2\sqrt{2} \alpha_{1} \beta_{3}-2\alpha_{2} \beta_{1} \\
 -\sqrt{2}\alpha_{1} \beta_{4}+\alpha_{2} \beta_{2}+3\alpha_{3} \beta_{1}
\end{pmatrix}
~ \\ \hline\hline
\end{array}
\end{eqnarray*}

\begin{eqnarray*}
\begin{array}{|c|c|} \hline\hline
~\bf{3}\otimes\bf{5}=\bf{3}\oplus\bf{3^\prime}\oplus\bf{4}\oplus\bf{5}~~ & ~\bf{3^\prime}\otimes\bf{5}=\bf{3}\oplus\bf{3^\prime}\oplus\bf{4}\oplus\bf{5}~ \\ \hline &  \\[-0.16in]
~\bf{3}:
\begin{pmatrix}
 -2 \alpha_{1} \beta_{1}+\sqrt{3}\alpha_{2} \beta_{5}+\sqrt{3}\alpha_{3} \beta_{2} \\
 \sqrt{3}\alpha_{1} \beta_{2}+\alpha_{2} \beta_{1}-\sqrt{6}\alpha_{3} \beta_{3} \\
 \sqrt{3}\alpha_{1} \beta_{5}-\sqrt{6}\alpha_{2} \beta_{4}+\alpha_{3} \beta_{1}
\end{pmatrix}
~~ &
~\bf{3}:
\begin{pmatrix}
 \sqrt{3} \alpha_{1} \beta_{1}+\alpha_{2}\beta_{4}+\alpha_{3} \beta_{3} \\
 \alpha_{1} \beta_{2}-\sqrt{2}\alpha_{2} \beta_{5} -\sqrt{2}\alpha_{3} \beta_{4}\\
 \alpha_{1} \beta_{5}-\sqrt{2} \alpha_{2} \beta_{3}-\sqrt{2} \alpha_{3} \beta_{2}
\end{pmatrix}
~ \\  &  \\[-0.16in]
~\bf{3^\prime}:
\begin{pmatrix}
 \sqrt{3} \alpha_{1} \beta_{1}+\alpha_{2} \beta_{5}+\alpha_{3} \beta_{2} \\
  \alpha_{1} \beta_{3}-\sqrt{2}\alpha_{2} \beta_{2}-\sqrt{2}\alpha_{3} \beta_{4} \\
 \alpha_{1} \beta_{4}-\sqrt{2}\alpha_{2} \beta_{3}-\sqrt{2}\alpha_{3} \beta_{5}
\end{pmatrix}
~~ & ~\bf{3^\prime}:
\begin{pmatrix}
 -2 \alpha_{1} \beta_{1}+\sqrt{3}\alpha_{2} \beta_{4} +\sqrt{3}\alpha_{3} \beta_{3}\\
 \sqrt{3}\alpha_{1} \beta_{3}+\alpha_{2} \beta_{1}-\sqrt{6}\alpha_{3} \beta_{5} \\
 \sqrt{3}\alpha_{1} \beta_{4}-\sqrt{6}\alpha_{2} \beta_{2}+\alpha_{3} \beta_{1}
\end{pmatrix}
~ \\  &  \\[-0.16in]
~\bf{4}:
\begin{pmatrix}
 2\sqrt{2} \alpha_{1} \beta_{2}-\sqrt{6} \alpha_{2} \beta_{1}+\alpha_{3} \beta_{3} \\
 -\sqrt{2}\alpha_{1} \beta_{3}+2\alpha_{2} \beta_{2}-3 \alpha_{3} \beta_{4} \\
 \sqrt{2}\alpha_{1} \beta_{4}+3\alpha_{2} \beta_{3}-2\alpha_{3} \beta_{5} \\
 -2\sqrt{2} \alpha_{1} \beta_{5}-\alpha_{2} \beta_{4}+\sqrt{6} \alpha_{3} \beta_{1}
\end{pmatrix}
~~ &  ~\bf{4}:
\begin{pmatrix}
 \sqrt{2} \alpha_{1} \beta_{2}+3 \alpha_{2}\beta_{5}-2\alpha_{3} \beta_{4} \\
 2\sqrt{2} \alpha_{1} \beta_{3}-\sqrt{6} \alpha_{2} \beta_{1}+\alpha_{3} \beta_{5} \\
 -2\sqrt{2} \alpha_{1} \beta_{4}-\alpha_{2} \beta_{2} +\sqrt{6} \alpha_{3} \beta_{1}\\
 -\sqrt{2} \alpha_{1}\beta_{5}+2 \alpha_{2}\beta_{3}-3 \alpha_{3} \beta_{2}
\end{pmatrix}
~ \\  &  \\[-0.16in]
~\bf{5}:
\begin{pmatrix}
 \sqrt{3} (\alpha_{2} \beta_{5}- \alpha_{3}\beta_{2}) \\
 -\alpha_{1} \beta_{2}-\sqrt{3} \alpha_{2} \beta_{1}-\sqrt{2}\alpha_{3} \beta_{3} \\
 -2 \alpha_{1} \beta_{3}-\sqrt{2}\alpha_{2} \beta_{2} \\
 2\alpha_{1} \beta_{4}+\sqrt{2}\alpha_{3} \beta_{5} \\
 \alpha_{1} \beta_{5}+\sqrt{2}\alpha_{2} \beta_{4}+ \sqrt{3} \alpha_{3} \beta_{1}
\end{pmatrix}
~~ &
~\bf{5}:
\begin{pmatrix}
 \sqrt{3} (\alpha_{2} \beta_{4}- \alpha_{3} \beta_{3}) \\
 2 \alpha_{1} \beta_{2}+\sqrt{2}\alpha_{3} \beta_{4} \\
 -\alpha_{1} \beta_{3}-\sqrt{3} \alpha_{2} \beta_{1}-\sqrt{2}\alpha_{3} \beta_{5} \\
\alpha_{1} \beta_{4}+\sqrt{2} \alpha_{2} \beta_{2} + \sqrt{3} \alpha_{3} \beta_{1}\\
 -2\alpha_{1} \beta_{5}-\sqrt{2} \alpha_{2} \beta_{3}
\end{pmatrix}
~ \\ \hline\hline
\end{array}
\end{eqnarray*}

\begin{eqnarray*}
\begin{array}{|c|c|} \hline\hline
\bf{4}\otimes\bf{4}=\bf{1_{S}}\oplus\bf{3_{A}}\oplus\bf{3^\prime_{A}}\oplus\bf{4_{S}}\oplus\bf{5_{S}} & \bf{4}\otimes\bf{5}=\bf{3}\oplus\bf{3^\prime}\oplus\bf{4}\oplus\bf{5}_1\oplus\bf{5}_2 \\ \hline  &  \\[-0.16in]
\bf{1_{S}}:\alpha_{1}\beta_{4}+\alpha_{2} \beta_{3}+\alpha_{3} \beta_{2}+\alpha_{4} \beta_{1}  &
\bf{3}:
\begin{pmatrix}
 2 \sqrt{2} \alpha_{1}\beta_{5}-\sqrt{2} \alpha_{2} \beta_{4}+\sqrt{2} \alpha_{3} \beta_{3}-2 \sqrt{2} \alpha_{4} \beta_{2}\\
 -\sqrt{6} \alpha_{1} \beta_{1}+2 \alpha_{2} \beta_{5}+3 \alpha_{3} \beta_{4}-\alpha_{4} \beta_{3} \\
 \alpha_{1} \beta_{4}-3 \alpha_{2}\beta_{3}-2\alpha_{3} \beta_{2}+\sqrt{6} \alpha_{4} \beta_{1}
\end{pmatrix}
 \\  &  \\[-0.16in]
\bf{3_{A}}:
\begin{pmatrix}
 -\alpha_{1} \beta_{4}+\alpha_{2}\beta_{3}-\alpha_{3}\beta_{2}+\alpha_{4} \beta_{1}\\
 \sqrt{2} (\alpha_{2} \beta_{4}- \alpha_{4} \beta_{2}) \\
 \sqrt{2} (\alpha_{1} \beta_{3}- \alpha_{3} \beta_{1})
\end{pmatrix}
 &
\bf{3^\prime}:
\begin{pmatrix}
 \sqrt{2} \alpha_{1} \beta_{5}+2\sqrt{2} \alpha_{2} \beta_{4}-2\sqrt{2} \alpha_{3} \beta_{3}-\sqrt{2} \alpha_{4} \beta_{2}   \\
 3\alpha_{1} \beta_{2}-\sqrt{6} \alpha_{2} \beta_{1}-\alpha_{3} \beta_{5}+2 \alpha_{4}\beta_{4} \\
 -2 \alpha_{1} \beta_{3}+\alpha_{2} \beta_{2}+\sqrt{6} \alpha_{3} \beta_{1}-3 \alpha_{4} \beta_{5}
\end{pmatrix}
 \\ &  \\[-0.16in]
\bf{3^\prime_{A}}:
\begin{pmatrix}
\alpha_{1} \beta_{4} +\alpha_{2} \beta_{3}-\alpha_{3}\beta_{2} -\alpha_{4} \beta_{1}\\
 \sqrt{2} (\alpha_{3} \beta_{4}- \alpha_{4} \beta_{3}) \\
 \sqrt{2} (\alpha_{1} \beta_{2}- \alpha_{2} \beta_{1})
\end{pmatrix}
 &
\bf{4}:
\begin{pmatrix}
 \sqrt{3} \alpha_{1} \beta_{1}-\sqrt{2} \alpha_{2} \beta_{5}+\sqrt{2} \alpha_{3} \beta_{4}-2\sqrt{2} \alpha_{4} \beta_{3}\\
 -\sqrt{2} \alpha_{1} \beta_{2}-\sqrt{3} \alpha_{2} \beta_{1}+2 \sqrt{2} \alpha_{3} \beta_{5}+\sqrt{2} \alpha_{4} \beta_{4} \\
 \sqrt{2} \alpha_{1} \beta_{3}+2\sqrt{2} \alpha_{2} \beta_{2}-\sqrt{3} \alpha_{3} \beta_{1}-\sqrt{2} \alpha_{4} \beta_{5}\\
 -2 \sqrt{2} \alpha_{1} \beta_{4}+\sqrt{2} \alpha_{2} \beta_{3}-\sqrt{2} \alpha_{3} \beta_{2}+\sqrt{3} \alpha_{4} \beta_{1}
\end{pmatrix}
  \\  &  \\[-0.16in]
\bf{4_{S}}:
\begin{pmatrix}
 \alpha_{2} \beta_{4}+\alpha_{3}\beta_{3}+\alpha_{4} \beta_{2} \\
 \alpha_{1} \beta_{1}+\alpha_{3} \beta_{4} +\alpha_{4} \beta_{3}\\
 \alpha_{1}\beta_{2}+\alpha_{2} \beta_{1}+\alpha_{4} \beta_{4} \\
 \alpha_{1} \beta_{3}+\alpha_{2}\beta_{2}+\alpha_{3} \beta_{1}
\end{pmatrix}
 &
\bf{5}_1:
\begin{pmatrix}
 \sqrt{2} \alpha_{1} \beta_{5}-\sqrt{2} \alpha_{2} \beta_{4}-\sqrt{2} \alpha_{3} \beta_{3}+\sqrt{2} \alpha_{4} \beta_{2} \\
 -\sqrt{2} \alpha_{1} \beta_{1}-\sqrt{3} \alpha_{3} \beta_{4} -\sqrt{3} \alpha_{4} \beta_{3}\\
 \sqrt{3} \alpha_{1} \beta_{2}+\sqrt{2} \alpha_{2} \beta_{1}+\sqrt{3} \alpha_{3} \beta_{5}\\
 \sqrt{3} \alpha_{2} \beta_{2}+\sqrt{2} \alpha_{3} \beta_{1}+\sqrt{3} \alpha_{4} \beta_{5}\\
 -\sqrt{3} \alpha_{1} \beta_{4}-\sqrt{3} \alpha_{2} \beta_{3}-\sqrt{2} \alpha_{4} \beta_{1}
\end{pmatrix}
  \\  &  \\[-0.16in]
\bf{5_{S}}:
\begin{pmatrix}
 \sqrt{3} (\alpha_{1} \beta_{4}- \alpha_{2} \beta_{3}-\alpha_{3} \beta_{2}+ \alpha_{4} \beta_{1}) \\
 -\sqrt{2} \alpha_{2} \beta_{4}+2 \sqrt{2} \alpha_{3} \beta_{3}-\sqrt{2} \alpha_{4} \beta_{2}\\
 -2 \sqrt{2} \alpha_{1} \beta_{1}+\sqrt{2} \alpha_{3} \beta_{4}+\sqrt{2} \alpha_{4} \beta_{3} \\
 \sqrt{2} \alpha_{1} \beta_{2}+\sqrt{2} \alpha_{2} \beta_{1}-2 \sqrt{2} \alpha_{4} \beta_{4} \\
 -\sqrt{2} \alpha_{1} \beta_{3}+2\sqrt{2} \alpha_{2} \beta_{2}-\sqrt{2} \alpha_{3} \beta_{1}
\end{pmatrix}
  &
\bf{5}_2:
\begin{pmatrix}
 2 \alpha_{1}\beta_{5}+4 \alpha_{2} \beta_{4}+4 \alpha_{3} \beta_{3} +2 \alpha_{4} \beta_{2}\\
 4 \alpha_{1} \beta_{1}+2 \sqrt{6} \alpha_{2} \beta_{5} \\
 -\sqrt{6} \alpha_{1}\beta_{2}+2 \alpha_{2} \beta_{1}-\sqrt{6} \alpha_{3} \beta_{5} +2 \sqrt{6} \alpha_{4} \beta_{4}\\
 2 \sqrt{6} \alpha_{1} \beta_{3}-\sqrt{6} \alpha_{2}\beta_{2}+2 \alpha_{3} \beta_{1}-\sqrt{6} \alpha_{4} \beta_{5} \\
 2 \sqrt{6} \alpha_{3} \beta_{2}+4 \alpha_{4} \beta_{1}
\end{pmatrix}
 \\ \hline\hline
\end{array}
\end{eqnarray*}

\begin{eqnarray*}
\begin{array}{|c|} \hline\hline
~\bf{5}\otimes\bf{5}=\bf{1_{S}}\oplus\bf{3_{A}}\oplus\bf{3^\prime_{A}}\oplus\bf{4_{S,1}}\oplus\bf{4_{A,2}}\oplus\bf{5}_{S,1}\oplus\bf{5}_{S,2}~ \\ \hline \\[-0.16in]
~\bf{1_{S}}:\alpha_{1}\beta_{1}+\alpha_{2}\beta_{5}+\alpha_{3}\beta_{4}+\alpha_{4}\beta_{3}+\alpha_{5}\beta_{2}~ \\
~\bf{3_{A}}:
\begin{pmatrix}
 \alpha_{2} \beta_{5}+2\alpha_{3} \beta_{4}-2 \alpha_{4} \beta_{3}-\alpha_{5} \beta_{2} \\
 -\sqrt{3} \alpha_{1}\beta_{2}+\sqrt{3} \alpha_{2} \beta_{1}+\sqrt{2}\alpha_{3} \beta_{5}-\sqrt{2} \alpha_{5} \beta_{3} \\
 \sqrt{3}\alpha_{1} \beta_{5}+\sqrt{2} \alpha_{2} \beta_{4}-\sqrt{2} \alpha_{4}\beta_{2}-\sqrt{3} \alpha_{5} \beta_{1}
\end{pmatrix}
~ \\
~\bf{3^\prime_{A}}:
\begin{pmatrix}
 2 \alpha_{2} \beta_{5}-\alpha_{3} \beta_{4}+\alpha_{4} \beta_{3}-2 \alpha_{5} \beta_{2}\\
 \sqrt{3} \alpha_{1}\beta_{3}-\sqrt{3} \alpha_{3} \beta_{1}+\sqrt{2}\alpha_{4} \beta_{5}-\sqrt{2} \alpha_{5} \beta_{4} \\
 -\sqrt{3}\alpha_{1} \beta_{4}+\sqrt{2} \alpha_{2} \beta_{3}-\sqrt{2} \alpha_{3}\beta_{2}+\sqrt{3} \alpha_{4} \beta_{1}
\end{pmatrix}
~ \\
~\bf{4_{S,1}}:
\begin{pmatrix}
 3 \sqrt{2} \alpha_{1} \beta_{2}+3 \sqrt{2} \alpha_{2} \beta_{1}-\sqrt{3}\alpha_{3}\beta_{5}+4 \sqrt{3} \alpha_{4} \beta_{4}-\sqrt{3}\alpha_{5} \beta_{3} \\
 3 \sqrt{2}\alpha_{1} \beta_{3}+4 \sqrt{3} \alpha_{2} \beta_{2}+3 \sqrt{2} \alpha_{3} \beta_{1}-\sqrt{3}\alpha_{4}\beta_{5}-\sqrt{3} \alpha_{5} \beta_{4} \\
 3 \sqrt{2} \alpha_{1} \beta_{4}-\sqrt{3}\alpha_{2} \beta_{3}-\sqrt{3} \alpha_{3} \beta_{2}+3 \sqrt{2} \alpha_{4} \beta_{1}+4 \sqrt{3}\alpha_{5}\beta_{5} \\
 3 \sqrt{2} \alpha_{1}\beta_{5}-\sqrt{3} \alpha_{2} \beta_{4}+4 \sqrt{3}\alpha_{3} \beta_{3}-\sqrt{3} \alpha_{4} \beta_{2}+3 \sqrt{2} \alpha_{5} \beta_{1}
\end{pmatrix}
 ~ \\
~\bf{4_{A,2}}:
\begin{pmatrix}
 \sqrt{2} \alpha_{1} \beta_{2}-\sqrt{2} \alpha_{2} \beta_{1}+\sqrt{3} \alpha_{3} \beta_{5}-\sqrt{3}\alpha_{5} \beta_{3} \\
 -\sqrt{2} \alpha_{1} \beta_{3}+\sqrt{2} \alpha_{3} \beta_{1}+\sqrt{3} \alpha_{4} \beta_{5}-\sqrt{3} \alpha_{5} \beta_{4} \\
 -\sqrt{2} \alpha_{1} \beta_{4}-\sqrt{3} \alpha_{2} \beta_{3}+\sqrt{3} \alpha_{3} \beta_{2}+\sqrt{2} \alpha_{4} \beta_{1}\\
 \sqrt{2} \alpha_{1} \beta_{5}-\sqrt{3}\alpha_{2} \beta_{4}+\sqrt{3} \alpha_{4} \beta_{2}-\sqrt{2} \alpha_{5} \beta_{1}
\end{pmatrix}
~  \\
~\bf{5}_{S,1}:
\begin{pmatrix}
 2 \alpha_{1} \beta_{1}+\alpha_{2} \beta_{5}-2 \alpha_{3} \beta_{4}-2 \alpha_{4} \beta_{3}+\alpha_{5} \beta_{2} \\
 \alpha_{1} \beta_{2}+\alpha_{2} \beta_{1}+\sqrt{6} \alpha_{3} \beta_{5}+\sqrt{6} \alpha_{5} \beta_{3} \\
 -2 \alpha_{1} \beta_{3}+\sqrt{6} \alpha_{2} \beta_{2}-2 \alpha_{3} \beta_{1} \\
 -2 \alpha_{1} \beta_{4}-2 \alpha_{4} \beta_{1}+\sqrt{6} \alpha_{5} \beta_{5} \\
 \alpha_{1} \beta_{5}+\sqrt{6} \alpha_{2} \beta_{4}+\sqrt{6} \alpha_{4} \beta_{2}+\alpha_{5} \beta_{1}
\end{pmatrix}
~ \\
~\bf{5}_{S,2}:
\begin{pmatrix}
 2 \alpha_{1} \beta_{1}-2 \alpha_{2} \beta_{5}+\alpha_{3} \beta_{4}+\alpha_{4}\beta_{3}-2 \alpha_{5} \beta_{2} \\
 -2 \alpha_{1} \beta_{2}-2 \alpha_{2} \beta_{1}+\sqrt{6} \alpha_{4} \beta_{4} \\
 \alpha_{1} \beta_{3}+\alpha_{3} \beta_{1}+\sqrt{6} \alpha_{4} \beta_{5}+\sqrt{6} \alpha_{5} \beta_{4} \\
 \alpha_{1} \beta_{4}+\sqrt{6} \alpha_{2} \beta_{3}+\sqrt{6} \alpha_{3} \beta_{2}+\alpha_{4} \beta_{1} \\
 -2 \alpha_{1} \beta_{5}+\sqrt{6} \alpha_{3} \beta_{3}-2 \alpha_{5} \beta_{1}
 \end{pmatrix}.~ \\ \hline\hline
\end{array}
\end{eqnarray*}


\section{\label{sec:B} Lepton flavor mixing from $A_5$ family symmetry without CP}

In this section, we investigate the possible lepton mixing patterns which can be derived from only $A_5$ family symmetry without CP symmetry imposed. As usual, the three generations of left-handed leptons are assigned to the triplet representation $\mathbf{3}$, and $A_5$ is broken into two different abelian subgroups $G_{l}$ and $G_{\nu}$ in the charged lepton and neutrino sector respectively. The residual flavor symmetry $G_{\nu}$ can only be a $Z_{2}$ or $K_{4}$ subgroup of $A_{5}$ since we assume neutrinos are Majorana particles here. In this approach, the PMNS matrix can be obtained by simply diagonalizing the representation matrices of the generators of $G_{l}$ and $G_{\nu}$ without resorting to the mass matrix~\cite{Lam:2007qc,Toorop:2011jn,deAdelhartToorop:2011re}.
For $G_{\nu}=K_{4}$ and $G_{l}$ is capable of distinguishing the three generations of charged lepton, i.e., the eigenvalues of the generators of $G_{l}$ aren't degenerate, the PMNS matrix would be completely fixed up to row and column permutations. However, only one column would be fixed by the remnant flavor symmetries $G_{l}$ and $G_{\nu}$ in case of $G_{\nu}=Z_{2}$. In the following, the scenario of $G_{l}=Z_2$ and $G_{\nu}=K_{4}$ shall be discussed as well, and one row would be fixed instead. It is noteworthy that two pairs of subgroups $\left(G_{l}, G_{\nu}\right)$ and $\left(G^{\prime}_{l}, G^{\prime}_{\nu}\right)$ lead to the same result for the PMNS matrix, if they are conjugate under an element of the $A_5$ group.

\subsection{\label{sec:B.1} $G_{\nu}=K_4$}

From Appendix~\ref{sec:A}, we know that $G_{l}$ can be a $Z_3$, $Z_5$ or $K_4$ subgroup of $A_5$. In case of $G_{l}=Z_5$, all $6\times5=30$ possible combinations of $G_{l}$ and $G_{\nu}$ give rise to the same mixing matrix
\begin{equation}
\label{eq:golden_ratio}
U_{PMNS}=\left(\begin{array}{ccc}
    -\sqrt{\frac{\kappa}{\sqrt{5}}}  & ~\sqrt{\frac{1}{\sqrt{5}\kappa}} ~ & 0  \\
    \sqrt{\frac{1}{2\sqrt{5}\kappa}}   & ~ \sqrt{\frac{\kappa}{2\sqrt{5}}} ~  & -\frac{1}{\sqrt{2}}  \\
   \sqrt{\frac{1}{2\sqrt{5}\kappa}}   & ~ \sqrt{\frac{\kappa}{2\sqrt{5}}} ~  & \frac{1}{\sqrt{2}}
  \end{array}\right)\equiv U_{GR}\,,
\end{equation}
which is the well-known golden ratio mixing pattern. The mixing angles are determined to be $\sin^2\theta_{12}=\left(3-\kappa\right)/5\simeq0.276$, $\sin^2\theta_{23}=1/2$ and $\sin^2\theta_{13}=0$. Obviously $\theta_{13}$ should acquire moderate corrections to accommodate the measured non-vanishing value of the reactor angle although $\theta_{12}$ and $\theta_{23}$ are in the experimentally favored $3\sigma$ ranges~\cite{Capozzi:2013csa}.

In case of $G_{l}=Z_3$, we find two mixing patterns can be obtained. For the representative symmetries $G_{l}=Z^{T^3ST^2S}_{3}$ and $G_{\nu}=K^{(ST^2ST^3S,TST^4)}_{4}$, the elements of $G_{l}$ and $G_{\nu}$ generate an $A_4$ subgroup instead of the full flavor symmetry group $A_5$. The resulting mixing matrix is given by the familiar democratic mixing in which all elements have the same absolute value~\cite{Cabibbo:1977nk}, i.e.,
\begin{equation}\label{eq:identical}
U_{PMNS}=\frac{1}{\sqrt{3}}\left(\begin{array}{ccc}
 1 &~ 1 ~& 1 \\
 e^{\frac{2\pi i}{3}} &~ 1 ~& -e^{\frac{\pi i}{3}} \\
 -e^{\frac{\pi i}{3}} &~ 1 ~& e^{\frac{2\pi i}{3}} \\
\end{array}\right)\equiv U_{DM}\,.
\end{equation}
The mixing angles are $\sin^2\theta_{12}=\sin^2\theta_{23}=1/2$ and $\sin^2\theta_{13}=1/3$. Large corrections to $\theta_{12}$ and $\theta_{13}$ are needed to be compatible with experimental data. For another representative symmetries $G_{l}=Z^{T^3ST^2S}_{3}$ and $G_{\nu}=K^{(S,T^3ST^2ST^3)}_{4}$, the parent group $A_5$ can be generated by $G_{l}$ and $G_{\nu}$. The mixing matrix is found to be of the form:
\begin{equation}\label{eq:same_row}
U_{PMNS}=\frac{1}{ \sqrt{6}}
\left(\begin{array}{ccc}
 \sqrt{2}\kappa & ~\sqrt{2}(1-\kappa)~ & 0 \\
 \kappa-1 & ~\kappa~ & - \sqrt{3} \\
 \kappa-1 &~ \kappa ~&  \sqrt{3} \\
\end{array}\right)\equiv U_{ST}\,,
\end{equation}
which leads to the following mixing angles: $\sin^2\theta_{12}=\left(2-\kappa\right)/3\simeq0.127$, $ \sin^2\theta_{23}=1/2$ and $\sin^2\theta_{13}=0$. Notice that both $\theta_{12}$ and $\theta_{13}$ are outside of the $3\sigma$ ranges~\cite{Capozzi:2013csa}. The same results have been obtained in Refs.~\cite{deAdelhartToorop:2011re,Lam:2011ag}. For the last case of $G_{l}=K_4$, where $G_{\nu}$ and $G_{l}$ are not the same Klein group, only one mixing pattern can be derived,
\begin{equation}\label{eq:same_column}
U_{PMNS}=\frac{1}{2}\left(\begin{array}{ccc}
 \kappa ~&~ -1 ~&~ \kappa-1 \\
 -1 ~&~ 1-\kappa ~&~ \kappa \\
\kappa-1 ~&~ \kappa ~&~ 1 \\
\end{array}\right)\equiv U_{RC}\,.
\end{equation}
We can extract the mixing angles:  $\sin^2\theta_{12}=\left(3-\kappa\right)/5\simeq0.276$,  $\sin^2\theta_{23}=\left(2+\kappa\right)/5\simeq0.724$ and  $\sin^2\theta_{13}=\left(2-\kappa\right)/4\simeq0.0955$. Both $\theta_{13}$ and $\theta_{23}$ are too large to be acceptable. This mixing pattern has also been found in Ref.~\cite{deAdelhartToorop:2011re}. In summary, no mixing matrix in agreement with experimental data can be obtained if the full Klein symmetry is preserved by the neutrino mass matrix. In the following, we consider the situation with a single residual $Z_2$ flavor symmetry in the neutrino sector or in the charged lepton sector.

\subsection{\label{sec:B.2}$G_{\nu}=Z_2$ or $G_{l}=Z_2$}

In this case, only one column or one row of the PMNS matrix would be determined up to permutations and phases of its elements by the remnant flavor symmetries $G_{l}$ and $G_{\nu}$~\cite{Hernandez:2012ra,Ge:2011ih}. This method generally allows us to obtain relations between mixing parameters and a non-zero $\theta_{13}$. We have scanned all independent combinations of $G_{l}$ and $G_{\nu}$, and the corresponding explicit forms of the fixed column or row vector are presented in Table~\ref{tab:fix_column_row}. Comparing with the present $3\sigma$ confidence level ranges of the moduli of the elements of the leptonic mixing matrix~\cite{Capozzi:2013csa}
\begin{equation}\label{eq:3sr}
|U_{PMNS}|_{3\sigma}=
\left(\begin{array}{ccc}
0.789\rightarrow0.853 ~&~ 0.501\rightarrow0.594 ~&~
0.133\rightarrow0.172\\
0.194\rightarrow0.558 ~&~ 0.408\rightarrow0.735 ~&~
0.602\rightarrow0.784 \\
0.194\rightarrow0.558 ~&~ 0.408\rightarrow0.735 ~&~
0.602\rightarrow0.784 \\
\end{array}\right)\,,
\end{equation}
we find that neither of the two possible row vectors can be accommodated by the data, and only four cases are viable. The remnant symmetries can be chosen to be $\left(G_{l}, G_{\nu}\right)=\left(Z^{T}_{5}, Z^{S}_{2}\right)$, $(Z^{T}_{5}, Z^{T^{3}ST^{2}ST^{3}}_{2})$, $(Z^{T^{3}ST^{2}S}_{3}, Z^{ST^{2}ST^{3}S}_{2})$ and $(K^{(ST^2ST^3S,TST^4)}_{4}, Z^{S}_{2})$ without loss of generality, and the fixed column are $(-\sqrt{\frac{\kappa}{\sqrt{5}}},\frac{1}{\sqrt{2\sqrt{5}\,\kappa}},\frac{1}{\sqrt{2\sqrt{5}\,\kappa}})^T$   , $(\sqrt{\frac{1}{\sqrt{5}\,\kappa}},\sqrt{\frac{\kappa}{2\sqrt{5}}},\sqrt{\frac{\kappa}{2\sqrt{5}}})^T$,  $\frac{1}{\sqrt{3}}(1, 1, 1)^T$ and  $\frac{1}{2}(\kappa, -1, \kappa-1)^T$ respectively. These column vectors can fit the first or the second column of the PMNS matrix. The resulting lepton mixing matrix can be obtained from $U_{GR}$, $U_{DM}$ and $U_{RC}$ by multiplying a unitary matrix $U_{23}$ or $U_{13}$ from the right-hand side with
\begin{table}[t!]
\centering
{\small
\begin{tabular}{|c|c|c|c|}
 \hline  \hline
 $G_{l}$  & $G_{\nu}$ & \texttt{Fixed column or row}  &    \\ \hline
 & & &   \\[-0.16in]
 \multirow{5}{*}{$Z^{T}_{5}$} & $Z^{S}_{2}$ & $(-\sqrt{\frac{\kappa}{\sqrt{5}}},\frac{1}{\sqrt{2\sqrt{5}\kappa}},\frac{1}{\sqrt{2\sqrt{5}\kappa}})^T$ & \cmark \\
  & & &   \\[-0.16in]
   \cline{2-4}
  & & &   \\[-0.16in]
  &  $Z^{T^3ST^2ST^3}_{2}$ & $(\sqrt{\frac{1}{\sqrt{5}\kappa}},\sqrt{\frac{\kappa}{2\sqrt{5}}},\sqrt{\frac{\kappa}{2\sqrt{5}}})^T$ & \cmark \\
  & & &   \\[-0.16in]
  \cline{2-4}
  & & &   \\[-0.16in]
  & $Z^{T^3ST^2ST^3S}_{2}$ & $(0,-\frac{1}{\sqrt{2}},\frac{1}{\sqrt{2}})^T$ & \xmark  \\
  & & &   \\[-0.16in] \hline
  & & &   \\[-0.16in]
  \multirow{4}{*}{$Z^{T^3ST^2S}_{3}$} & $Z^{S}_{2}$ & $(0,-\frac{1}{\sqrt{2}},\frac{1}{\sqrt{2}})^T$ & \xmark  \\
  & & &   \\[-0.16in] \cline{2-4}
  & & &   \\[-0.16in]
   & $Z^{T^3ST^2ST^3}_{2}$ & $(\frac{1-\kappa}{\sqrt{3}},\frac{\kappa}{\sqrt{6}},\frac{\kappa}{\sqrt{6}})^T$ & \xmark  \\
  & & &   \\[-0.16in] \cline{2-4}
  & & &  \\[-0.16in]
   & $Z^{T^3ST^2ST^3S}_{2}$¡¡& $(\frac{\kappa}{\sqrt{3}},\frac{\kappa-1}{\sqrt{6}},\frac{\kappa-1}{\sqrt{6}})^T$ & \xmark \\
  & & &  \\[-0.16in] \hline
  & & &   \\[-0.16in]
   $Z^{T^3ST^2S}_{3}$ & $Z^{ST^2ST^3S}_{2}$ & $(\frac{1}{\sqrt{3}},\frac{1}{\sqrt{3}},\frac{1}{\sqrt{3}})^T$ & \cmark \\
   & & &   \\[-0.16in] \hline
   & & &   \\[-0.16in]
  $K^{(ST^2ST^3S,~TST^4)}_{4}$ & $Z^{S}_{2}$ & $(\frac{\kappa}{2},-\frac{1}{2},\frac{\kappa-1}{2})^T$ & \cmark \\
  & & &   \\[-0.16in] \hline
  & & &   \\[-0.16in]
  $K^{(ST^2ST^3S,~TST^4)}_{4}$ & $Z^{TST^4}_{2}$ & $(1,0,0)^T$ & \xmark \\
  & & &   \\[-0.16in] \hline
  & & &   \\[-0.16in]
  $Z^{S}_{2}$ & $K^{(ST^2ST^3S,~TST^4)}_{4}$ &  $(\frac{\kappa}{2},\frac{1}{2},\frac{\kappa-1}{2})$ & \xmark \\
  & & &   \\[-0.16in] \hline
  & & &   \\[-0.16in]
  $Z^{TST^4}_{2}$ & $K^{(ST^2ST^3S,~TST^4)}_{4}$ &  $(1,0,0)$ & \xmark \\
   & & &   \\[-0.16in] \hline \hline
\end{tabular}}
\caption{\label{tab:fix_column_row} The possible form of one column (row) of the PMNS matrix determined by the residual flavor symmetry $G_{\nu}=Z_2$ ($G_{l}=Z_2$) within the framework of $A_5$ flavor symmetry. The notation ``\cmark'' denotes that the relevant lepton mixing is compatible with the experimental data at $3\sigma$ level~\cite{Capozzi:2013csa}. The notation ``\xmark'' implies the resulting mixing is not viable.}
\end{table}
\begin{equation}
U_{13}=\left(\begin{array}{ccc}
    \cos\theta  ~&~ 0  ~&~  \sin\theta e^{-i\delta}  \\
    0  ~&~  1  ~&~  0  \\
    -\sin\theta e^{i\delta}  ~&~  0  ~&~  \cos\theta
  \end{array}\right),\qquad U_{23}=\left(\begin{array}{ccc}
    1  ~&~ 0  ~&~  0  \\
    0  ~&~  \cos\theta  ~&~  \sin\theta e^{-i\delta}  \\
    0  ~&~  -\sin\theta e^{i\delta}  ~&~  \cos\theta
  \end{array}\right)\,,
\end{equation}
where $\theta$ and $\delta$ are real, and a arbitrary phase matrix in the right-hand side of $U_{13}$ and $U_{23}$ is omitted, since they can be absorbed into the Majorana phases which are not constrained by flavor symmetry. The multiplication of $U_{13}$ ($U_{23}$) corresponds to performing a unitary linear transformation of the 1st (2nd) and 3rd columns. In the following, we shall discuss the predictions for the PMNS matrix and lepton mixing parameters in each case.

\subsubsection{\label{sec:B.2.1} $G_{l}=Z^{T}_{5},G_{\nu}=Z^{S}_{2}$}

\begin{figure}[t!]
\begin{center}
\includegraphics[width=0.98\textwidth]{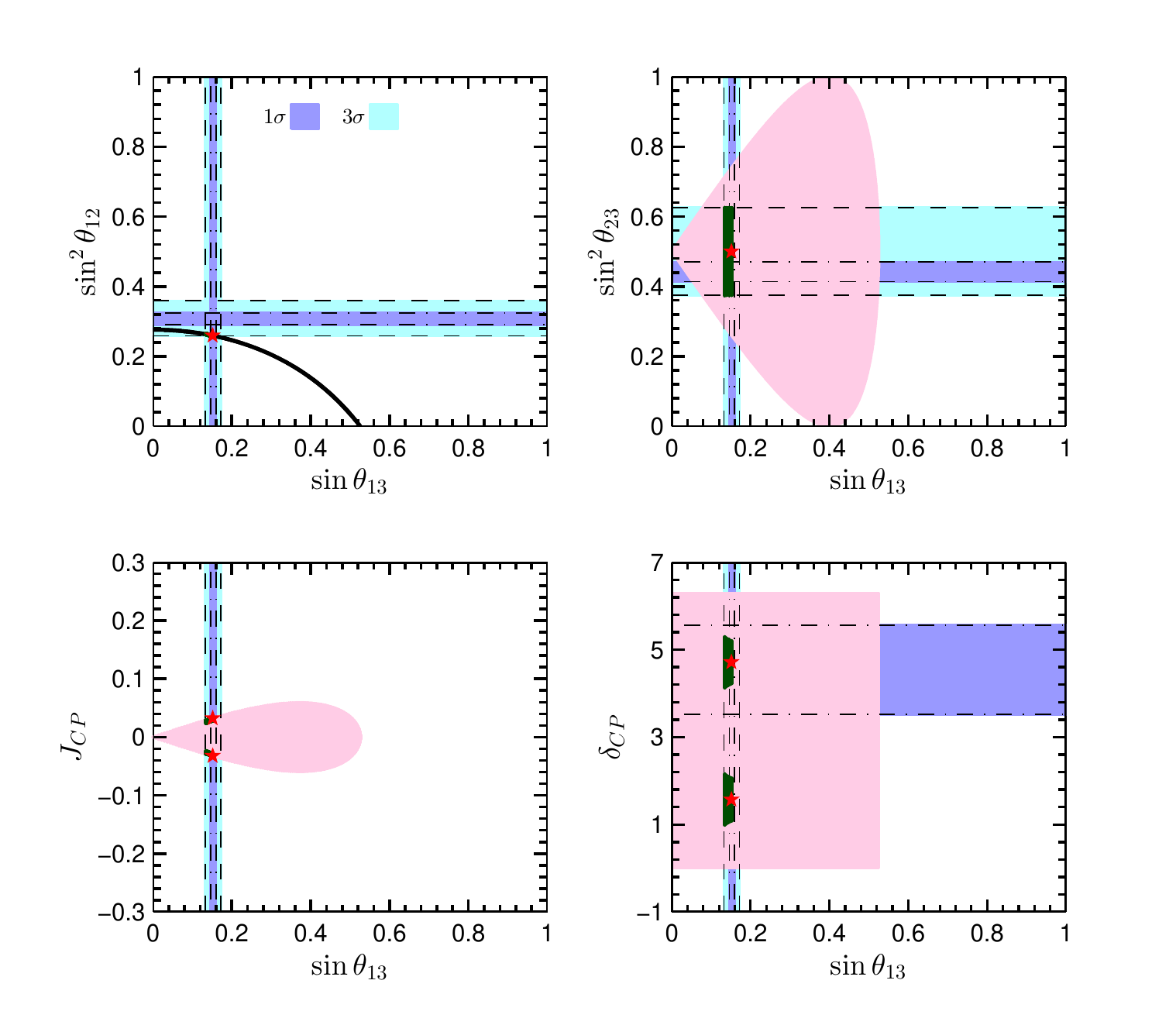}
\caption{\label{fig:deviation_Rotation1} Predictions for the mixing parameters $\sin^{2}\theta_{12}$, $\sin^{2}\theta_{23}$, $J_{CP}$ and $\delta_{CP}$ with respect to $\sin\theta_{13}$ when the remnant flavor symmetries are $G_{l}=Z^{T}_{5}$ and $G_{\nu}=Z^{S}_{2}$. The corresponding PMNS matrix is given by Eq.~\eqref{eq:PMNS_deviation_one}. The pink regions denote the possible values of the parameters when both $\theta$ and $\delta$ freely vary in the whole region of $[0, 2\pi]$. The dark green areas represent the regions allowed by the current experimental data for three neutrino mixing angles at $3\sigma$ level~\cite{Capozzi:2013csa}. The red pentagrams refer to the best fitting values of case II discussed in section \ref{sec:3.2.1}, after the generalized CP is imposed.}
\end{center}
\end{figure}

The lepton mixing matrix $U_{PMNS}$ is predicted to have one column $(-\sqrt{\frac{\kappa}{\sqrt{5}}}, \frac{1}{\sqrt{2\sqrt{5}\kappa}}, \frac{1}{\sqrt{2\sqrt{5}\kappa}})^T$, which coincides with the first column of the GR mixing. The other two columns should be orthogonal to it, and they can be obtained by making a unitary rotation of the 2nd and 3rd columns of $U_{GR}$.
\begin{equation}\label{eq:PMNS_deviation_one}
U_{PMNS}=U_{GR}U_{23}=
\left(\begin{array}{ccc}
 -\sqrt{\frac{\kappa}{\sqrt{5}}} ~&~ \sqrt{\frac{1}{\sqrt{5}\kappa}} \cos\theta ~&~ \sqrt{\frac{1}{\sqrt{5}\kappa}}  \sin\theta e^{-i \delta }\\
 \frac{1}{\sqrt{2\sqrt{5}\kappa}} ~&~ \sqrt{\frac{\kappa}{2\sqrt{5}}}\cos\theta+\frac{\sin\theta}{\sqrt{2}}e^{i\delta} ~&~ \sqrt{\frac{\kappa}{2\sqrt{5}}}\sin\theta e^{-i\delta }-\frac{\cos\theta}{\sqrt{2}} \\
 \frac{1}{\sqrt{2\sqrt{5}\kappa}} ~&~ \sqrt{\frac{\kappa}{2\sqrt{5}}}\cos\theta-\frac{ \sin\theta}{\sqrt{2}}e^{i \delta } ~&~ \sqrt{\frac{\kappa}{2\sqrt{5}}}\sin\theta e^{-i \delta }+\frac{\cos\theta}{\sqrt{2}}
\end{array}\right)\,.
\end{equation}
This form of modification to the GR mixing has been discussed in a phenomenological way in Ref.~\cite{Wang:2013wya,Varzielas:2013hga,Petcov:2014laa}. Here we show that this mixing pattern can be naturally reproduced from the $A_5$ flavor symmetry. The mixing angles can be straightforwardly extracted as follows,
\begin{eqnarray}
\nonumber &&  \sin^{2}\theta_{13}=\frac{3-\kappa}{5}\sin^{2}\theta\,, \quad  \sin^{2}\theta_{12}=\frac{2\cos^2\theta}{3+2\kappa+\cos2\theta}\,, \\
  && \sin^{2}\theta_{23}=\frac{1}{2}-\frac{\sqrt{3+4\kappa}\sin 2\theta\cos\delta  }{3+2\kappa+\cos2\theta}\,.
\end{eqnarray}
Wee see that the solar and reactor mixing angles are related by
\begin{equation}\label{eq:dv_mix_re_two}
5\cos^{2}\theta_{12}\cos^{2}\theta_{13}=2+\kappa\,.
\end{equation}
Given the $3\sigma$ ranges $1.76\times 10^{-2}\leq\sin^{2}\theta_{13}\leq2.95\times10^{-2}$ and $0.259\leq\sin^{2}\theta_{12}\leq0.359$ from global analysis~\cite{Capozzi:2013csa}, $\theta_{13}$ and $\theta_{12}$  are further constrained to be in the intervals of $1.76\times 10^{-2}\leq\sin^{2}\theta_{13}\leq2.35\times10^{-2}$ and $0.259\leq\sin^{2}\theta_{12}\leq0.263$ by this correlation. The well-known Jarlskog invariant $J_{CP}$~\cite{Jarlskog:1985ht}, which measures the size of the CP violation, is written as
\begin{eqnarray}
J_{CP}=-\frac{\sqrt{3-\kappa}}{20}\sin2\theta\sin\delta\,.
\end{eqnarray}
The Dirac CP violating phase $\delta_{CP}$ is expressed in terms of $\theta$ and $\delta$ as
\begin{equation}
\sin\delta_{CP}=-\frac{\sqrt{2}(3+2\kappa+\cos2\theta)\text{sign}(\sin2\theta)\sin\delta}{\sqrt{4(3+2\kappa)\cos2\theta+(7+8\kappa)(3+\cos4\theta)-4(3+4\kappa)\cos2\delta\sin^22\theta}}\,,
\end{equation}
In order to see how well the lepton mixing angles can be described by this mixing pattern and its prediction for $\delta_{CP}$, we perform a numerical analysis. The free parameters $\theta$ and $\delta$ are scattered in their whole allowed ranges of $0\leq\theta<2\pi$ and $0\leq\delta<2\pi$. The correlations and the possible values of the mixing parameters are plotted in Fig.~\ref{fig:deviation_Rotation1}. Furthermore, the experimental data of three mixing angles $\theta_{12}$, $\theta_{13}$ and $\theta_{23}$ at $3\sigma$ level~\cite{Capozzi:2013csa} are considered, accordingly the allowed values of the mixing parameters would generically be constrained in small regions. Here and herafter, we perform numerical analysis and present results only for normal ordering neutrino mass spectrum. The results would change a little bit for the inverted ordering case. From Fig.~\ref{fig:deviation_Rotation1}, we can read that $\sin^2\theta_{12}$ is predicted to be around 0.26, any value of $\theta_{23}$ within the $3\sigma$ range can be achieved and $\delta_{CP}$ is restricted in the range of $[0.990,2.152]\cup[4.131,5.293]$. Recalling that if both $A_5$ family symmetry and generalized CP are imposed, as discussed in section~\ref{sec:3.2.1}, the parameter $\delta$ can only be $\pi/2$ (case II) rather than free. Note that case I is not viable. As a consequence, the Dirac CP $\delta_{CP}$ would be maximal. Therefore we conclude that the generalized CP symmetry is a quite effective method of predicting the CP violating phases.

\subsubsection{\label{sec:B.2.2}$G_{l}=Z^{T}_{5},G_{\nu}= Z^{T^{3}ST^{2}ST^{3}}_{2}$}

In this case, one column of $U_{PMNS}$ is determined to be  $(\sqrt{\frac{1}{\sqrt{5}\kappa}}, \sqrt{\frac{\kappa}{2\sqrt{5}}}, \sqrt{\frac{\kappa}{2\sqrt{5}}})^T$ which is exactly the second column of the GR mixing. The corresponding PMNS matrix can be obtained from $U_{GR}$
by multiplying $U_{13}$ from right-hand side,
\begin{equation}\label{eq:PMNS_deviation_two}
U_{PMNS}=U_{GR}U_{13}=\left(\begin{array}{ccc}
-\sqrt{\frac{\kappa}{\sqrt{5}}} \cos\theta ~&~ \sqrt{\frac{1}{\sqrt{5}\kappa}} ~&~ -\sqrt{\frac{\kappa}{\sqrt{5}}}  \sin\theta~ e^{-i \delta }\\
 \frac{\cos\theta}{\sqrt{2\sqrt{5}\kappa}}+\frac{\sin\theta}{\sqrt{2}}e^{i \delta } ~&~ \sqrt{\frac{\kappa}{2\sqrt{5}}} ~&~ -\frac{\cos\theta}{\sqrt{2}}+\frac{\sin\theta}{\sqrt{2\sqrt{5}\kappa}}e^{-i \delta } \\
 \frac{\cos\theta}{\sqrt{2\sqrt{5}\kappa}}-\frac{\sin\theta}{\sqrt{2}}e^{i \delta } ~&~ \sqrt{\frac{\kappa}{2\sqrt{5}}} ~&~ \frac{\cos\theta}{\sqrt{2}}+\frac{ \sin\theta}{\sqrt{2\sqrt{5}\kappa}}e^{-i \delta }
\end{array}\right)\,.
\end{equation}
\begin{figure}[t!]
\begin{center}
\includegraphics[width=0.98\textwidth]{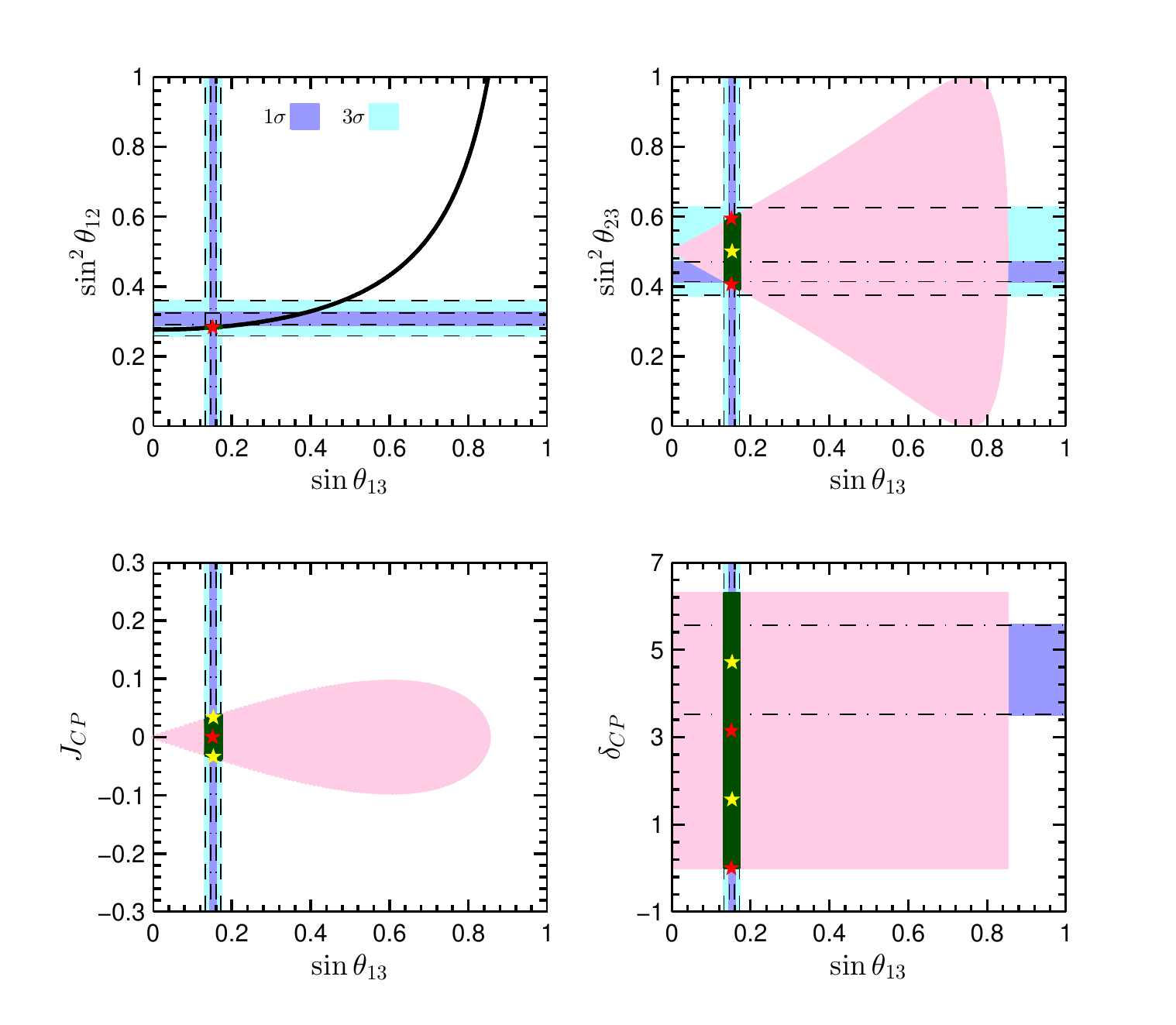}
\caption{\label{fig:deviation_Rotation2}
Predictions for the mixing parameters $\sin^{2}\theta_{12}$, $\sin^{2}\theta_{23}$, $J_{CP}$ and $\delta_{CP}$ with respect to $\sin\theta_{13}$ when the remnant flavor symmetries are $G_{l}=Z^{T}_{5}$ and $G_{\nu}= Z^{T^{3}ST^{2}ST^{3}}_{2}$. The corresponding PMNS matrix is given by Eq.~\eqref{eq:PMNS_deviation_two}. The pink regions denote the possible values of the parameters when both $\theta$ and $\delta$ freely vary in the whole region of $[0, 2\pi]$. The dark green areas represent the regions allowed by the current experimental data for three neutrino mixing angles at $3\sigma$ level~\cite{Capozzi:2013csa}. The red and yellow pentagrams denote the best fitting values of case III and case IV discussed in section \ref{sec:3.2.2}, where the generalized CP symmetry is considered. Notice that the red pentagrams almost coincides with the yellow one in the first panel, since the best fitting values of $\sin^2\theta_{12}$ and $\sin\theta_{13}$ are nearly the same in case III and case IV.}
\end{center}
\end{figure}
The lepton mixing parameters read
\begin{eqnarray}
\nonumber && \sin^{2}\theta_{13}=\frac{2+\kappa}{5}\sin^{2}\theta\,, \quad  \sin^{2}\theta_{12}=\frac{2}{3+\kappa+(1+\kappa)\cos2\theta}\,, \\
\nonumber&&  \sin^{2}\theta_{23}=\frac{1}{2}-\frac{\sqrt{2+\kappa} \sin 2 \theta \cos \delta}{3+\kappa+(1+\kappa)\cos2\theta},\quad J_{CP}=\frac{\sqrt{2+\kappa}}{20}\sin2\theta\sin\delta,\\
&&\sin\delta_{CP}=\frac{\sqrt{2(2+\kappa)}  \left(3 \kappa-2+\kappa \cos2 \theta\right)\text{sign}(\sin 2\theta)\sin\delta }{ \sqrt{(13+4\kappa) (3+\cos4\theta)+4(7+6\kappa) \cos 2 \theta-20\sin^22\theta\cos2\delta }}\,.
\end{eqnarray}
We have a relation between $\theta_{12}$ and $\theta_{13}$,
\begin{equation}\label{eq:dv_mix_re_one}
5\sin^{2}\theta_{12}\cos^{2}\theta_{13}=3-\kappa\,.
\end{equation}
The solar mixing angle $\theta_{12}$ is restricted by the observed value of $\theta_{13}$ such as $0.281\leq\sin^{2}\theta_{12}\leq0.285$ which is in the $3\sigma$ range~\cite{Capozzi:2013csa}. We display the allowed regions of the mixing angles, $J_{CP}$ and $\delta_{CP}$ in Fig.~\ref{fig:deviation_Rotation2}. No dependence of $\delta_{CP}$ on $\sin\theta_{13}$ is observed, and $\delta_{CP}$ can take any value in the whole range of $[0, 2\pi]$. However, $\delta_{CP}$ can only be conserved or maximally broken if generalized CP is considered, as shown in section~\ref{sec:3.2.2}. Note that the mixing pattern in Eq.~\eqref{eq:PMNS_deviation_two} has been discussed in Ref.~\cite{Wang:2013wya,Petcov:2014laa}.

\subsubsection{\label{sec:B.2.3} $G_{l}=Z^{T^{3}ST^{2}S}_{3},G_{\nu}=Z^{ST^{2}ST^{3}S}_{2}$}

\begin{figure}[t!]
\begin{center}
\includegraphics[width=0.98\textwidth]{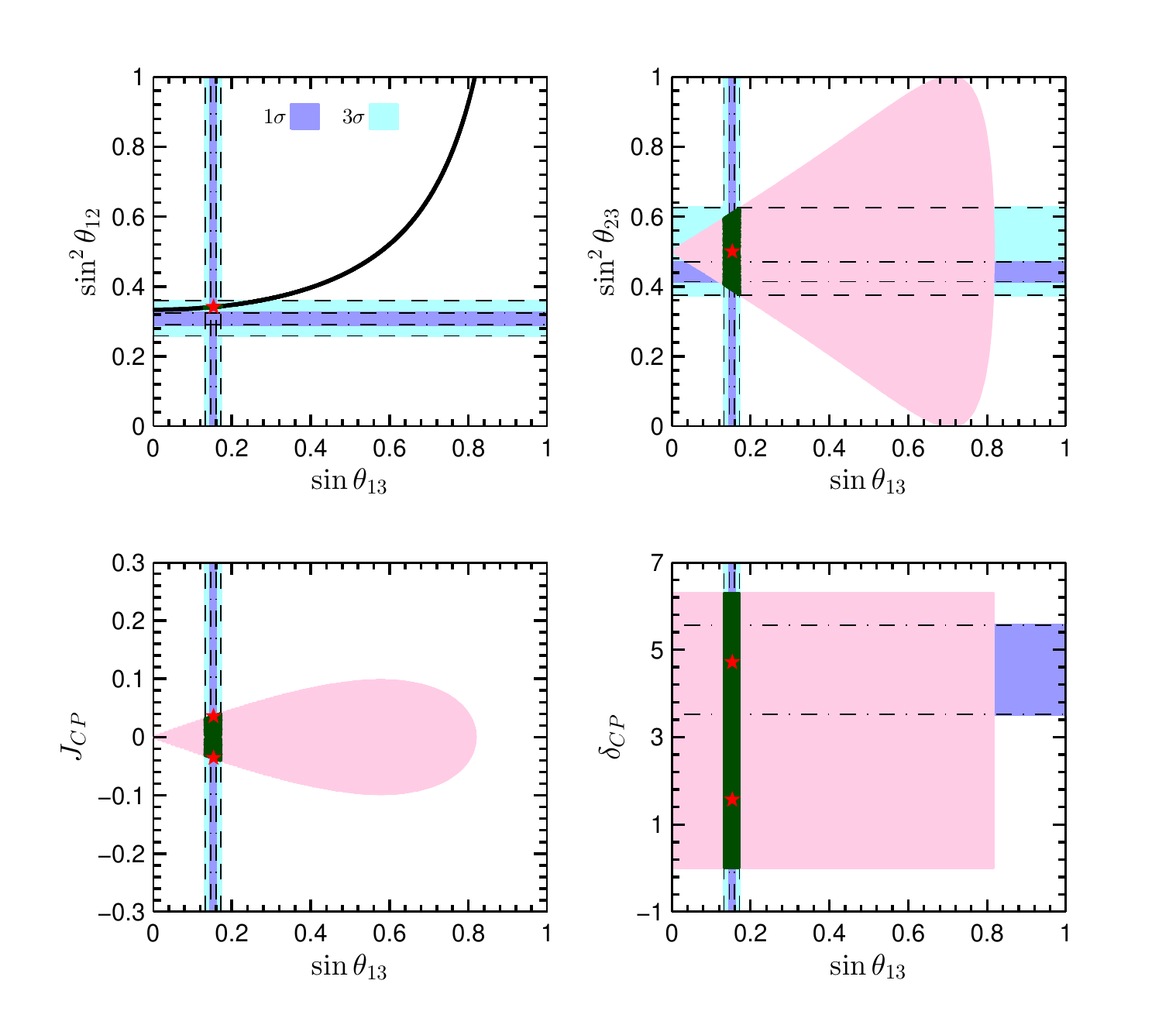}
\caption{\label{fig:deviation_Rotation3}Predictions for the mixing parameters $\sin^{2}\theta_{12}$, $\sin^{2}\theta_{23}$, $J_{CP}$ and $\delta_{CP}$ with respect to $\sin\theta_{13}$ when the remnant flavor symmetries are $G_{l}=Z^{T^{3}ST^{2}S}_{3}$ and $G_{\nu}=Z^{ST^{2}ST^{3}S}_{2}$. The corresponding PMNS matrix is given by Eq.~\eqref{eq:PMNS_deviation_three}. The pink regions denote the possible values of the parameters when both $\theta$ and $\delta$ freely vary in the whole region of $[0, 2\pi]$. The dark green areas represent the regions allowed by the current experimental data for three neutrino mixing angles at $3\sigma$ level~\cite{Capozzi:2013csa}. The red pentagrams refer to the best fitting values of case V discussed in section \ref{sec:3.2.3}, after the generalized CP is imposed.}
\end{center}
\end{figure}

The chosen remnant symmetry leads to a trimaximal column $\frac{1}{\sqrt{3}}\left(1, 1, 1\right)^{T}$, and $U_{PMNS}$ takes the form
\begin{equation}
\label{eq:PMNS_deviation_three}
U_{PMNS}=U_{DC}U_{13}=\frac{1}{\sqrt{3}}
\left(\begin{array}{ccc}
 \cos \theta-e^{i \delta } \sin \theta  & ~1~ & \cos \theta+e^{-i \delta } \sin \theta  \\
 e^{\frac{2\pi i}{3}} \cos \theta + e^{i (\frac{\pi}{3}+\delta) } \sin \theta & ~1~ &  e^{i(\frac{2\pi}{3}- \delta)}\sin\theta-e^{\frac{\pi i}{3}}\cos\theta \\
 -e^{\frac{\pi i}{3}}\cos\theta- e^{i(\frac{2\pi}{3}+\delta)} \sin \theta  & ~1~ & e^{\frac{2\pi i}{3}}\cos \theta-e^{i (\frac{\pi}{3}-\delta) } \sin \theta  \\
\end{array}\right)\,.
\end{equation}
Such a mixing pattern as a minimal modification to the tri-bimaximal has been widely discussed in the literature~\cite{Wang:2013wya,Petcov:2014laa,Albright:2008rp}, and it can also be naturally reproduced from simple flavor symmetries $A_4$~\cite{Ding:2013bpa,King:2011zj} and $S_4$~\cite{Feruglio:2012cw,Ding:2013hpa,King:2011zj}. The predictions for the lepton mixing parameters are given by
\begin{eqnarray}
\nonumber && \sin^{2}\theta_{13}=\frac{1}{3} (1+\sin 2 \theta\cos \delta ), \qquad \sin^{2}\theta_{12}=\frac{1}{2-\sin2\theta\cos\delta}\,,  \\
\nonumber&& \sin^{2}\theta_{23}=\frac{1}{2}-\frac{\sqrt{3} \sin2\theta\sin\delta}{4-2\sin2\theta\cos\delta},\qquad  J_{CP}=-\frac{\cos2\theta }{6\sqrt{3}},\\
&&\sin\delta_{CP}=\frac{-\sqrt{2}\cos2\theta(2-\sin2\theta\cos\delta)}{\sqrt{(1-\sin 2\theta\cos\delta) (5+3\cos4\theta+2\sin^32\theta\cos3\delta)}}\,.
\end{eqnarray}
As expected, the following relation is fulfilled,
\begin{equation}\label{eq:dv_mix_re_three}
3\sin^{2}\theta_{12}\cos^{2}\theta_{13}=1\,,
\end{equation}
which generically holds true for trimaximal mixing. Inserting the experimental bound of $\theta_{13}$~\cite{Capozzi:2013csa}, we obtain $0.339\leq\sin^2\theta_{12}\leq0.343$. A numerical analysis similar to previous cases is performed, as shown in Fig.~\ref{fig:deviation_Rotation3}. We see that no prediction for $\delta_{CP}$ can be made. Recalling that $\delta_{CP}$ would be constrained to be maximal by generalized CP symmetry discussed in section~\ref{sec:3.2.3}.

\subsubsection{\label{sec:B.2.4} $G_{l}=K^{(ST^{2}ST^{3}S,TST^{4})}_{4},G_{\nu}=Z^{S}_{2}$}

\begin{figure}[t!]
\begin{center}
\includegraphics[width=0.98\textwidth]{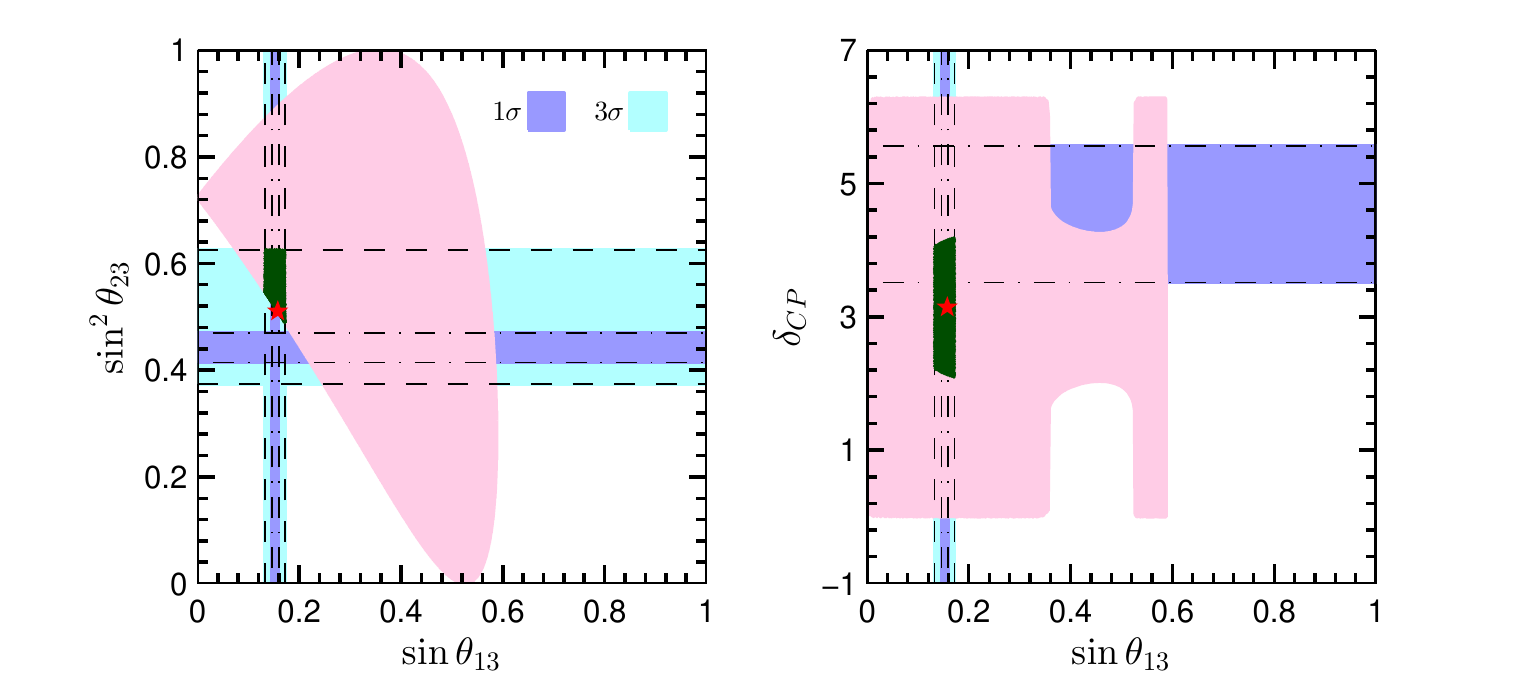}
\caption{\label{fig:deviation_Rotation5} The correlations of $\sin^{2}\theta_{23}$ and $\delta_{CP}$ with respect to $\sin\theta_{13}$, where the PMNS matrix arises from an exchange of the second and third rows in the pattern in Eq.\eqref{eq:PMNS_deviation_four}. The pink regions denote the possible values of the parameters when both $\theta$ and $\delta$ freely vary in the whole region of $[0, 2\pi]$. The dark green areas represent the regions allowed by the current experimental data for three neutrino mixing angles at $3\sigma$ level~\cite{Capozzi:2013csa}. The red pentagrams refer to the best fitting values of case VII with $\theta_{23}(\theta_{bf})>45^{\circ}$ discussed in section \ref{sec:3.2.4}, after the generalized CP is imposed.}
\end{center}
\end{figure}

One column is fixed to be $\frac{1}{2}\left(\kappa, -1, \kappa-1\right)^{T}$ in this case, and it can only be the first column of the PMNS matrix in order to be consistent with the experimental data. As a result, $U_{PMNS}$ is of the form
\begin{equation}
\label{eq:PMNS_deviation_four}
 U_{PMNS}=U_{RC}U_{23}=\frac{1}{2}\left(\begin{array}{ccc}
 \kappa  &~ -\cos\theta-\kappa^{-1}\sin\theta e^{i\delta} ~&~  \kappa^{-1}\cos\theta-\sin\theta e^{-i\delta}  \\
 -1 &~ -\kappa^{-1}\cos\theta-\kappa\sin\theta e^{i \delta} ~&~  \kappa \cos\theta-\kappa^{-1}\sin\theta e^{-i\delta}\\
 \kappa-1  &~ \kappa\cos\theta-\sin\theta e^{i\delta }  ~&~ \cos\theta+\kappa\sin\theta e^{-i\delta}
\end{array}\right)\,.
\end{equation}
\begin{figure}[t!]
\begin{center}
\includegraphics[width=0.98\textwidth]{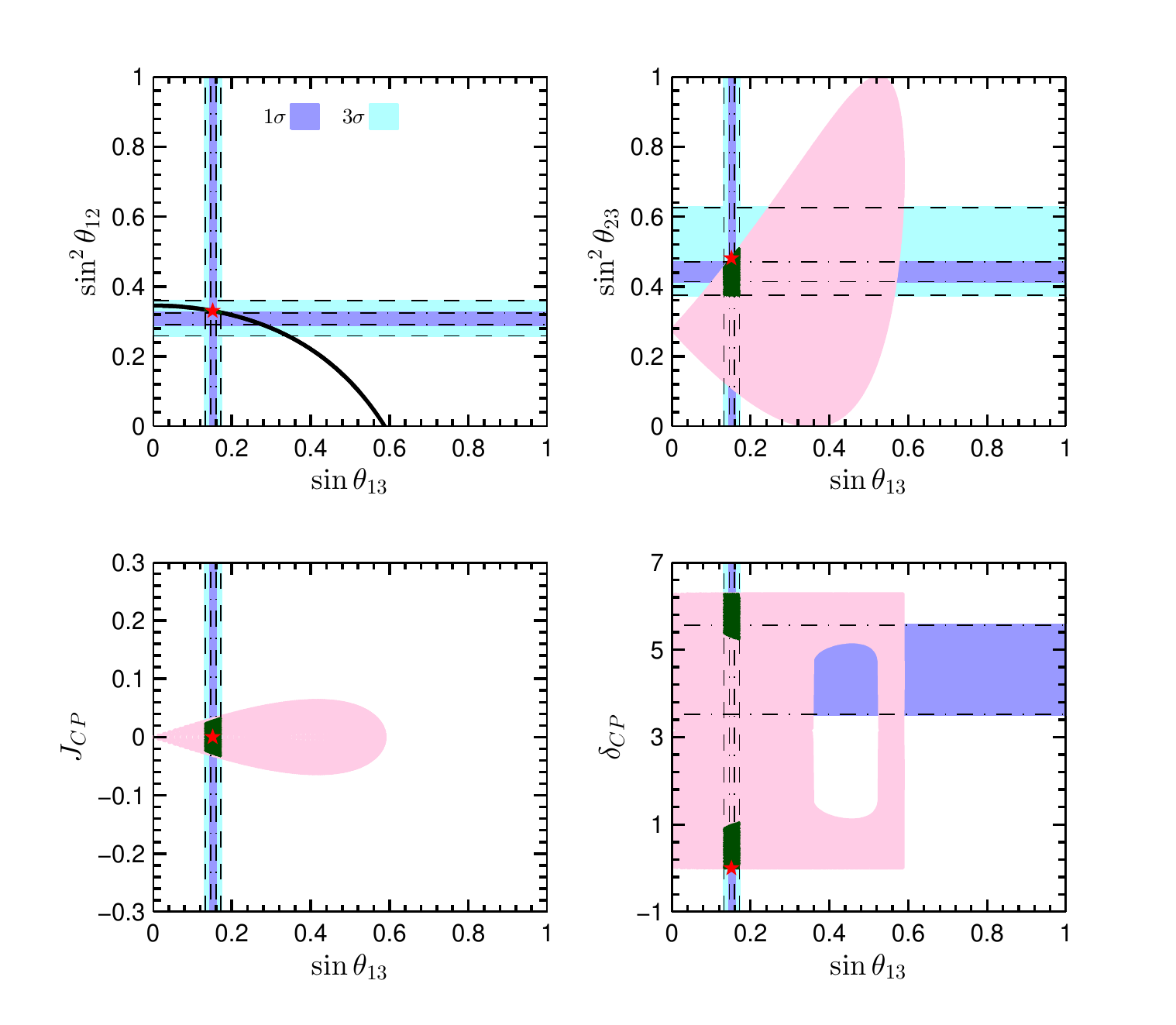}
\caption{\label{fig:deviation_Rotation4}
Predictions for the mixing parameters $\sin^{2}\theta_{12}$, $\sin^{2}\theta_{23}$, $J_{CP}$ and $\delta_{CP}$ with respect to $\sin\theta_{13}$ when the remnant flavor symmetries are $G_{l}=K^{(ST^{2}ST^{3}S,TST^{4})}_{4}$ and $G_{\nu}=Z^{S}_{2}$. The corresponding PMNS matrix is given by Eq.~\eqref{eq:PMNS_deviation_four}. The pink regions denote the possible values of the parameters when both $\theta$ and $\delta$ freely vary in the whole region of $[0, 2\pi]$. The dark green areas represent the regions allowed by the current experimental data for three neutrino mixing angles at $3\sigma$ level~\cite{Capozzi:2013csa}. The red pentagrams refer to the best fitting values of case VII with $\theta_{23}(\theta_{bf})<45^{\circ}$ discussed in section \ref{sec:3.2.4}, after the generalized CP is imposed.}
\end{center}
\end{figure}
Then the three mixing angles read
\begin{eqnarray}
\label{eq:deviate_case4_mixing_angles}
\nonumber && \sin^{2}\theta_{13}=\frac{\kappa-1}{8}(\sqrt{5}-\cos2\theta-2\sin2\theta\cos\delta)\,,\\
\nonumber && \sin^{2}\theta_{12}=\frac{3-\kappa+(\kappa-1)(\cos2\theta+2\sin2\theta\cos\delta)}{5+\kappa+(\kappa-1)(\cos2\theta+2\sin2\theta\cos\delta)}\,,\\
&&  \sin^{2}\theta_{23}=\frac{3+\sqrt{5}\cos2\theta-2\sin2\theta\cos\delta}{5+\kappa+(\kappa-1)(\cos2\theta+2\sin2\theta\cos\delta)}\,.
\end{eqnarray}
A relation between $\theta_{12}$ and $\theta_{13}$ follows immediately
\begin{equation}
\label{eq:dv_mix_re_four}
4\cos^{2}\theta_{12}\cos^{2}\theta_{13}=1+\kappa\,.
\end{equation}
The solar mixing angle is predicted as $0.326\leq\sin^2\theta_{12}\leq0.334$ which is in the experimental $3\sigma$ bound~\cite{Capozzi:2013csa}. The Jarlskog invariant $J_{CP}$ is given by
\begin{equation}
J_{CP}=-\frac{1}{16}\sin2\theta\sin\delta\,.
\end{equation}
The Dirac CP violating phase $\delta_{CP}$ is
{\small
\begin{equation}
\sin\delta_{CP}=\frac{-\sqrt{2\kappa-3}(6\kappa+1+\cos2\theta+2\sin2\theta\cos\delta)\sin2\theta\sin\delta}
 {\sqrt{\left[5-(\cos2\theta+2\sin2\theta\cos\delta)^2\right](\sqrt{5}-\cos2\theta+2\sin2\theta\cos\delta)(3+\sqrt{5}\cos2\theta-2\sin2\theta\cos\delta)}}\,.
\end{equation}}
The numerical results are displayed in Fig.~\ref{fig:deviation_Rotation4}.  We see that $\delta_{CP}$ is predicted to be in the range of $[0,1.043]\cup[5.240,2\pi]$, and the atmospheric mixing angle $\theta_{23}$ mostly is less than $45^{\circ}$ (i.e., in the first octant) in order to be compatible with experimental data of $\theta_{13}$. The scenario of $\theta_{23}$ in the second octant can be achieved, if the second and third rows of the PMNS matrix in Eq.~\eqref{eq:PMNS_deviation_four} are exchanged. Then the predictions for the solar and reactor mixing angles in Eq.~\eqref{eq:deviate_case4_mixing_angles} remain, $\delta_{CP}$ becomes $\pi+\delta_{CP}$, and $\theta_{23}$ becomes $\pi/2 -\theta_{23}$. Consequently both $J_{CP}$ and $\sin\delta_{CP}$ change into their opposite, and the expression of $\sin^{2}\theta_{23}$ in Eq.~\eqref{eq:deviate_case4_mixing_angles} is replaced by
\begin{equation}
\sin^{2}\theta_{23}=\frac{\kappa(\sqrt{5}-\cos2\theta+2\sin2\theta\cos\delta)}{5+\kappa+(\kappa-1)(\cos2\theta+2\sin2\theta\cos\delta)}\,.
\end{equation}
The predictions for $\sin^2\theta_{23}$ and $\delta_{CP}$ versus $\sin\theta_{13}$ are shown in Fig.~\ref{fig:deviation_Rotation5}. As expected, $\theta_{23}$ is really larger than $45^{\circ}$ to accommodate the measured values of $\theta_{13}$, and the CP phase $\delta_{CP}$ is in the range of $[2.099,4.185]$. Notice that generalized CP would constrain $\delta_{CP}$ to be trivial, as studied in section~\ref{sec:3.2.4}. In summary, if a single $Z_2$ subgroup of the $A_5$ flavor symmetry is preserved by the neutrino mass matrix, only one column of the PMNS matrix can be determined and agreement with experimental data can be achieved. However, the Majorana phases cannot be predicted by flavor symmetry, and the Dirac phase $\delta_{CP}$ is constrained very weakly. On the other hand, if we extend the $A_5$ family symmetry to include the generalized CP, $\delta_{CP}$ is predicted to be trivial or maximal and Majorana phases are trivial.

\end{appendix}

\end{document}